%% file: medicaid_hte_paper.tex
\documentclass[letterpaper,12pt]{article}
\usepackage[english]{babel}
\usepackage[utf8]{inputenc}
\usepackage[left=2.5cm,right=2.5cm,top=2.5cm,bottom=2.5cm]{geometry}
\usepackage{enumerate}
\usepackage{graphicx,subcaption}
\usepackage{amssymb,amsthm,amsmath}
\usepackage{istgame}
\usepackage{epstopdf}
\usepackage{latexsym}
\theoremstyle{plain}
\usepackage{ragged2e}
\usepackage{pdflscape}

\def\esp{\mathbb{E}}

\usepackage[title]{appendix}
\usepackage{tikz}
\usetikzlibrary{trees,calc,arrows.meta,positioning,decorations.pathreplacing,bending}
\usepackage{array}
\newcolumntype{H}{>{\setbox0=\hbox\bgroup}c<{\egroup}@{}}
\newcolumntype{P}[1]{>{\centering\arraybackslash}p{#1}}
\usepackage{footnote}
\usepackage[flushleft]{threeparttable}
\usepackage{cancel,color}
\usepackage[autostyle]{csquotes} %For proper quotation marks
\usepackage[font=bf]{caption} %Bold Table Caption
\usepackage{booktabs,multirow,stackengine} % Lines in Tables and other features
\usepackage{setspace}
\usepackage{har2nat} % loads "natbib" automatically
\usepackage[linguistics]{forest}
\usepackage[hyphens]{url}
\usepackage{hyperref}
\definecolor{darkblue}{rgb}{0.0,0.0,0.3}
\hypersetup{colorlinks,breaklinks=true,linkcolor=darkblue,urlcolor=darkblue,anchorcolor=darkblue,citecolor=darkblue}
\usepackage{titlesec}
\titlelabel{\thetitle.\quad}

\usepackage{bm}
\usepackage{bbm}
\usepackage{newtxtext,newtxmath}
\usepackage{nicefrac}
\usepackage[hang,flushmargin]{footmisc}
\usepackage{comment}
\usepackage[detect-all]{siunitx}
\usepackage{float}
\def\citeapos#1{\citeauthor{#1}'s \citeyear{#1}}

\begin{document}

%================================== TITLE PAGE =================================

\begin{titlepage}
  \title{\Large \textbf{Who Increases Emergency Department Use?
      New Insights from the Oregon Health Insurance Experiment%
      \footnote{We thank Amy Finkelstein, Anirban Basu, Paul Goldsmith-Pinkham, Tymon Słoczy\'nski, Timothy Layton, Jacob Wallace, Sherri Rose, Jason Fletcher, Carlos Lamarche, Mary Olson, Makayla Lavender, Prithvijit Mukherjee,
        Samuel Asare, Nicholas A. Wright, Anthony Strittmatter, Michael C. Knaus, and seminar/conference participants
        at Tulane University, University of Washington, University of Kentucky, Institute for Research on Poverty (University of Wisconsin-Madison), Swiss Health Economics Workshop, and The American Society of Health Economists (ASHEcon) for helpful comments.}}
    \vspace{1.5cm}}
  \author{%
    Augustine Denteh%
    \footnote{Correspondence to Department of Economics, Tulane University,
      6823 St.\ Charles Avenue, New Orleans, LA 70118;
      \href{mailto:adenteh@tulane.edu}{adenteh@tulane.edu}.}
    \and
    Helge Liebert%
    \footnote{Department of Economics, University of Zurich,
      Schönberggasse 1, 8001 Zürich, Switzerland;
      \href{mailto:helge.liebert@econ.uzh.ch}{helge.liebert@econ.uzh.ch}.}
  }%
  \date{\today}
  \maketitle

%=================================== ABSTRACT ==================================

  %\centering \large{Please do not cite or circulate.}
  \begin{singlespace}
    \begin{abstract}
      \noindent
      % 100 words for aer/aej
We provide new insights regarding the headline result that Medicaid increased emergency department (ED) use from the Oregon experiment. We find meaningful heterogeneous impacts of Medicaid on ED use using causal machine learning methods. The individualized treatment effect distribution includes a wide range of negative and positive values, suggesting the average effect masks substantial heterogeneity. A small group---about 14\% of participants---in the right tail of the distribution drives the overall effect. We identify priority groups with economically significant increases in ED usage based on demographics and previous utilization. Intensive margin effects are an important driver of increases in ED utilization.
      \vspace{0.5cm} 
      \noindent \textbf{JEL No:} H75, I13, I38
    \end{abstract}
  \end{singlespace}

\end{titlepage}

%\tableofcontents
%\listoffigures
%\listoftables

%\setstretch{2} %For spacing
\doublespacing
\pagenumbering{arabic} %restarts page numbering from 1

%==================================== PAPER ====================================

\section{Introduction}\label{intro}

% motivation
The finding that Medicaid increased emergency department (ED) utilization in the 2008 Oregon Health Insurance Experiment drew widespread national attention \cite{taubman2014medicaid}. Economic theory predicts that health insurance coverage reduces the out-of-pocket cost of care, leading to increased ED use. However, Medicaid may reduce ED use if people substitute primary care services for ED visits \citep{sommers2017health}. Empirically, the positive effect of Medicaid on ED use in Oregon contradicts most quasi-experimental studies that find a reduction in ED utilization following health insurance expansions \cite{chen2011perspective,miller2012effect,sommers2016changes,chou2020medicaid,giannouchos2022association}.\footnote{\citet{garthwaite2019all} is a notable exception that finds that the Affordable Care Act's Medicaid expansion increased ED use for ``deferrable'' conditions.} Recently, \citet{kowalski2021reconciling} attempts to reconcile the contradictory findings in the Oregon experiment and the Massachusetts reform. Nonetheless, the effect of Medicaid on ED use, especially for non-urgent conditions, remains a crucial consideration in health insurance expansions because of the continued rise in ED visits, the declining number of emergency departments, and the effects of ED crowding on health \citep{tang2010trends, moore2020costs,woodworth2020swamped,sabbatini2022medicaid}.

% research question
This paper provides new insights into how Medicaid affects ED utilization when we go beyond the \emph{average} effect. Using the Oregon experiment, we estimate the heterogeneous impacts of Medicaid on ED use and characterize those driving the positive average effects. In 2008, Oregon randomly assigned the opportunity to apply for spots in its Oregon Health Plan Standard (OHP Standard). Unlike OHP Plus, which serves Oregon's typical Medicaid population, OHP Standard was newly offered to a group of uninsured adults who were categorically ineligible for Medicaid under federal guidelines \cite{allen2013oregon}. The Medicaid expansion increased utilization for most types of ED visits, including those not requiring immediate care, for up to two years \cite{taubman2014medicaid,finkelstein2016effect}. We do not revisit the average impacts of Medicaid that are well-documented in this literature. Instead, we extend the literature by investigating how Medicaid's effect on ED use varies by covariates, which helps to isolate the risk factors that matter for increased utilization. 

%contributions

% first contribution
This paper makes three salient contributions. First, we nonparametrically estimate the heterogeneous treatment effects of Medicaid on ED utilization using causal machine learning methods. We estimate various types of heterogeneous effects, including \emph{individualized} treatment effects that condition on individual covariate values and higher-level \emph{group} average treatment effects using generalized random forests \cite{athey2019generalized}. Previous studies report a small number of subgroup effects for a subset of the Oregon experiment who completed a follow-up survey \citep{taubman2014medicaid}. Subgroup analysis can miss interesting patterns of heterogeneity due to non-linear interactions between multiple covariates \cite{heckman1997making,angrist2004treatment,deaton2010instruments}. In addition to the need to pre-specify the covariates, subsample heterogeneity analysis is prone to multiple hypothesis testing problems \citep[e.g.][]{List2019,Young2019}. Considering that Oregon expanded Medicaid to a population not typically covered by it, insights from examining its heterogeneous effects can help interpret its findings and inform policy-making. 
%Our results are consistent with the fact that the average treatment effect does not represent the individualized treatment effect in the population when there is substantial effect heterogeneity \cite{heckman1997making,angrist2004treatment,deaton2010instruments}.

% second contribution
Second, we document the risk factors for increasing ED use upon Medicaid coverage, opening the doorway for identifying priority groups for targeted policy interventions. Unlike \citet{kowalski2021reconciling}, we do not attempt to reconcile Oregon's ED finding with the Massachusetts health insurance reform results. In particular, Kowalski attributes the discrepancy to different local average treatment effects (LATE) arising from the same marginal treatment effect (MTE) function. In doing so, Kowalski estimates an MTE function for Oregon, assuming that treatment effects vary linearly with the fraction treated. The MTE function shows that treatment effects are positive for compliers but negative for never-takers. \citet{kowalski2021reconciling} then extrapolates the estimated MTE function to Massachusetts, finding that ED utilization is predicted to decrease for Massachusetts compliers comparable to a subset of Oregon never-takers. However, a related study, \citet{marx2020sharp}, shows that a different set of distribution assumptions within the same MTE framework predicts a positive treatment effect for Oregon's never-takers. Nonetheless, a key contribution of \citet{kowalski2021reconciling} is showing that treatment effect heterogeneity is crucial for understanding Medicaid's impact on ED use and transporting those results to new environments. Given the importance of heterogeneity, our paper complements the above studies by directly linking effect heterogeneity in Oregon to several observed characteristics in a data-driven manner without making additional assumptions beyond those needed to recover the LATE. Doing so sheds light on the drivers of Medicaid's positive average impact on ED use, providing new insights into the complex relationship between insurance coverage and ED use. 

%third contribution
Third, we discuss the potential to leverage the heterogeneous effects of Medicaid to propose simple decision rules to identify enrollees for targeted educational/outreach interventions. Reducing ``unnecessary'' ED use (i.e., ED use for non-emergent conditions) is a high priority policy consideration for Medicaid policymakers. This is partly because one-third of ED visits are preventable and it costs up to five times less to treat the same health problems in a doctor's office \cite{adams2013emergency,giannouchos2022association}. As such, policymakers are experimenting with various policies to curb unnecessary ED use among Medicaid recipients. For instance, as of 2020, about 14 states use blunt tools such as higher copayments to dissuade non-emergency ED use.%
\footnote{See list of states at \href{https://www.kff.org/health-reform/state-indicator/premium-and-cost-sharing-requirements-for-selected-services-for-medicaid-expansion-adults/?currentTimeframe=0&sortModel={"colId":"Location","sort":"asc"}}{https://www.kff.org}.}
Evidence of Medicaid's heterogeneous effects is essential for exploring targeted policy options (e.g., outreach programs) for reducing non-emergency ED use. However, policymakers might need guidance on how to use such evidence. Therefore, we illustrate one way of using heterogeneous treatment effects to estimate decision rules based on recently developed policy learning algorithms \citep{athey2021policy,kitagawa2018should}. Such decision trees that optimize non-emergent ED use may be a promising way to identify enrollees at risk of greater non-urgent ED use for targeted outreach efforts.
%One advantage of such decision rules is that fairness or legal considerations can be accounted for by excluding variables prohibited by discriminatory laws from the decision rule estimation.

% detailed findings
Several broad findings emerge from our study. First, we find substantial heterogeneity in the impact of Medicaid coverage on ED visits. Individualized treatment effects span a wide range of negative and positive changes in ED utilization. Moreover, the statistically significant increases in overall ED use are concentrated in the right tail of the distribution. Specifically, for ED use on the extensive margin, a small group of people---about 14\% of recipients---with statistically significant increases in ED visits appear to drive the positive average impact of Medicaid. We find similar patterns of heterogeneity for ED use on the intensive margin. These findings persist for various types of ED visits. In some cases, we find that average null effects hide important countervailing forces---reductions in ED use by some people offset increases in utilization by others. For instance, \citet{taubman2014medicaid} find no effects on average for conditions classified as ``emergent, not preventable'' (i.e., illnesses that require immediate medical care and could not have been prevented). In this case, the individualized treatment effects show a large share of both statistically significant increases and decreases in ED use, culminating in the average null effect.

%Second, when we aggregate the individualized treatment effects, we find weaker extensive margin results in magnitude and statistical significance. The aggregate estimate indicates that Medicaid increased the probability of any ED use by 4.5 percentage points ($p$-value = 0.111), which is about 65 percent of the magnitude from the linear instrumental variable (IV) method. For comparison, \citeasnoun{taubman2014medicaid} find that Oregon's Medicaid expansion increased the probability of an ED visit by 7 percentage points (a 20 percent increase). We also find a smaller extensive margin impact of a statistically significant 5.7 percentage points for outpatient ED visits (90 percent of total visits), representing 70 percent of the linear IV method's effect. Our results suggest that the average impact of Medicaid on ED use poorly approximates the individualized treatment effect. 
%The aggregate heterogeneous effects suggest that the average impact of Oregon's Medicaid expansion on ED use did not primarily occur through newly insured people using ED services for the first time.
%This pattern is consistent with selection on moral hazard \cite{einav2013selection}.

Second, the heterogeneous effects of Medicaid yield insights into who drives the average impact when we aggregate the individualized effects to higher group levels. The main risk factors for increased ED use are gender, age, participation in other safety net programs, and past ED use for conditions not requiring emergency care. Based on those factors, we identify four subgroups estimated to have statistically significant increases in ED use of at least twice the magnitude of the average effect. These groups are men, prior SNAP participants, younger adults under 50 years, and those with pre-lottery ED use classified as primary care treatable. We do not find statistically significant group-level effects defined by other pre-lottery ED use variables or information collected during the lottery sign-up. Overall, the results suggest that the increases in ED use occur primarily for newly insured people with a history of using the emergency department for conditions that do not require it and could be treated in primary care settings. 

%The importance of previous ED use for effect heterogeneity suggests a central role of intensive margin effects. Much of the overall positive Medicaid effect stems from increases in ED use by those accustomed to using the emergency department for their health care needs rather than people initiating ED use after obtaining coverage. Notably, the increases in ED use occur primarily for newly insured people with a history of using the emergency department for conditions that do not require it.

% I RE-WROTE THIS SECTION BUT STILL DECIDED TO COMMENT IT OUT 
%Third, we find noteworthy differences in subgroup effects between the nonparametric and linear IV methods, suggesting a cautionary tale on ad hoc subgroup exploration of heterogeneity. While the forest-based and linear IV methods yield similar treatment effects for some subgroups, other subgroup effects identified by the latter do not show up in the former case. This is likely because the forest-based method is able to better account for the non-linear ways in which multiple variables may drive heterogeneous treatment effects. For instance, the linear IV method finds significant effects for subgroups defined by lottery list variables collected during the sign-up process. \textcolor{red}{Given the same set of covariates for exploring heterogeneity}, the forest-based method shows no differential effects by those lottery list variables. Our results suggest that the nonparametric approach is more conservative at identifying group-level effects since it flexibly targets effect heterogeneity.

Finally, we illustrate one way to use the heterogeneous effects to estimate decision rules identifying those predicted to increase non-emergent ED use. Based on a small set of covariates, the policy learning algorithm identifies two groups of people at risk of increased ED use. The first group consists of people with minimal prior outpatient ED use but who did so for non-emergent conditions. The second group is those with higher previous outpatient ED visits who rarely use the ED for conditions requiring inpatient care. Policymakers can use those assignment rules to prioritize outreach to new recipients at risk of unnecessary ED use. While our decision rule reasonably identifies those unlikely to use ED care for emergencies, we caution that this exercise is illustrative. Policymakers must tailor policy-relevant decision rules to meet their objectives subject to legal and other practical constraints. 

The rest of the paper is organized as follows. Section~\ref{data} describes the context of Oregon's health insurance experiment and the data. Section~\ref{methods} presents the causal machine learning framework. Section~\ref{results} presents and discusses the results. Section~\ref{conclusion} concludes. Additional results are collected in Appendices \ref{appendixA} and \ref{appendixB}.

\section{The Oregon Health Insurance Experiment}\label{data}

\subsection{Background}

We provide a brief overview of the essential features of the Oregon Health Insurance Experiment. Previous studies contain detailed institutional information \cite{finkelstein2012oregon,baicker2013oregon,allen2013oregon,taubman2014medicaid,finkelstein2016effect}. Oregon split its Medicaid program into two due to a budget shortfall in 2003. The first program (OHP Plus) served categorically eligible people under federal rules---low-income income children, pregnant women, blind or disabled people, and Temporary Assistance to Needy Families (TANF) recipients. The second program (OHP Standard) served low-income and non-disabled adults not categorically eligible for the OHP Plus program. Oregon's philosophy was to provide more people with coverage for fewer services rather than limiting participation. Thus, OHP Standard covered fewer services and came with higher premium and cost-sharing requirements, resulting in a decline in enrolment \cite{allen2013oregon}. The resulting attrition in OHP Standard participation led to an accumulated budgetary surplus by 2007. Then, Oregon decided to expand its OHP Standard program by offering about 10,000 spots using a lottery to solicit or invite applications.

Lottery winners applied for Medicaid, and those determined to be eligible received coverage. The main eligibility criteria were as follows---being 19–64 years of age; being an Oregon resident who is not otherwise eligible for public insurance; being a U.S. citizen or legal immigrant; being uninsured for the previous six months, and having income below the federal poverty level with assets not exceeding \$2,000. Of the 89,824 individuals who signed up for the lottery, 35,169 won, and about 30\% of the winners enrolled in Medicaid.

When interpreting the Oregon experiment results, the low-income, uninsured population served by OHP Standard should be considered. \citet{finkelstein2012oregon} elaborates the difficulties with extrapolating Oregon's finding to other Medicaid expansions. The population served by OHP Standard is neither representative of the typical Medicaid population served by OHP Plus nor the low-income uninsured population in the United States. For example, compared to the low-income U.S. population, OHP Standard's target population has more whites (84\%), fewer Blacks (2\%), is older, and reports being in worse health \cite{allen2010oregon}. Also, people voluntarily signed up for the lottery. These issues do not threaten the internal validity of the Oregon experiment. However, the inherent sample selection on observed (and unobserved dimensions) suggests a crucial role for effect heterogeneity in understanding its findings. 

\subsection{Data}
% \citep{Finkelstein2013} - data source citation. 
We use the publicly available data files from \citeasnoun{taubman2014medicaid}. The sample contains ED visit information for 24,646 individuals from 2007 to 2009. The pre-lottery period spans January 1, 2007, through March 9, 2008. The study period is 18 months from the earliest notification date on March 10, 2008, through September 30, 2009. Medicaid coverage is constructed from state administrative records and defined as any receipt during the study period.

We analyze fourteen ED visit measures. These variables measure overall ED use and three categories of ED visits. The first category groups ED use by hospital admission, including outpatient and inpatient ED visits. The second category groups ED visits by the time of occurrence and consists of those occurring during on-time (7 am to 8 pm on Monday to Friday) and off-time hours (nights and weekends).

The final category groups ED visits by whether they required immediate care or not. We adopt the classification of ED visits based on the primary ICD-9 diagnosis code \cite{taubman2014medicaid,billings2000emergency}. While we refer the reader to \citet{taubman2014medicaid} for a detailed description, the algorithm assigns to each ED visit a probability that it is one of four types. The first type---emergent, non-preventable---includes unpreventable illnesses requiring immediate ED care (e.g., heart attacks and nonspecific chest pain). The second type---emergent, preventable---includes ED visits that require immediate ED care but are avoidable (e.g., asthma attacks and urinary tract infections). The third type---primary care treatable---includes ED visits requiring immediate care but not through the emergency department (e.g., sprains, strains, and abdominal pain). The final type---non-emergent---contains ED visits that do not require immediate care (e.g., headaches and back problems). Each visit is assigned a probability of being in all four categories. The number of visits for each type is then obtained by summing the assigned probabilities across all visits \citep{taubman2014medicaid}. We do not analyze unclassified visits not assigned to any of the above four categories. 
%For instance, the number of emergent, non-preventable ED visits is the sum of probabilities that each visit is of this type across all visits.

We analyze both the binary and continuous versions of these types of ED visits. While \citet{taubman2014medicaid} only analyzes the continuous versions, we create their binary counterparts as follows. Since the number of ED visits for each type is computed as a sum of probabilities, we conservatively classify the individual as having only one type of visit for people with one ED visit, which is set equal to the visit type with the highest probability. For those with two ED visits, we classify the individual as having at most two types of visits, that is, visit types with the top two highest probabilities. We classify the individual as having at most three visit types for those with three ED visits, with the types being those with the top three highest probabilities. Finally, we classify individuals with four or more ED visits as having all visit types with non-zero assigned probabilities.

%\footnote{The measures used in the original analysis are sums of probabilities, not discrete counts.}
%Here is a more detailed description of the types from the Appendix to the Science paper.
%The categories are: non-emergent cases where care was not required within 12 hours (e.g. a toothache), primary care treatable cases where care was needed within 12 hours but could be provided in a primary care setting (e.g. a lumbar sprain), emergent, preventable cases that the doctors judge could
%have been avoided with proper primary or ambulatory care (e.g. an asthma attack), or emergent, nonpreventable cases that could not have been avoided with primary care (e.g. a heart attack). An emergency department admission is marked as unclassified if the emergency department algorithm did not assign it a probability weight or if the primary diagnosis code was missing

All our analyses include variables needed to ensure unbiased estimation of treatment effects, such as household size and an indicator for which of the eight lottery draws between March and September 2008 the individual was assigned. Given our central focus on heterogeneous effects along observed dimensions, we include two additional groups of variables which have not previously being used to study effect heterogeneity in the Oregon experiment studies---lottery list variables and baseline characteristics measured before the lottery. We focus on these group of variables in exploring heterogeneity to ensure that they are not endogenous to Medicaid receipt.\footnote{We deliberately omit other potential  heterogeneity variables available from the pre-lottery survey. Those variables are only available for a small number of people, limiting the sample considerably and potentially subject to attrition bias.}

The lottery list contains eight variables constructed from the lottery sign-up sheet---age; sex; indicators for whether English is the preferred language for receiving materials; whether the individuals signed themselves up for the lottery; whether an individual provided a phone number on the sign-up form; whether the individuals listed their address as a P.O.\ Box; whether the individual signed up on the first day of the lottery, and the median household income in the applicant's zip code from the 2000 decennial census. We are only missing the last variable compared to the original analysis because it is not publicly available, but this does not affect the internal validity of the treatment effects.

The baseline characteristics are various pre-lottery ED use measures, participation in the Supplemental Nutrition and Assistance Program (SNAP), and receiving TANF in the pre-lottery period. We include twenty pre-lottery ED measures covering overall ED use, on-/off-time use, ED visits resulting in hospital admission, visits for each type described above, and visits for specific types of injuries and health conditions. We include this expanded list of pre-lottery ED visit information to capture historical emergency care demand. The pre-lottery SNAP/TANF participation variables, which proxy for one's experience with the social safety net system, come from state administrative data. Those variables include indicators of program receipt and the total household benefit amounts received between January 1, 2007, through the individual's notification date.

\begin{singlespace}
  \input{Tables/Tables_final/tdescriptives.tex}
\end{singlespace}

We restrict the sample to those with pairwise non-missing observations in the covariates and each outcome. Our baseline sample consists of 24,615 individuals with non-missing information on all covariates (i.e., 99.9 percent of \citeapos{taubman2014medicaid} sample). We use the baseline sample to analyze our primary outcome---whether the participant had any ED visits. The analysis samples for other outcomes are slightly smaller due to missing data. However, the difference is negligible; the smallest sample comprises 24,588 observations---only 27 fewer people relative to the baseline sample. Our results are unaffected by conditioning on a uniform sample of non-missing observations across all outcomes.

Table~\ref{tab:desc.base} presents the summary statistics for our baseline sample. The sample is 55 percent female with an average age of 40 years. About 54 percent of the sample received SNAP with an average SNAP benefit amount of $\$ 1,332$ in the pre-lottery period. TANF receipt is much lower at 2 percent of the sample with an average benefit of $\$ 96$. The sample averaged 0.77 ED visits in the pre-lottery period, with the total emergency department facility charges being $\$ 895$ on average.\footnote{The continuous measures of ED utilization have been top-coded in the public-use Oregon experiment files such that each value has as least ten observations. Nonetheless, this does not matter practically given \citeapos{taubman2014medicaid} online appendix states that ``[t]his means that for many of the number-of-visit outcomes, the publicly available data will not directly replicate the results presented in the main text and supplementary materials, although our findings are robust to the censoring we imposed in the public use data.''}

\section{Methods}\label{methods}

Our analysis of the impacts of Medicaid on ED use is twofold. First, we estimate different types of heterogeneous effects of Medicaid on ED utilization using generalized random forests (GRF). GRF is an extension of random forests that can estimate various population quantities identified as solutions to local moment conditions \cite{wager2018estimation,athey2019generalized}. In the second part of our analysis, we use the heterogeneous treatment effects to estimate assignment rules using policy learning algorithms \citep{athey2021policy,kitagawa2018should}. 
%We then describe the population that would be prioritized for coverage under optimal assignment given an objective function and constraints on program size.

\subsection{Treatment effect heterogeneity using generalized random forests}

We represent the treatment effect parameters in the potential outcomes framework. Let $D \in \{0,1\}$ denote the individual's Medicaid coverage. Also, let $Y(D)$ represent the potential outcomes associated with Medicaid receipt $(D=1)$ or otherwise $(D=0)$. Since Medicaid coverage is endogenous, we pursue instrumental variable estimation using Oregon's lottery as an instrument, denoted by $Z \in \{0,1\}$. As such, let $D(Z)$ be the potential treatment status corresponding with the instrument, with $D(1)$ being the potential treatment with winning the lottery and $D(0)$ denoting the potential treatment with losing the lottery. Suppose that the outcome is given by a structural model of the form $Y = \mu(X) +\tau(X)D +\varepsilon$, where $X$ is a vector of covariates, $ \mu(X)$ is the mean outcome, $\tau(X)$ is the heterogeneous treatment effect, and $\varepsilon$ is the error term such that $\esp[\varepsilon|X,Z]=0$. 

\citet{abadie2003semiparametric} shows that $\tau(X)$ can be expressed as the simple IV estimand, which identifies the \emph{conditional} LATE (i.e., the conditional average treatment effect for the sub-population of compliers):\footnote{See Proposition 5.3 in \citet{abadie2003semiparametric}. The identification result are based on the standard assumptions needed for the IV estimand to identify the LATE---instrument independence, exclusion of the instrument, relevance of the instrument and monotonicity \citep{imbens1994identification,angrist1996identification,abadie2003semiparametric}.}
\begin{align}\label{clate}
	\tau(X) = \frac{\esp[Y|X, Z=1] - \esp[Y|X, Z=0]}{\esp[D|X, Z=1] - \esp[D|X, Z=0]} = \esp[Y(1)- Y(0) | X, D(1)=1, D(0)=0]. 
\end{align}

%It is well-known that the individual treatment effect, $\tau_i = Y_{i}(1)- Y_{i}(0)$ is not identified because of the fundamental problem of causal inference. Under additional assumptions, we can estimate aggregate estimands such as the Average Treatment Effect (ATE).

As captured by $\tau(X)$, the heterogeneous effects with respect to covariates refers to various types of conditional average treatment effects (CATE) for compliers at different covariate levels.\footnote{For easier exposition, we refer to $\tau(X)$ as CATE in the rest of the paper although a more appropriate acronym is CLATE when performing IV estimation.} At the most granular level, $\tau(X)$ is an estimate of the individualized treatment effect (ITE) in the smallest covariate partition defined by an individual's covariate values. At higher levels, $\tau(X)$ represents the average treatment effect for well-defined subgroups. We refer to such aggregated parameters as the group average treatment effects (GATE) \cite{lechner2020heterogeneous}. Several machine learning methods for causal inference have been proposed for estimating heterogeneous treatment effects. For a recent review, see \citet{knaus2021machine}. 

In this paper, we estimate $\tau(X)$ using generalized random forests \cite{athey2019generalized}. To motivate how the forest-based estimation works, consider the case of a completely randomized experiment where we are interested in the CATE for some specified covariates. \citet{wager2018estimation} proposed \emph{causal forest} for this scenario.\footnote{Causal forests are an extension of the idea of random forests to causal inference problems \cite{breiman2001random}. Traditional random forests aim to predict an outcome based on some observed covariates. Random forests are, in turn, an average of the so-called regression trees for predicting a continuous outcome. A regression tree is built by splitting the sample into partitions to minimize the mean-squared error of predictions. After recursively partitioning the sample into a tree structure, the predictions are obtained as the outcome means in each leaf. Each tree is grown on a random sub-sample and a random subset of the variables. Since regression trees are unbiased but exhibit high variance, random forests average the final predictions across several regression trees, leading to stable forest-based predictions. For an introduction to random forests for prediction tasks, see \citet{hastie2009elements}.} The basic building block of causal forests is a causal tree. Causal trees recursively split the sample into small leaves, $L(x)$, for any given point $x$ defined by the vector of covariates. For any leaf, we can then estimate the CATE as the difference in means between treatment and control units. The causal forest is then an average of several such causal trees.

While causal forests are appealing, they do not work for observational studies such as those relying on instrumental variables. \citet{athey2019generalized} introduced generalized random forests as an adaptive nearest-neighbor estimator that solves a locally weighted version of a population moment condition. As such, causal forests can be viewed as a special type of generalized random forests. In the case of estimating the conditional LATE, generalized random forests apply because we can view $\tau(X)$ as a parameter identified by the population moment conditions: 
\begin{align}\label{momcon}
  \esp\bigg[\bigg(Y_i - \mu(X) - \tau(X)D_i \bigg)\bigg(1\ Z_i^\prime \bigg)^\prime | X_i=x\bigg] = 0.
\end{align}

To proceed with estimation, \citet{athey2019generalized} first define a similarity weight, $\alpha_i(X)$, which captures the importance of unit $i$ to estimating $\tau(X)$ at point $x$. The estimates $\big(\widehat{\tau}(x),\ \widehat{\mu}(x)\big)$ are then the solutions to the local estimating equations based on \eqref{momcon}: 
\begin{align}\label{localnn}
  \sum_{1}^{n}\alpha_i(x)\bigg[\bigg(Y_i - \widehat{\mu}(x) - \widehat{\tau}(x)D_i\bigg)\bigg(1\ Z_i^\prime \bigg)^\prime\bigg] = 0.
\end{align}
Part of the innovation in \citet{athey2019generalized} is to use random forests to estimate the weight, $\alpha_i(X)$, needed for the solution to the heterogeneous estimating equation in \eqref{localnn}. The weights are obtained by first growing $B$ trees using the standard regression tree algorithm. However, unlike traditional regression trees, the splits at the parent nodes are chosen to maximize heterogeneity in $\tau(X)$. For each tree,  the observations that fall in the same leaf as $x$ are marked and given a positive weight. The forest-based algorithm then averages over the $B$ tree-based neighborhoods to obtain $\alpha_{i}(X)$, which are then used to fit the estimating equations in \eqref{localnn}. Thus, the similarity weight $\alpha_i(X)$ measures how often an observation falls in the same neighborhood (leaf) as $x$. \citet{athey2019generalized} show that using the forest-based weights in equation \eqref{localnn} yields consistent and asymptotically normal estimates of $\tau(X)$.

We note some key implementation details. We grow all forests using 100,000 trees and randomly sample half the data to build each tree. Tree building relies on further subsample splitting based on the honesty principle, which is the idea that different sub-samples should be used to determine tree splits and estimate the within-leaf treatment effects to reduce bias \cite{wager2018estimation}. Here, we use half of the sampled data to determine the tree splits and the other half for estimation. 
%Since this may result in empty leaves, we prune these empty leaves so that each tree can handle all test points. 
For each randomly drawn subsample, we must also randomly select some covariates to build each tree. We use the default rule-of-thumb to determine the number of variables considered at each split point. That is, if $p$ is the number of covariates, then the number of variables used in each split is the minimum of $(\sqrt{p} + 20)$ and $p$. For inference, the variance of the forest-based $\widehat{\tau}(X)$ is constructed using the bootstrap of little bags, which amounts to training trees in small groups and comparing $\widehat{\tau}(X)$  predictions within and across those groups \citep{athey2019generalized,SextonLaake2009}. We cluster standard errors at the household level.

To quantify which covariates are most important for driving treatment effect heterogeneity in growing the forest, we report a variable importance ranking of all covariates used in growing the forest. This variable importance measure is based on the number of times the variable is used for splitting, weighted by the tree depth of the split. In addition to the nonparametric estimates of $\tau(X)$, we can obtain average treatment effects by averaging the individualized treatment effects. We plug the $\tau(X)$ estimates into the standard augmented inverse probability weighting (AIPW) estimator to obtain more accurate average estimates. Doing so allows the average estimates to retain the desirable doubly robust property \citep{chernozhukov2018double}. Similarly, we can obtain doubly robust GATE estimates for specific subgroups defined by one or more covariates.

\subsection{Policy learning exercises}\label{policytree}
%Academic researchers and policy leaders usually promote randomization schemes to allocate scarce resources. Using a lottery seems fair and politically feasible. However, lotteries are one of many options. The Oregon Medicaid Director, Jim Edge, summarizes the difficulty of the treatment assignment problem in the New York Times: ``[w]e thought about other options, such as should we try to pick all of the sickest people or the kids or the people with cancer or heart disease. But the Feds won’t allow that, and there’s just no way to guarantee the fairness of that. Why would cancer be more deserving than heart disease?'' \cite{yardley2008drawing}. Based on our heterogeneous effects, we discuss alternative treatment assignment schemes. These alternative assignments schemes are necessarily normative. Suppose that administrators are interested in maximizing some notion of welfare conditional on observed characteristics. Then random assignment may be inferior to a policy based on the empirical distribution of treatment effects. 
%In the context of Medicaid, we might be interested in soliciting applications from individuals in a way that minimizes undesirable utilization, such as ED visits for non-emergent conditions. Moreover, devising a rule-based allocation scheme may also improve targeting and free up funds to extend benefits to a larger pool of people. 
%(1) How should a social planner who wants to minimize unnecessary ED use optimally choose people to treat, and how would the assigned population differ from the lottery population?

Estimating heterogeneous treatment effects opens the doorway for performing interesting policy learning exercises that help understand alternative treatment assignment regimes and target program recipients with educational interventions. We perform two exercises leveraging the estimated heterogeneous effects in the intent-to-treat (ITT) scenario. We focus on the ITT context for the policy learning exercises because it more closely aligns with the practical problem of allocating Medicaid spots in Oregon's expansion of OHP Standard to a new population. Both exercises draw heavily on the literature on statistical decision rules that uses the distribution of treatment effects to design treatment assignment rules that maximize mean social welfare, typically defined as the mean post-treatment outcome \citep{manski2004statistical,dehejia2005program,hirano2009asymptotics,stoye2009minimax,stoye2012minimax,kitagawa2018should,athey2021policy}. 
%We rely on the theoretical guarantees from this literature to conduct the policy learning exercises. 

The first exercise is a theoretical one that estimates an alternative treatment assignment scheme selecting whom to solicit applications from to achieve a well-defined objective. This alternative treatment assignment scheme is necessarily infeasible because state Medicaid administrators cannot use it in practice to solicit applications. However, we estimate the alternative assignment scheme to contrast those selected under it with the observed random assignment scheme in the data. The question we seek to answer here is, how should a social planner who wants to optimize some objective function (e.g., minimize unnecessary ED use) optimally choose people to treat (i.e., solicit applications from in the Oregon context), and how would the assigned population differ from the lottery population?

In this exercise, we compute the optimal allocation scheme based on a specified objective function to solicit Medicaid applications using integer programming \citep{kitagawa2018should}. The integer programming approach lets us incorporate capacity constraints, allowing us to restrict the selected sample to equal the number of lottery winners in the Oregon experiment. We then compare the population selected based on the optimal linear program to those who won the Oregon lottery.
%We do not impose additional constraints on the budgetary costs of the program, but incorporating them is straightforward. 

The second exercise is more practical and aims to identify enrollees at risk of increased ED utilization that may be deemed unnecessary by administrators. As mentioned above, states are constantly exploring policy options to curb the non-emergent use of emergency departments by Medicaid recipients. In this case, how would a simple decision rule identifying recipients most likely to increase unnecessary ED visits look? We use \citeapos{athey2021policy} policy learning algorithm to estimate a simple depth-2 (shallow) decision tree. Administrators can use such decision rules to select and target enrollees (who already received Medicaid) for specific educational interventions.

%In other words, the social planner can only offer the opportunity to apply for coverage but not forcibly assign people to the program. Assignment rules for inviting applications for coverage may improve post-treatment outcomes relative to the lottery system. Of course, due to concerns regarding external validity, the assignment rules learned from Oregon may not directly apply to other states but rather provide a blueprint for similar expansions.

% I rewrote this footnote (was commented) but maybe we should remove it entirely? (yes, I removed entirely)
%\footnote{Given that Medicaid is an entitlement program, the state Medicaid administrator would be unable to use treatment assignment rules to determine who should receive benefits conditional on being eligible. Instead, the decision problem should be thought of as expanding a (Medicaid-like) state insurance program with a limited budget to as many people as possible from a population which is not categorically eligible for Medicaid.}

Estimating the assignment rules requires calculating doubly robust scores targeting the estimand of interest---the ITT effect of winning the lottery effect in this case. Then we choose a policy $\pi$ in the finite-depth class of policies $\Pi$ to minimize regret, i.e., the loss resulting from not choosing the ideal policy \citep{athey2021policy}. Previous work has shown that this optimization problem is equivalent to a weighted binary classification problem, effectively predicting the sign of the treatment effect, where the weights are given by the size of the conditional treatment effects \citep[e.g.][]{zhao2012estimating}.

The above policy learning exercises require three practical choices. The first is to define the post-treatment outcome or utility constituting the objective function. We use the number of non-emergent ED visits described in the data section. Such visits do not require immediate care and are most likely deemed unnecessary (i.e., not requiring service in the ED) by state administrators. %Our learned assignment rule minimizes the number of non-emergent visits.

Second, we choose which covariates to include in the decision rule estimation. We exclude variables that interfere with fairness, ethical and legal considerations. While we use the complete set of covariates to estimate the heterogeneous treatment effects, we limit the variables used to estimate the assignment rules to nine pre-lottery ED utilization variables. Specifically, we include the overall number of pre-lottery ED visits, the number of inpatient ED visits, the number of outpatient ED visits, the number of on-hour ED visits, and the number of off-hours ED visits. The remaining variables indicate the type of pre-lottery ED utilization---indicators of any pre-lottery emergent ED use, any pre-lottery primary care treatable ED use, any pre-lottery non-emergent ED use, and the pre-lottery sum of total ED facility charges. Finally, we estimate the assignment rules as decision trees of depth 2. Since the algorithm obtains the decision rule by exact (exhaustive) tree search, the required time for estimating more complex decision rules is exponential in tree depth.

\section{Results}\label{results}

\subsection{Distribution of individualized treatment effects of Medicaid on ED use}

We first present the heterogeneous impacts of Medicaid using generalized random forests as described in Section~\ref{methods}. We mainly focus on the extensive margin results of receiving Medicaid on binary indicators of ED use in the paper. We briefly refer to additional intensive margin results of Medicaid on continuous ED visit measures in Appendix \ref{appendixA}. Also, we provide the intent-to-treat heterogeneous effects of winning the lottery on ED utilization in Appendix~\ref{appendixB}.

Figure \ref{fig:any-visit-combined-unif} displays the individualized treatment effects of Medicaid on the probability of any visit, with the darker blue shade indicating statistical significance at the 10\% level. The subplots in Figure \ref{fig:any-visit-combined-unif} display results for any overall visits (panel a), any outpatient visits (panel b), and any inpatient visits (panel c). These results come from $\tau(X)$ estimates at the smallest covariate partition, with each point $X$ denoting each individual's covariate values. Table \ref{tab:qte.base} complements the graphical results by reporting selected quantiles of the empirical distribution of the ITE estimates for all outcomes. Table \ref{tab:qte.base} also reports the average effect in the first column (discussed momentarily in Section \ref{ate}) and the share of the heterogeneous effects that are non-positive in the last column. 

We find substantial and meaningful heterogeneous effects of Medicaid across all ED visit outcomes. Specifically, for any ED visits, the heterogeneous effects of Medicaid span a wide range of negative and positive values. Although the average effect of Medicaid turns to be positive, about one-third of the ITE estimates are negative, with the 25\textsuperscript{th} percentile being a reduction of 3.1 percentage points. Moreover, there is considerable variation in the magnitude of the of ITE estimates when we focus on the positive heterogeneous effects. For instance, Table \ref{tab:qte.base} shows that the median heterogeneous effect is a 7 percentage points increase, which is larger than the mean effect of 4.5 percentage points.

In addition, the statistically significant increases in ED use at the 10\% level are limited to about 14.2\% of the sample (with 8\% being significant at the 5\% level). Interestingly, there is no overlap between the statistically significant ITE estimates and the overall mean effect, further suggesting that the average effect masks heterogeneity. To put this non-overlapping finding in context, at the 10\% level, the smallest ITE estimate among those who significantly increase ED usage is 4.9 percentage points, about 10\% larger then the mean effect.\footnote{At the 5\% level, the smallest ITE estimate among those who significantly increase ED usage is 7.6 percentage points, close to 1.7 times the mean effect.}
When we incorporate  statistical significance in discussing the  heterogeneous effects, we continue to find evidence of decreases in ED use. While most of the significant ITE estimates are positive, we find that about  2\% of all individualized effects are statistically significant reductions in ED use.

\begin{singlespace}
	\begin{figure}[H]
		\vspace{-1cm}
		\centering
		\caption{Distribution of individualized treatment effects of Medicaid on the propensity of any ED visit}
		\label{fig:any-visit-combined-unif}
		\begin{minipage}{\linewidth}
			\makebox[\textwidth][c]{
				\includegraphics[width=\linewidth]{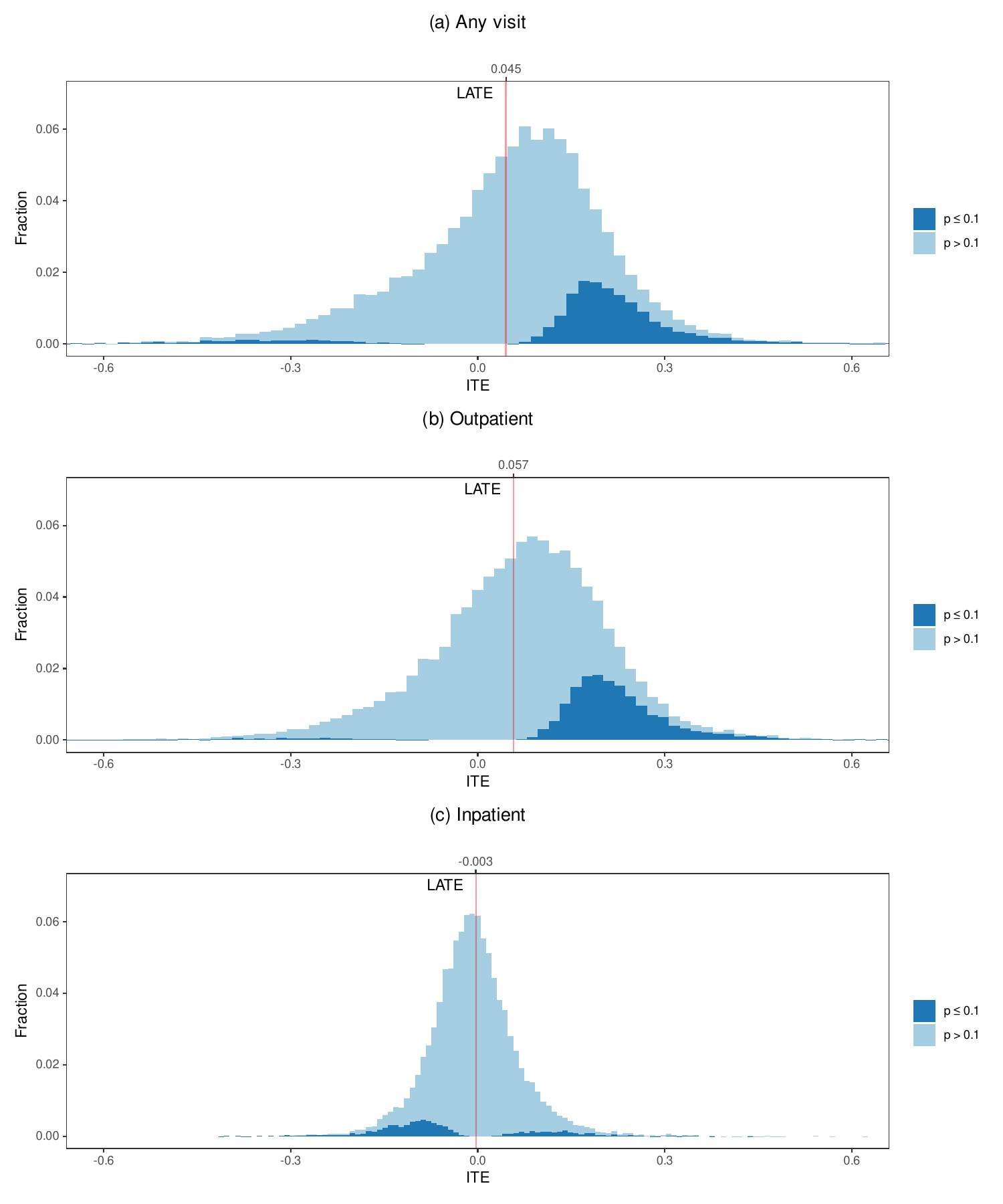}
			}
			\footnotesize
			\emph{Notes}: This figure plots the individualized treatment effects of Medicaid on any overall ED visit (panel a), any outpatient ED visits (panel b), and any inpatient ED use (panel c) based on generalized random forests. The darker shade denotes statistical significance at the 10\% level. The red vertical line indicates the local average treatment effect. The baseline sample consists of 24,615 individuals in the \citeasnoun{taubman2014medicaid} sample with non-missing information on pre-lottery emergency department utilization and SNAP/TANF receipt. The estimates displayed exclude less than half a percentile at the top and bottom of the distribution, resulting in the axes corresponding approximately to the percentile range $[0.5\%, 99.5\%]$. Bin size is chosen according to the Freedman-Diaconis rule.
		\end{minipage}
	\end{figure}
\end{singlespace}

\input{Tables/Tables_final/tqte.tex}

An important distinction when examining ED visits is whether they are for outpatient or inpatient care since the latter signifies visits for conditions severe enough to warrant hospitalization. Regarding average effects, \citet{taubman2014medicaid} find a statistically significant effect for outpatient visits and null effects for inpatient ED use. Both types of ED visits exhibit substantial effect heterogeneity, but there are some crucial differences. A comparison of panels (a) and (b) in Figure \ref{fig:any-visit-combined-unif} reveals that Medicaid's impacts on ED utilization are concentrated among outpatient visits. Moreover, the pattern of the distribution of ITE estimates for outpatient visits is similar to any overall ED visit, with negative and positive values. For instance, the 25\textsuperscript{th} percentile for outpatient visits is a reduction of 1.3 percentage points. However, the median ITE estimate of 7.8 percentage points is larger than the mean effect of 5.7 percentage points. Again, we find that the statistically significant increases in outpatient ED use are concentrated in the right tail of the distribution. About 16\% of all estimated outpatient effects are positive and statistically significant, compared to 0.8\% which are negative and significant.

For inpatient ED use, the distribution of the individualized treatment effects is relatively tighter than those for any ED or outpatient ED use. 
Notably, the distribution of the ITE estimates shows that the null effects of Medicaid on inpatient ED use mask an important heterogeneity. Panel (c) of Figure \ref{fig:any-visit-combined-unif} shows sizable proportions of statistically significant negative and positive effects, with a slightly bigger mass for the negative ITEs. Thus, Medicaid coverage appears to have significantly reduced inpatient ED use for a reasonable share of participants. In contrast to outpatient visits, statistically significant and negative effects exceed significant and positive effects (6\% vs.\ 2\%). The significant positive and negative heterogeneous effects appear to have counterbalanced and produced a null effect on average.

For the intensive margin, we obtain the same pattern of results for the number of ED visits. The corresponding graphs for the number of visits are shown in Figure \ref{fig:num-visit-combined-unif}  of Appendix~\ref{appendixA}, with Table \ref{tab:qte.base} showing selected quantiles of the ITE distribution. Again, we find meaningful heterogeneity with individualized treatment effects ranging from a reduction of 1.50 to an increase of 5.94 ED visits. Unlike the extensive margin, the median of 0.27 is slightly smaller than the mean effect of 0.35 ED visits. However, we still find a concentration of statistically significant increases in the number of ED visits in the distribution's right tail. On the intensive margin, we also see the similarities and differences mentioned above for the outpatient and inpatient care. First, the results for outpatient ED visits closely mirror the overall number of ED visits. The outpatient ITE estimates span a decrease of 1.20 to an increase of 5.83 ED visits, with the median and mean effects being 0.29 and 0.36, respectively. Second, the inpatient ITE estimates are more tightly distributed. The distribution has substantially more mass to the left, outweighing the effect of high-use individuals in the right tail, which results in a statistically insignificant mean effect.

\begin{singlespace}
  \begin{figure}
    \centering
    \caption{Distribution of individualized treatment effects of Medicaid on the propensity of any ED visit by type of condition}
    \label{fig:any-visit-type}
    \begin{minipage}{\linewidth}
      \makebox[\textwidth][c]{
        \includegraphics[width=\linewidth]{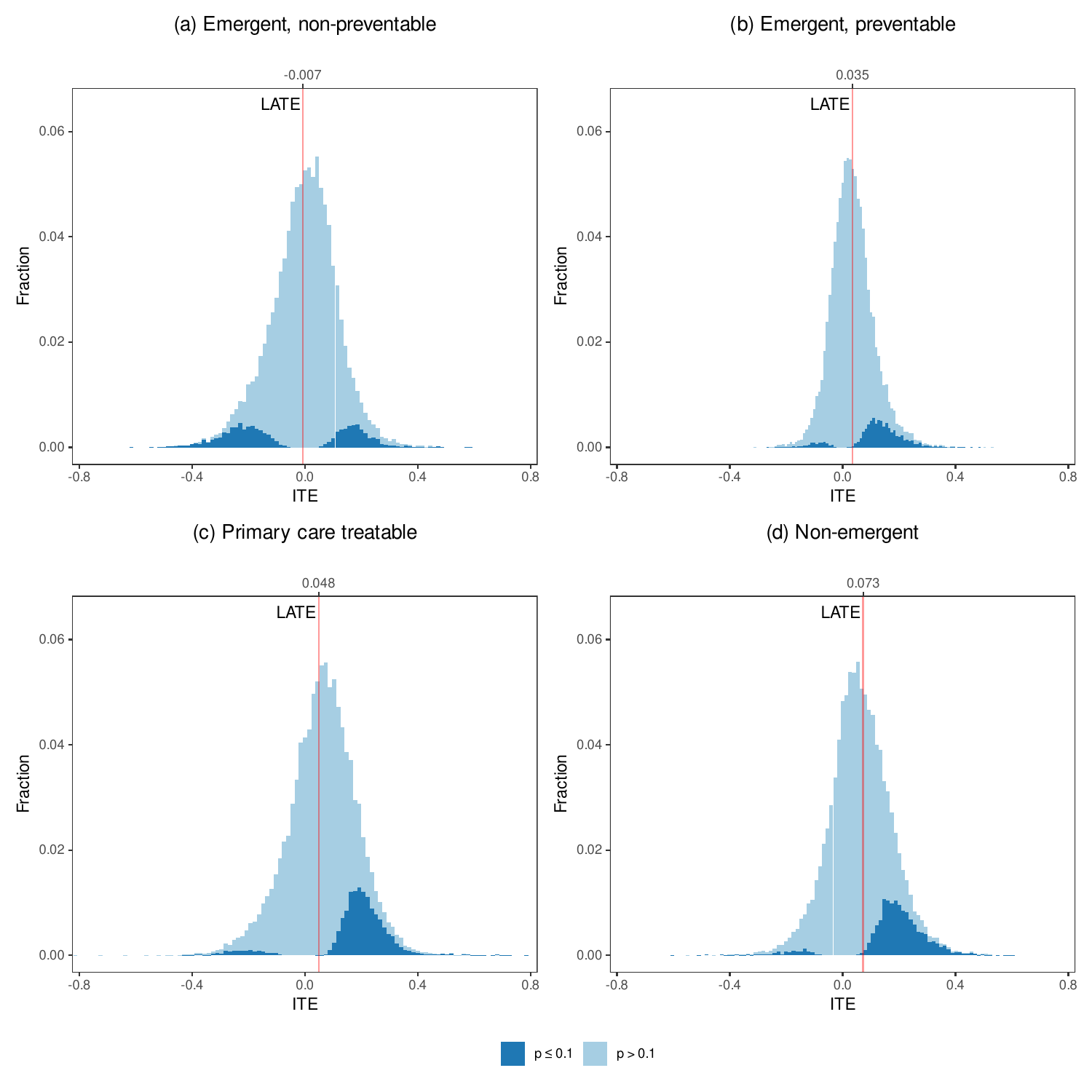}
      }
      \footnotesize
      \emph{Notes}: This figure plots the individualized treatment effects of Medicaid by type of ED visit based on generalized random forests for any emergent, non-preventable visit (panel a), any emergent, preventable visit (panel b), any primary care treatable visit (panel c), and any non-emergent visit (panel d). Measures of the type of ED visit are based on \citeapos{billings2000emergency} algorithm described in \citet{taubman2014medicaid}. We use these measures to construct binary indicators of ED visits by type of condition as described in the main text. The darker shade denotes statistical significance at the 10\% level. The red vertical line indicates the local average treatment effect. The baseline sample consists of 24,615 individuals in the \citeasnoun{taubman2014medicaid}  sample with non-missing information on pre-lottery emergency department utilization and SNAP/TANF receipt. The estimates displayed exclude less than half a percentile at the top and bottom of the distribution, resulting in the axes corresponding approximately to the percentile range $[0.5\%, 99.5\%]$. Bin size is chosen according to the Freedman-Diaconis rule.
    \end{minipage}
  \end{figure}
\end{singlespace}

We turn our attention to the heterogeneous results by type of ED visit presented in Figure \ref{fig:any-visit-type} for the extensive margin. Previous studies have highlighted the importance of looking beyond overall ED visits and categorizing them by medical urgency \citep{,taubman2014medicaid,giannouchos2022association,garthwaite2019all}. Consistent with the overall ED visit results, individualized treatment effects of Medicaid for all types of ED visits exhibit substantial heterogeneity. All ED visit types show a wide range of ITE estimates, including negative and positive values. Moreover, the share of significant ITE estimates differs across the ED visit types. For any emergent, non-preventable visits, panel (a) in Figure \ref{fig:any-visit-type} shows negative and positive statistically significant effects of sizable proportions. These two opposing forces are consistent with the average null effect in \citet{taubman2014medicaid}. This finding accords with the large share of negative ITE estimates discussed above for inpatient ED care. Across all visit types, we find the largest share of significant reductions in ED uses for emergent, non-preventable visits. A similar pattern holds for the number of emergent, non-preventable visits, with the corresponding graph in Figure \ref{fig:num-visit-type} of Appendix~\ref{appendixA}.

Figure \ref{fig:any-visit-type} shows some differences in the heterogeneous impacts for the remaining three types of ED visits. First, we find a larger share of positive and statistically significant individualized effects for primary care treatable and non-emergent visits than emergent visits (i.e., non-preventable and preventable) on both extensive and intensive margins. The two largest increases on the extensive and intensive margins occur for primary-care treatable and non-emergent visits. However, the increase in primary care treatable visits is more pronounced on the intensive margin, whereas the increase in non-emergent visits is larger on the extensive margin.

In summarizing, the pattern of heterogeneous effects we uncover offers a more nuanced interpretation of Medicaid's effect on ED utilization that is not discernible by focusing on the mean effect alone. For instance, the large fraction of statistically significant increases for primary care treatable and non-emergent visits suggests that gaining Medicaid coverage increased ED use in these categories for most people, not just on average. %Thus, this finding does not support the hypothesis that the average effects may have missed subpopulations who might have reduced ED use because they substituted primary care services to address those conditions. 
In contrast, we observe a sizable amount of negative ITEs for the emergent, non-preventable category. While the average effect is not statistically significant, this finding suggests that some people reduced ED usage by not seeking treatment for conditions that may require emergency department treatment. 

\begin{figure}
		\centering
		\caption{Variable importance scores in growing the causal forest (Any Visit)}
		\label{fig:varimp-any-visit-ed}
		\begin{minipage}{\linewidth}
			\makebox[\textwidth][c]{
				\includegraphics[width=\linewidth]{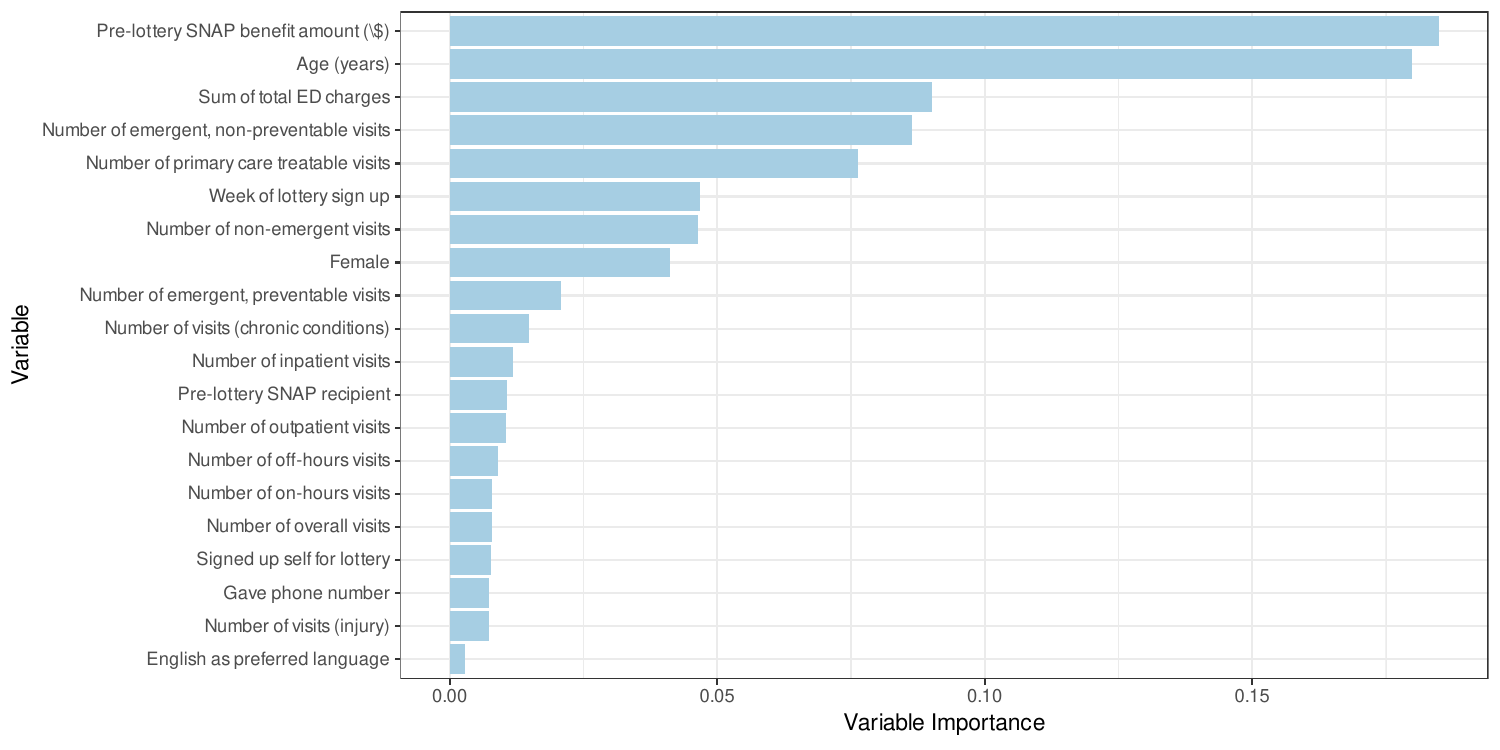}
			}
			\footnotesize
			\emph{Notes:} This figure shows the variable importance scores of the top 20 characteristics in growing the generalized random forests used to estimate the individualized treatment effects of Medicaid for any overall ED visit. The variable importance measure is a simple weighted sum of the proportion of times a variable is used in a tree splitting step at each depth in growing the forest. The scores roughly capture how important a variable is for driving treatment effect heterogeneity. The baseline sample consists of 24,615 individuals in the \citeasnoun{taubman2014medicaid}  sample with non-missing information on pre-lottery emergency department utilization and SNAP/TANF receipt.
		\end{minipage}
	\end{figure}

It is also instructive to highlight the most important covariates driving these heterogeneous effects. Put differently, what are the most used individual characteristics for tree splitting in growing the causal forests? Table \ref{tab:desc.base} lists the covariates utilized in growing the causal forests---pre-randomization ED use (overall use and by visit type), SNAP and TANF receipt measures, and lottery list variables. Figure \ref{fig:varimp-any-visit-ed} displays the variable importance scores for the top 20 characteristics. We present the corresponding variable importance plot for the number of ED visits and the scores for all variables in Appendix \ref{appendixA}. For the extensive margin, Figure \ref{fig:varimp-any-visit-ed} shows that the causal forest splits the most on pre-lottery SNAP benefits followed by age, the sum of total ED facility charges, the number of pre-lottery emergent but non-preventable ED visits, and the number of pre-lottery primary care treatable ED visits. 
%It is reassuring that the forest rarely splits on variables that we do not expect to drive treatment effect heterogeneity (e.g., lottery list variables such as whether the individual requested English language materials).
%The variable importance score is computed as the tree depth weighted sum of the proportion of times each variable is used as a split candidate in the causal forest.

\begin{singlespace}
\end{singlespace}

\subsection{Aggregate effects of Medicaid on ED use}\label{ate}

In this section, we present doubly robust estimates of the \emph{average} Medicaid effect (i.e., the LATE). Although not the main focus of our analysis, a computationally cheap by-product of estimating heterogeneous effects is that we can aggregate them to produce the average effects. These average effects are obtained by averaging the doubly robust scores targeting the LATE instead of computing a simple average of the individualized treatment effects, $\widehat{\tau}(X)$. 
%By so doing, the forest-based LATE estimator retains the desirable double robust property of being consistent if at least one component of the score is consistent \cite{chernozhukov2018double,athey2019generalized}. 
Given that the standard linear IV estimator also recovers the LATE, we only discuss the forest-based LATE to replicate the results in \citet{taubman2014medicaid} and show that the qualitative average findings are unchanged.\footnote{See \citet{sloczynski2020should} for a recent discussion on when the linear IV estimand recovers the LATE when covariates are included in the IV estimation.}  

\begin{singlespace}
	\input{Tables/Tables_final/tmain.tex}
\end{singlespace}

Table \ref{tab:main.base} presents the nonparametric and linear IV estimates of Medicaid's effect on all measures of ED utilization.
The linear IV estimates confirm that we closely replicate \citeapos{taubman2014medicaid} main findings, with minor differences in magnitude plausibly due to rounding and the top-coding in the public-use Oregon data files. The fact that we replicate their findings is not surprising given the negligible difference between our samples when we restrict the sample to observations with non-missing pre-lottery utilization.\footnote{In their online appendix, \citet{taubman2014medicaid} note using ``a missing indicator to handle the small number of missing prerandomization observations.''}  Therefore, we are confident that our linear IV results are practically the same as those in \citet{taubman2014medicaid} and serve as a useful baseline for comparing our nonparametric results.

Except for some differences in magnitudes in a few cases of the binary outcomes, the nonparametric LATE estimates are qualitatively the same as those based on the linear IV method. On the extensive margin, the nonparametric method appears slightly inefficient (higher standard errors), but there is no difference in statistical significance between the estimates from the two methods. On the intensive margin, the nonparametric estimates more closely align with the linear method. Overall, our nonparametric estimates of the average effect comport with previous studies that Medicaid increased ED use while allowing us to explore heterogeneity at a more granular level. 

\subsection{Risk factors associated with changes in ED use}

This section investigates the risk factors associated with significant changes in ED usage by examining treatment effects for selected subgroups. We summarize treatment effects for selected subgroups by aggregating the individualized treatment effects discussed above. Table \ref{tab:gates.any} presents group average treatment effects for an extended list of characteristics from the causal forest.

There are two notable findings. First, we find statistically significant GATE estimates for four subgroups---men, prior SNAP participants, young and middle-aged adults below age 50, and individuals with any previous ED use classified as primary care treatable. These groups exhibit treatment effects between 8 and 10 percentage points, at least twice the average effect. If we zero in on smaller subgroups at the intersection of these characteristics, we find even larger effects (not reported). For instance, the GATE we estimate for men below age 50 with any pre-lottery primary care treatable ED use implies a 21 percentage point increase in the likelihood of using the emergency department upon gaining coverage. The forests do not reveal statistically significant group effects for other pre-lottery ED utilization or lottery list variables. 

% These are some of the largest effects we find among those subgroups

% any_visit_ed, simple def of 'any' use
% group                                                                                             estimate pval
% Any primary care treatable ED visit prior to study period: Yes, Male: Yes, Age ≥ 50: No, Addition 0.210509 0.0124666

% num_visit_cens_ed, simple def of 'any' use
% group                                                                                             estimate pval
% Any primary care treatable ED visit prior to study period: Yes, Male: Yes, Age ≥ 50: No, Addition 0.970851 0.11016

% any_visit_ed, new def of 'any' use
% group                                                                                                        estimate pval
% Any primary care treatable ED visit prior to study period (new def): Yes, Male: Yes, Age ≥ 50: No, Addition  0.22056 0.0224604

% num_visit_cens_ed, new def of 'any' use
% group                                                                                                        estimate pval
% Any primary care treatable ED visit prior to study period (new def): Yes, Male: Yes, Age ≥ 50: No, Addition  1.22996 0.110271

\begin{figure}%[H]
  \centering
  \caption{Distribution of individualized treatment effects of Medicaid by selected groups (Any visit)}
  \label{fig:fcombined-any-visit-ed}
  \begin{minipage}{\linewidth}
    \makebox[\textwidth][c]{
      \includegraphics[width=0.8\linewidth]{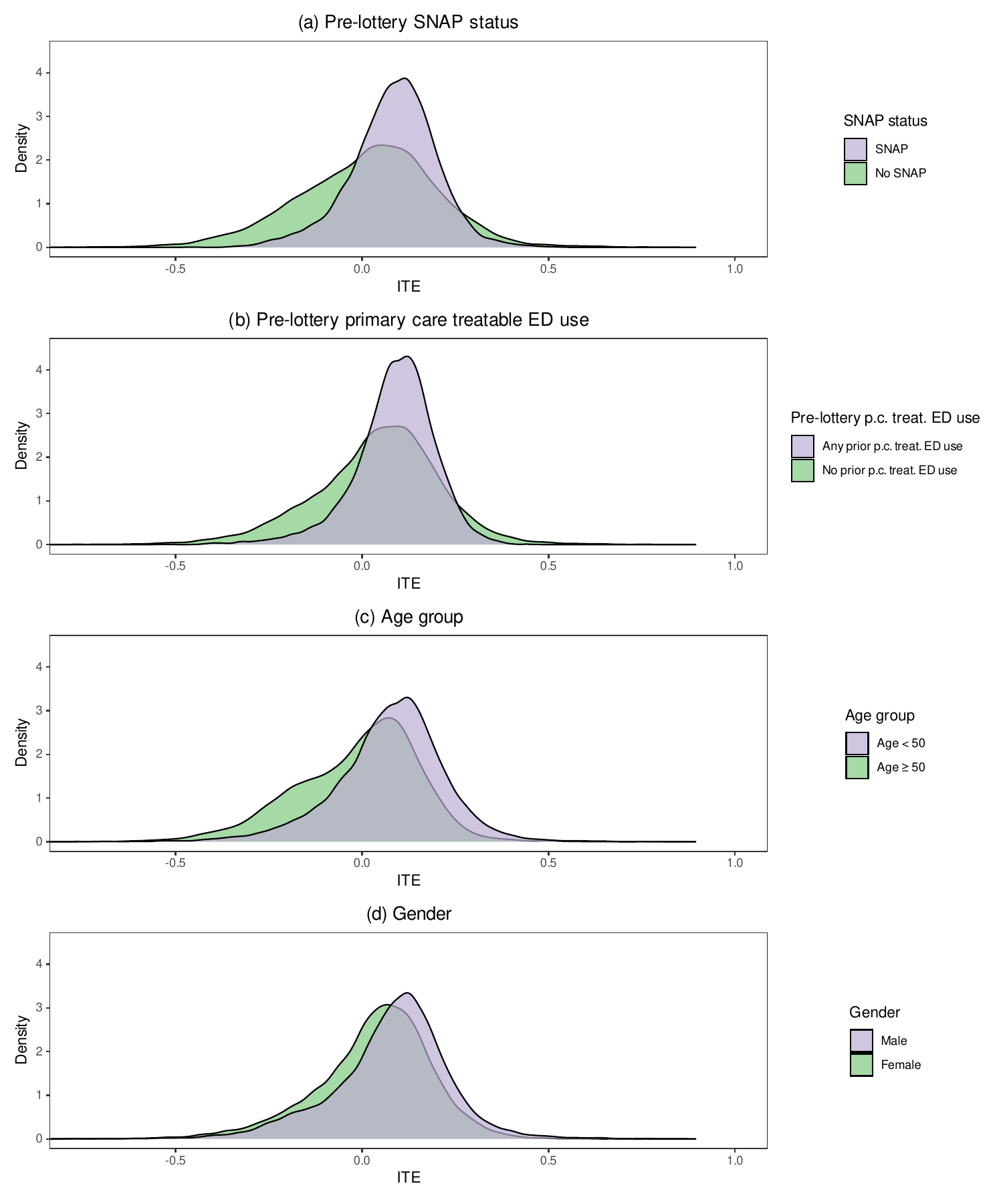}
    }
    \footnotesize
    \emph{Notes}: This figure plots the individualized treatment effects of Medicaid on any overall ED visit for the four major groups identified with substantial group average effects---pre-lottery SNAP receipt (panel a), pre-lottery primary care treatable ED visit (panel b), age group (panel c), and gender (panel d). The baseline sample consists of 24,615 individuals in the \citeasnoun{taubman2014medicaid}  sample with non-missing information on pre-lottery emergency department utilization and SNAP/TANF receipt.
  \end{minipage}
\end{figure}

Figure \ref{fig:fcombined-any-visit-ed} contrasts the distributions for the four main groups identified by the causal forest in more detail. The graphs highlight the differences in the empirical effect distribution that translate into the average group effects. Panel (a) shows a noticeable difference in the empirical distribution of the individualized treatment effects by pre-lottery SNAP receipt. The distribution of the ITEs for SNAP recipients is narrower and exhibits a denser mass in the positive effects region. This difference in the distribution of the heterogeneous effects reflects in the aggregated group average effect. Previous SNAP recipients are estimated to increase ED use by a statistically significant 9 percentage points, while there is no effect for SNAP non-recipients.

Interestingly, panel (b) of Figure \ref{fig:fcombined-any-visit-ed} shows that the ITE distribution by pre-lottery primary care treatable ED use exhibits a similar pattern to prior SNAP receipt, but with a bigger group effect for those with any prior visit of 13 percentage points. Finally, Panels (c) and (d) display the heterogeneous effects for age group (age $<$ 50 vs.\ age $\geq$ 50 years) and gender (male vs.\ female). In contrast to the previous groups, the distributions for partitioning these groups are similar in shape but shifted to the right for the younger age group and men.

Second, the only subgroups with sizable negative effects are older adults (aged 50 and above) and those with any pre-lottery emergent, preventable ED visits. The effect for older adults is a 6-percentage-point reduction in the probability of using the ED. However, the estimate is insignificant ($p$-value=0.24), possibly due to a lack of power. Further analyses in this direction suggest that those who reduce inpatient visits are mostly older people---individuals in the left tail of Figures 1c and 2a are older.

\begin{singlespace}
  \input{Tables/Tables_final/tgates-any-visit.tex}
\end{singlespace}

\input{Tables/Tables_final/tcomp-any-visit.tex}

Finally, we divide the sample based on the individualized treatment effects to contrast the characteristics of those who increase and decrease usage ex-post. Table \ref{tab:comp.any} presents the mean characteristics of individuals identified to increase and reduce emergency department utilization on the extensive margin.\footnote{The corresponding summary of characteristics for the intensive margin of ED use are available in Appendix \ref{appendixA}.} In almost all cases, we find that those who increase and decrease are different on observed dimensions. To highlight a few key differences, those estimated to increase the likelihood of any ED use are almost three years younger (39.8 vs.\ 41.5 years), less likely to be female (0.51 vs.\ 0.63), more likely to have previously received SNAP (0.62 vs.\ 0.35) with higher total household benefits averaging (\$1,607 vs.\ \$717), and more likely to have been a previous TANF recipient (0.03 vs.\ 0.01) with higher benefits (\$111 vs.\ \$63). In terms of prior ED visits, those estimated to increase use have a higher number of visits across all visits. The largest mean differences are for the overall number of ED visits and outpatient ED visits.

\subsection{Results of policy learning exercises}

This section presents the results of the policy learning exercises described in Section \ref{policytree} using the estimated heterogeneous effects. Until now, we have focused on the effects of Medicaid coverage (i.e., the conditional LATE). However, the policy learning exercises are based on the intent-to-treat effects of winning the lottery reported in Appendix~\ref{appendixB}. The heterogeneous intent-to-treat effects follow a similar pattern as the LATE coverage effects of Medicaid but the former is more natural for the policy learning exercises because it is the lottery and not the actual coverage that is random. 

\subsubsection{Comparing the observed lottery to an alternative assignment scheme}

In this theoretical exercise, our objective is to estimate an alternative allocation scheme that prioritizes those least likely to increase unnecessary ED use, defined as the number of non-emergent ED visits. Doing so allows us to describe how the sample selected by such an optimal allocation scheme would compare to the sample of observed lottery winners. Again, while this exercise is infeasible from the point of policy-making, we use it to illustrate the value of the heterogeneous effects and verify that, given a well-defined objective, a random assignment scheme may be inferior to one based on the empirical distribution of treatment effects. 

%We then estimate a simple decision rule to select such a sample as an alternative to a lottery. 
%The allocation could be used to solicit applications in an insurance expansion or to target educational interventions. 
%Until now, we have focused on the effects of Medicaid. However, we conduct the policy learning exercises using the intent-to-treat effects of winning the lottery. One can think of our assignment rule for soliciting OHP Standard applications as the equivalent of inviting lottery winners to apply for the program. The algorithms proposed by \citet{athey2021policy}, \citet{kitagawa2018should} and \citet{zhao2012estimating} rely on first estimating the heterogeneous treatment effects of winning the lottery and then optimizing an objective function based on the doubly robust scores targeting the treatment effect estimand of interest. We estimate an optimal allocation and an assignment rule that satisfy the objective of minimizing undesirable ED use, defined as the number of non-emergent ED visits. 

\input{Tables/Tables_final/tcomp-lp.tex}

Following \citet{kitagawa2018should}, we first compute the optimal treatment allocation scheme that minimizes the objective function of the number of non-emergent ED utilization using integer programming. We constrain the number of people selected by the allocation scheme to be the same number that won the Oregon lottery. We then compare the selected sample to the original lottery winners. As expected in Table \ref{tab:comp.rule}, the selected sample based on the optimal allocation scheme has substantially lower ex-post ED utilization for non-emergent causes (0.18 fewer visits). Other ED use measures following treatment are also substantially lower. The same finding holds for pre-lottery ED use, suggesting that ED use patterns are complex and treatment effects are driven by a population that uses the ED for most of their health care needs (urgent or non-urgent). Comparing the magnitude of pre-and post-randomization ED use outcomes within the same category suggests larger differences between the two samples, indicating that the estimated optimal assignment scheme successfully selects those who react to obtaining coverage by reducing ED usage. For instance, the difference in outpatient use between the two samples in the pre-lottery period is 0.24 visits, while the same difference in the post-lottery period is 0.51 visits.  
%We do so while imposing a capacity constraint on the total number of people treated. 
%Based on the double robust scores, we compute the optimal allocation scheme which minimizes non-emergent ED use using integer programming. 

\subsubsection{Identifying people at risk of non-emergent ED use via decision rules}

A more practical way to use the estimated heterogeneous effects for policy-making is to design simple decision rules identifying people at greater risk of increasing non-emergent ED use among Medicaid enrollees. Here, the administrator may still use a lottery to solicit Medicaid applications but target a subset of new enrollees with an outreach or educational intervention. For instance, policymakers may want to identify some enrollees who may be required to establish primary care upon receiving Medicaid coverage. Of course, a straightforward decision rule might be to target the intervention based on their predicted individualized treatment effects, such as those with $\widehat{\tau}(X)\geq 0$. However, \citet{athey2019machine} points out that such decision rules are not always optimal in the sense of minimizing the loss from not using the ideal assignment rule or policy.

Thus, we estimate the depth-2 decision trees to identify those most likely to increase non-emergent ED visits using only a subset of pre-lottery ED use variables \citep{athey2021policy}. Figure \ref{fig:optimal-policy-d2} presents the estimated decision rule. The first node of the decision tree partitions the sample based on a threshold of two prior outpatient ED visits. On the one hand, for people with two or fewer outpatient ED visits, the decision rule selects them if they had no previous ED use classified as emergent. This decision rule is interesting because, given the objective of maximizing non-emergent use, it identifies those with no prior emergent use as those likely to increase non-emergent use. 

On the other hand, for those with more than two previous outpatient ED visits (potentially excessive ED use), those with at most one prior ED visit resulting in hospitalization are selected. Again, this rule reasonably targets those who rarely used the ED for conditions severe enough to warrant inpatient care in the past.

% From Figure \ref{fig:optimal-policy-d2}, the decision rule selects a total of 2,258 people, making up about 9 percent of the entire sample. In our estimation sample, there are 9,607 randomly selected lottery winners in the Oregon experiment, with 5,912 of them ultimately receiving Medicaid.

\input{Figures/Figures_final/optimal-policy-d2.tex}

We emphasize that our estimated decision rule is not set in stone and may be impractical, especially if the previous ED utilization variables are not easily accessible. Policymakers could also specify a different objective function to align with their goals or accommodate additional constraints. Moreover, the set of variables included in the estimation of the decision rule can be modified to respect fairness or other political/legal requirements.

%Optimal policy rule estimation presents an alternative approach for policy administrators who are faced with constraints in allocating scarce resources, especially with public assistance programs relative to the random or automatic assignment status quo. Even when program spots are not scarce, optimal policy learning might help allocate recipients to different insurance plan options in programs such as Medicaid, where managed care is on the rise. About 25 million people will be auto-assigned into a Medicaid managed care plan as states continue contracting with risk-based managed care organizations to deliver benefits to Medicaid recipients \cite{ndumele_improving_2020}. Policymakers have an opportunity to utilize advances in statistical treatment assignment rule estimation to assign people to health care plans to improve health and reduce spending.
%Alternative welfare functions could place extra weight on targeting people with severe conditions who increase non-preventable ED use. Further capacity constraints can restrict the maximum allowable cost
%

\section{Conclusion}\label{conclusion}

This paper estimates the heterogeneous impacts of Medicaid coverage on emergency department visits using records from 12 hospitals in Portland, Oregon, matched to the Oregon Health Insurance Experiment. The result that Medicaid increased ED utilization in the 2008 Oregon Health Experiment made headlines because of the hypothesis that insurance coverage should make it easier for recipients to access primary care and reduce the need to use the emergency department. However, this is an empirical question, given that insurance coverage also reduces out-of-pocket healthcare spending. We provide new insights into the ED results of the Oregon experiment when we go beyond the average impacts of Medicaid coverage. We do so by estimating the individualized treatment effects of Medicaid using nonparametric machine learning methods. We then leverage the heterogeneous effects to estimate decision rules to illustrate their usefulness for policymakers in similar settings.

We find substantial treatment effect heterogeneity in the impacts of Medicaid on emergency department utilization. The individualized treatment effects of Medicaid on ED use indicate positive and negative effects, suggesting a more nuanced interpretation of Medicaid's average impacts. We also find that coverage effects for different types of ED use exhibit meaningful heterogeneity. Due to the variation in the individualized effect distribution, the average treatment effect is not typical but is driven by sections of the sample in the right tail of the distribution. A small proportion of high-use individuals drive the positive (average) Medicaid effect. Despite the positive average effects, the predicted treatment effects for many individuals are negative. 

%On the extensive margin, we estimate a weaker Medicaid effect on the probability of any overall ED use of a 4.5 percentage points increase, which is statistically insignificant and about 65 percent of the linear IV estimation method's magnitude. 

Our findings suggest that the average effect sometimes masks countervailing forces. In some instances, although many people increase ED usage due to obtaining coverage, a sizable proportion of people either decrease usage or do not respond. We find that reductions in ED use are pronounced for inpatient visits and emergent, non-preventable conditions. The overall null effects for these outcomes hide substantial opposing effects of obtaining coverage.

We also find substantial treatment effect heterogeneity on the intensive margin. Overall, the results suggest that increased ED usage is likely driven by intensive margin effects rather than extensive margin effects. Individuals who already utilize the emergency department further increase their usage after receiving Medicaid. 
%Intensive margin effects and patterns of ED usage across various types of ED use suggest the average effect is largely driven by people who indiscriminately utilize the ED for all their care needs.(repetitive?)

We also characterize the groups of people mainly driving the positive average Medicaid effects in the right tail of the distribution of the heterogeneous effects. Those with positive heterogeneous effects are predominantly younger, more likely to be men, more likely to receive SNAP and TANF in the pre-lottery period, and more likely to have higher baseline ED use. In particular, we identify four groups estimated to have statistically significant increases in ED use that are at least twice as large as the magnitude of the average effect. These groups are men, prior SNAP participants, younger adults under 50 years old, and people with pre-lottery ED use classified as primary care treatable. 
%We also find that the nonparametric approach finds no effect for other groups that the linear model's subsample analysis finds statistically significant effects.

Finally, we illustrate one way to use the estimated heterogeneous effects to estimate decision rules that may help to identify and target enrollees likely to increase non-emergent emergency department utilization. For people with minimal previous outpatient ED use, the decision rule identifies those with no previous emergent use as likely to increase non-emergent ED use. For those with higher previous outpatient ED use, the decision rule selects those whose ED visit resulted in hospitalization only once (i.e., those who have rarely presented with severe health problems justifying inpatient treatment). While our decision rules are illustrative, they highlight the potential for using statistical decision rules to guide policymakers to achieve context-specific objectives.

%================================== References =================================

\clearpage
\begin{singlespace}
  \bibliography{biblio_oregon.bib}
\end{singlespace}
%\bibliographystyle{jpe}

%=================================== Appendix ==================================
% \newpage
\appendix
\setcounter{table}{0}
\setcounter{figure}{0}
\renewcommand\thetable{\Alph{section}.\arabic{table}}
\renewcommand\thefigure{\Alph{section}.\arabic{figure}}
\renewcommand{\thepage}{\roman{page}}
\setcounter{page}{0}

\section{Supplementary Material for Medicaid Coverage Effects}\label{appendixA}
% \clearpage
%\subsection{Additional Extensive Margin Results}

%Graphs
%\subsection{Intensive Margin Results}
\begin{singlespace}
  \begin{figure}[H]
  	\vspace{-1cm}
    \centering
    \caption{Distribution of  individualized treatment effects  of Medicaid  on the number of ED visits}
    \label{fig:num-visit-combined-unif}
    \begin{minipage}{1\linewidth}
      \makebox[\textwidth][c]{
        \includegraphics[width=0.9\linewidth]{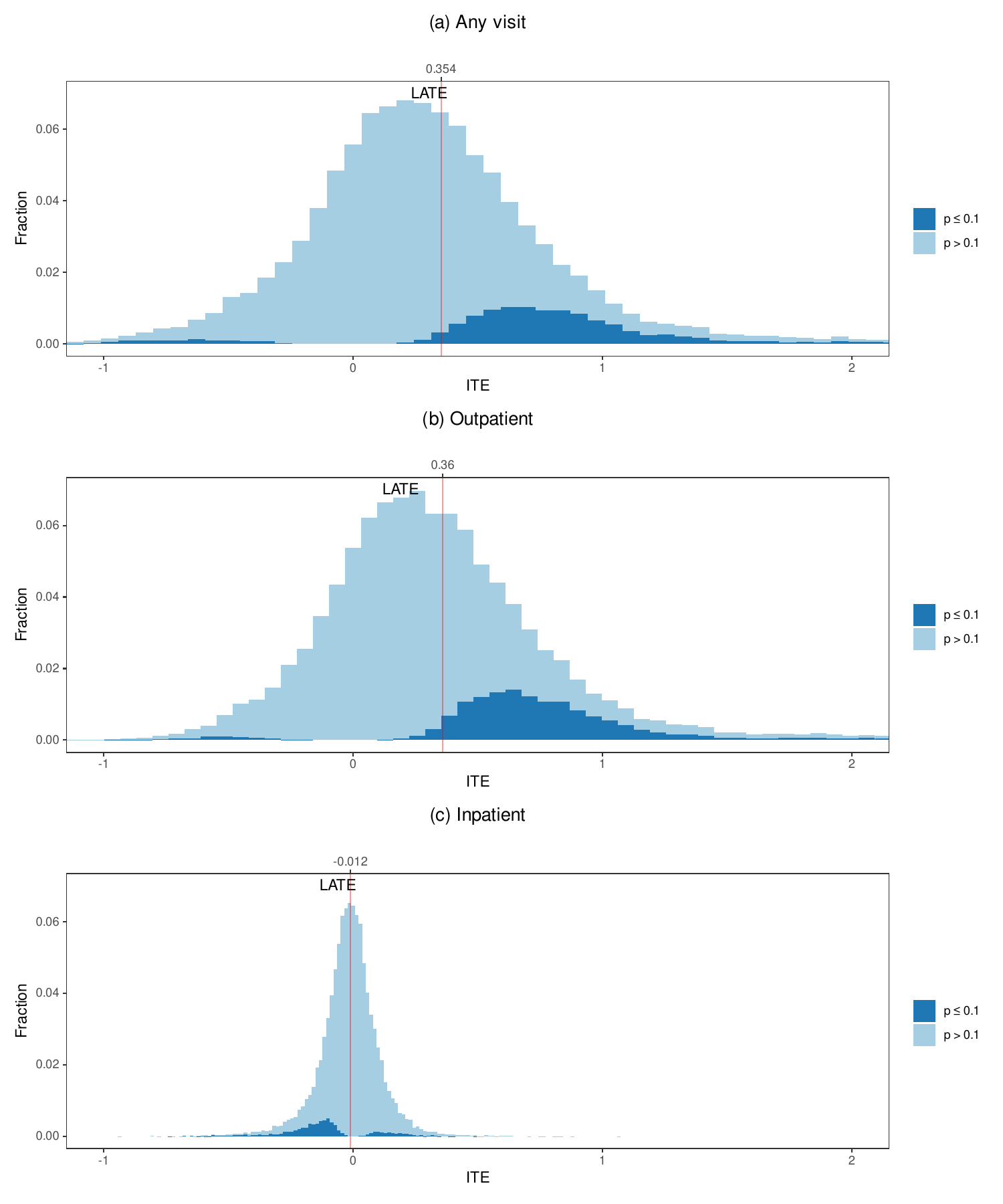}
      }
      \footnotesize
      \emph{Notes}: This figure plots the individualized treatment effects of Medicaid on the number of overall ED visit (panel a), the number of outpatient ED visits (panel b), and the number of inpatient ED use (panel c) based on generalized random forests. The darker shade denotes statistical significance at the 10\% level. The red vertical line indicates the local average treatment effect. The baseline sample consists of 24,615 individuals in the \citeasnoun{taubman2014medicaid} sample with non-missing information on pre-lottery emergency department utilization and SNAP/TANF receipt. The estimates displayed exclude less than half a percentile at the top and bottom of the distribution, resulting in the axes corresponding approximately to the percentile range $[0.5\%, 99.5\%]$. Bin size is chosen according to the Freedman-Diaconis rule.
    \end{minipage}
  \end{figure}
\end{singlespace}

\begin{singlespace}
  \begin{figure}
    \centering
    \caption{Distribution of  individualized treatment effects  of Medicaid on the number of ED visits by type of condition}
    \label{fig:num-visit-type}
    \begin{minipage}{\linewidth}
      \makebox[\textwidth][c]{
        \includegraphics[width=\linewidth]{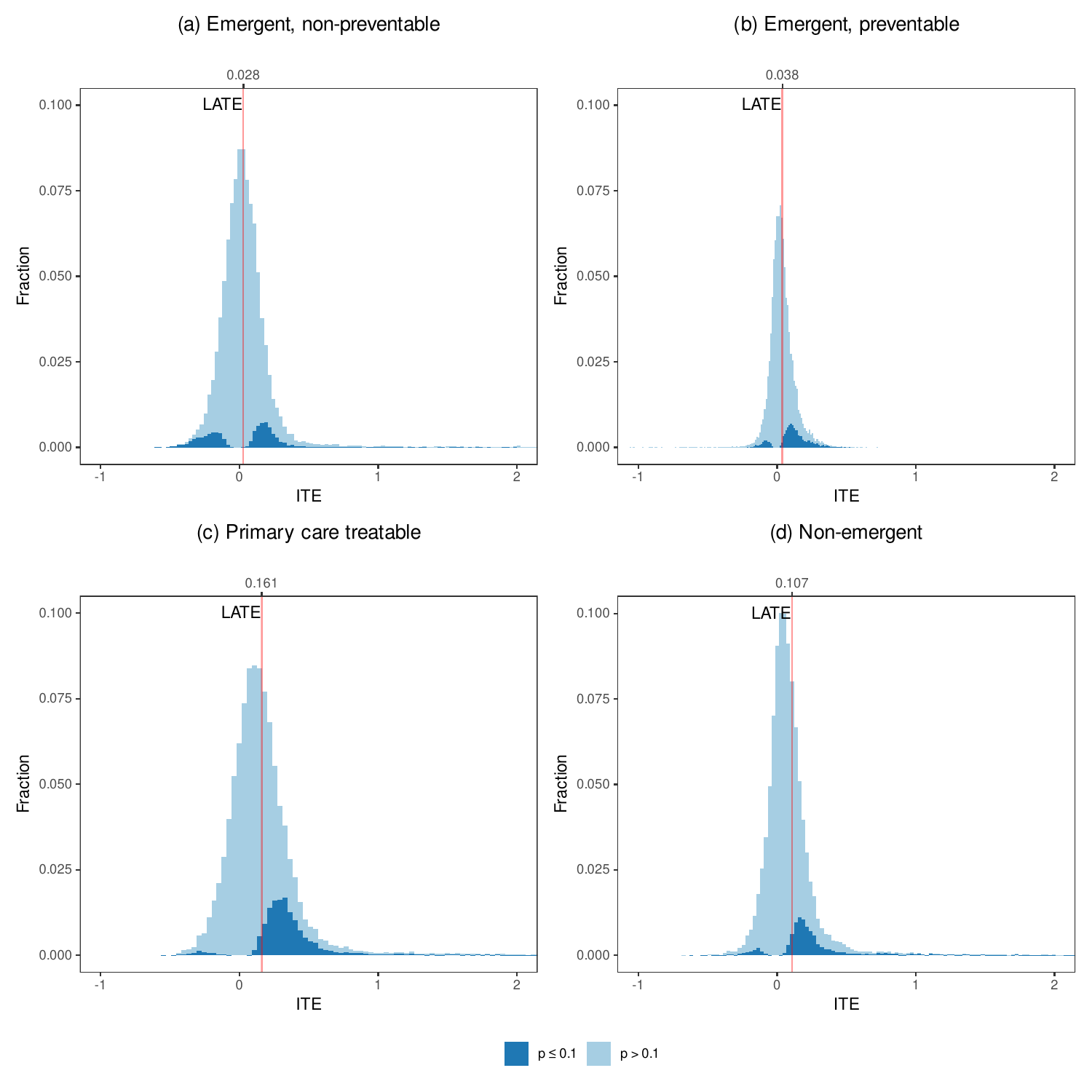}
      }
      \footnotesize
      \emph{Notes}: This figure plots the  individualized treatment effects  of Medicaid  by type of ED visit based on generalized random forests for the number of emergent, non-preventable visits (panel a), the number of emergent, preventable visits (panel b) , the number of primary care treatable visits (panel c), and the number of non-emergent visits (panel d). Measures of the type of ED visit are based on \citeapos{billings2000emergency} algorithm described in \citet{taubman2014medicaid}.  The number of visits of each type are then obtained by summing the probabilities across all visits for an individual. The darker shade denotes statistical significance at the 10\% level. The red vertical line indicates the local average treatment effect. The baseline sample consists of 24,615 individuals in the \citeasnoun{taubman2014medicaid}  sample with non-missing information on pre-lottery emergency department utilization and SNAP/TANF receipt. The estimates displayed exclude less than half a percentile at the top and bottom of the distribution, resulting in the axes corresponding approximately to the percentile range $[0.5\%, 99.5\%]$. Bin size is chosen according to the Freedman-Diaconis rule.
    \end{minipage}
  \end{figure}
\end{singlespace}

\begin{singlespace}
  \begin{figure}
    \centering
    \caption{Variable importance scores in growing causal forest (Number of visits)}
    \label{fig:num-visit-ed}
    \begin{minipage}{\linewidth}
      \makebox[\textwidth][c]{
        \includegraphics[width=\linewidth]{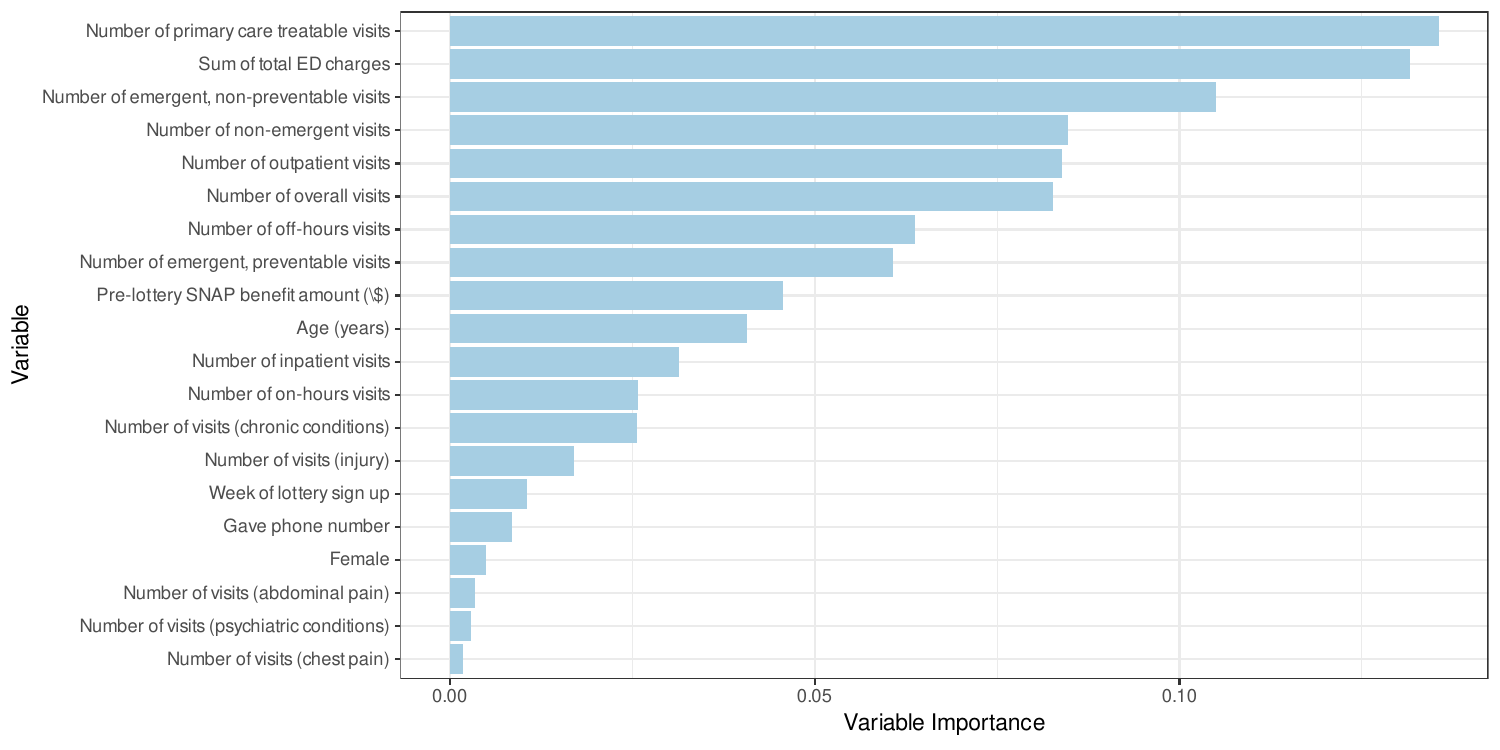}
      }
      \footnotesize
      \emph{Notes} This figure shows the variable importance scores of the top 20 characteristics in growing the generalized random forests used to estimate the individualized treatment effects of Medicaid for the number of overall ED visit. The variable importance measure is a simple weighted sum of the proportion of times a variable is used in a tree splitting step at each depth in growing the forest. The scores roughly capture how important a variable is for driving treatment effect heterogeneity. The baseline sample consists of 24,615 individuals in the \citeasnoun{taubman2014medicaid}  sample with non-missing information on pre-lottery emergency department utilization and SNAP/TANF receipt.
    \end{minipage}
  \end{figure}
\end{singlespace}
\clearpage

% Tables
\begin{singlespace}
  \input{Tables/Tables_final/tvarimp-visit.tex}
  \input{Tables/Tables_final/tgates-num-visit.tex}
  \input{Tables/Tables_final/tcomp-num-visit.tex}
\end{singlespace}
\clearpage

\section{Supplementary Material for Intent-to-treat effects}\label{appendixB}
\setcounter{table}{0}
\setcounter{figure}{0}
\renewcommand{\thepage}{\roman{page}}
\setcounter{page}{0}

%================================== Figures =================================

%=============================== Binary Outcomes ===============================

\begin{singlespace}
\begin{figure}[H]
	\vspace{-1cm}
  \centering
  \caption{Distribution of individualized treatment effects of winning the lottery on the propensity of ED use}
  \label{fig:any-visit-combined-unif-itt}
  \begin{minipage}{\linewidth}
    \makebox[\textwidth][c]{
      \includegraphics[width=0.9\linewidth]{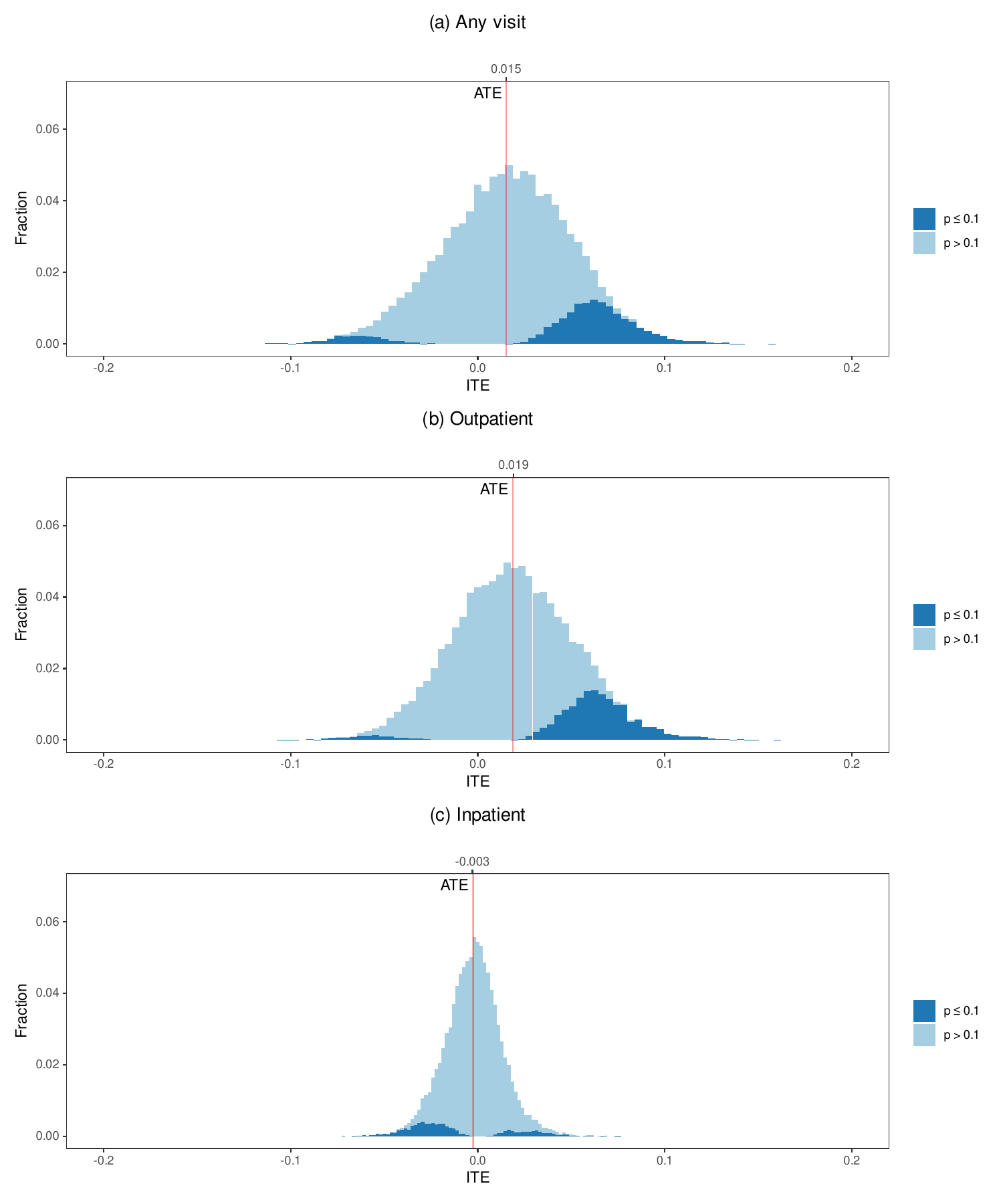}
    }
    \footnotesize
    \emph{Notes}: This figure plots the individualized treatment effects of winning the lottery (and being invited to apply for Medicaid) on any overall ED visit (panel a), any outpatient ED visits (panel b), and any inpatient ED use (panel c) based on generalized random forests. The darker shade denotes statistical significance at the 10\% level. The red vertical line indicates the local average treatment effect. The baseline sample consists of 24,615 individuals in the \citeasnoun{taubman2014medicaid} sample with non-missing information on pre-lottery emergency department utilization and SNAP/TANF receipt. The estimates displayed exclude less than half a percentile at the top and bottom of the distribution, resulting in the axes corresponding approximately to the percentile range $[0.5\%, 99.5\%]$. Bin size is chosen according to the Freedman-Diaconis rule.
  \end{minipage}
\end{figure}
\end{singlespace}

\begin{figure}
  \centering
  \caption{Distribution of  individualized treatment effects of winning the lottery on the propensity of ED use by type of condition}
  \label{fig:any-visit-type-itt}
  \begin{minipage}{\linewidth}
    \makebox[\textwidth][c]{
      \includegraphics[width=\linewidth]{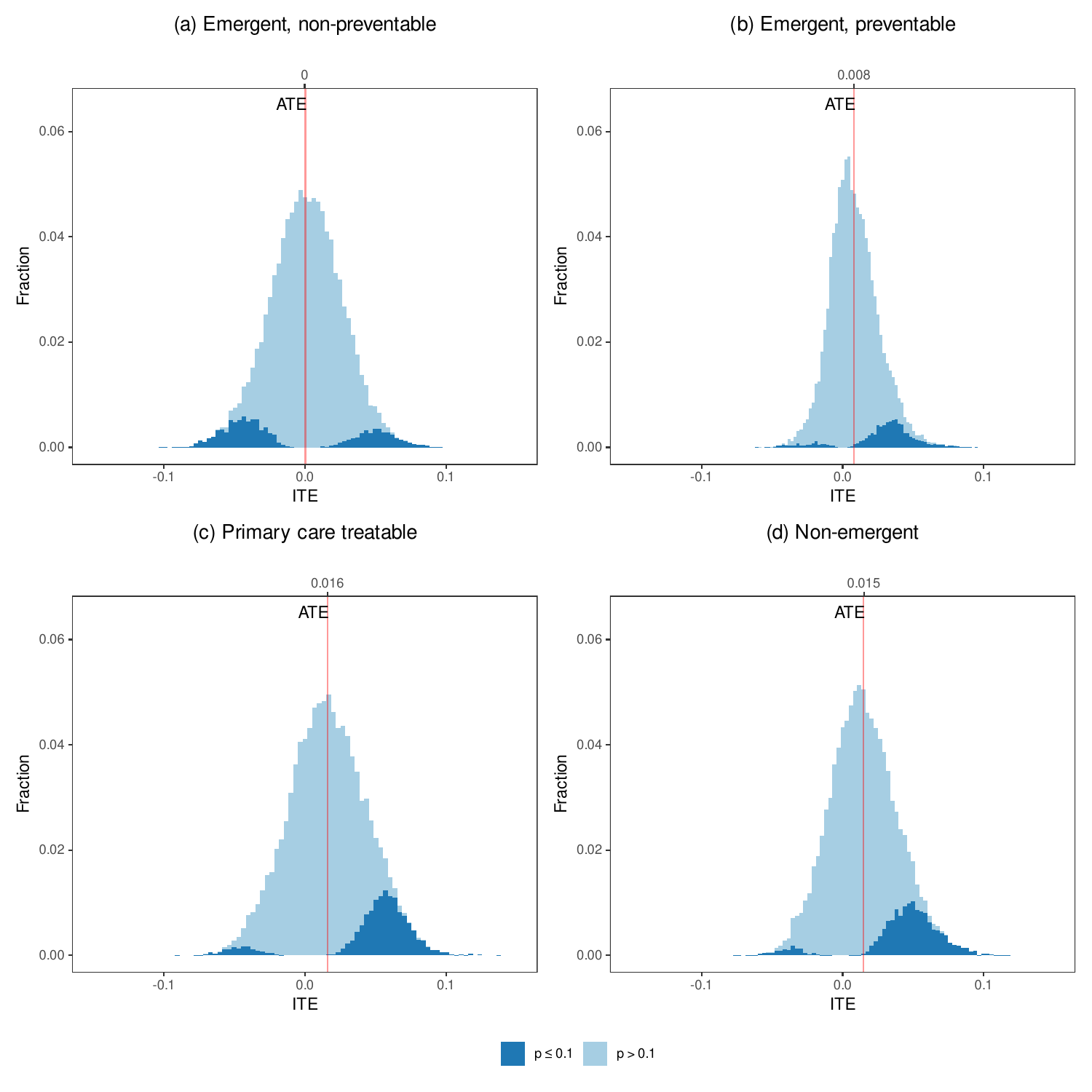}
    }
    \footnotesize
    \emph{Notes}: This figure plots the individualized treatment effects of winning the lottery (and being invited to apply for Medicaid) by type of ED visit based on generalized random forests for any emergent, non-preventable visit (panel a), any emergent, preventable visit (panel b), any primary care treatable visit (panel c), and any non-emergent visit (panel d). Measures of the type of ED visit are based on \citeapos{billings2000emergency} algorithm described in \citet{taubman2014medicaid}. We use these measures to construct binary indicators of ED visits by type of condition as described in the main text. The darker shade denotes statistical significance at the 10\% level. The red vertical line indicates the local average treatment effect. The baseline sample consists of 24,615 individuals in the \citeasnoun{taubman2014medicaid}  sample with non-missing information on pre-lottery emergency department utilization and SNAP/TANF receipt. The estimates displayed exclude less than half a percentile at the top and bottom of the distribution, resulting in the axes corresponding approximately to the percentile range $[0.5\%, 99.5\%]$. Bin size is chosen according to the Freedman-Diaconis rule.
  \end{minipage}
\end{figure}

%\begin{figure}
%  \centering
%  \caption{Group average treatment effects of winning the lottery (Any visit) }
%  \label{fig:any-visit-gate-itt}
%\begin{minipage}{\linewidth}
%    \makebox[\textwidth][c]{
%      \includegraphics[width=1.2\textwidth]{Figures/Figures_final/itt/fgate-any-visit-ed-itt.pdf}
%    }
%	\footnotesize
%    \emph{Notes}: This figure plots the group average treatment effects of  winning the lottery (and being invited to apply for Medicaid)  on any overall ED visit based on generalized random forests for binary indicators of gender, whether the individual provided a phone number, whether the individual requested English language materials, whether the individual signed up in the first week, pre-lottery SNAP receipt, an indicator of household size of at least two members, whether individual is at least 50 years old, and whether the individual had any pre-lottery primary care treatable ED visits. The darker tone colors denote statistical significance at the 10\% level. The  red horizontal line indicates the average effect. The baseline sample consists of 24,615 individuals in the \citeasnoun{taubman2014medicaid}  sample with non-missing information on pre-lottery emergency department utilization and SNAP/TANF receipt.
%  \end{minipage}
%\end{figure}

\begin{figure}
  \centering
  \caption{Variable importance scores in growing causal forest (Any visit) }
  \label{fig:varimp-any-visit-ed-itt}
  \begin{minipage}{\linewidth}
    \makebox[\textwidth][c]{
      \includegraphics[width=\linewidth]{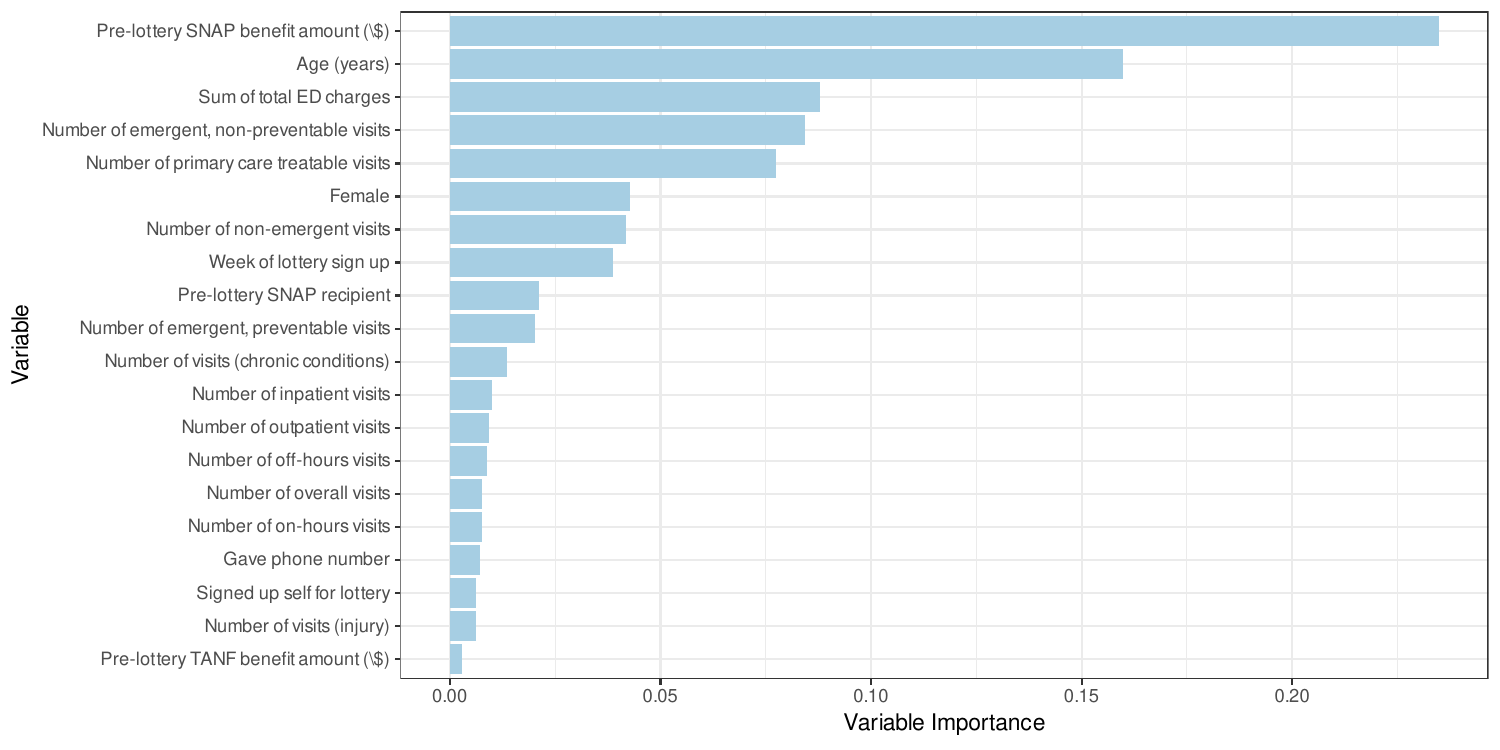}
    }
    \footnotesize
    \emph{Notes} This figure shows the estimates of variable importance for the top 20 characteristics used in growing the generalized random forests in estimating the CATE of winning the lottery (and being invited to apply for Medicaid) for any overall ED visit. The variable importance measure is a simple weighted sum of the proportion of times a variable is used in a splitting step at each depth in growing the causal forest, thus, capturing how important a variable is for driving treatment effect heterogeneity. The sample consists of 24,615 individuals in the \citeasnoun{taubman2014medicaid}  sample with non-missing information on pre-lottery emergency department utilization and SNAP/TANF receipt.
  \end{minipage}
\end{figure}

\clearpage

%=============================== Numeric Outcomes ==============================

\begin{figure}
  \centering
  \caption{Distribution of individualized treatment effects of winning the lottery on the number of ED visits}
  \label{fig:num-visit-combined-unif-itt}
  \begin{minipage}{\linewidth}
    \makebox[\textwidth][c]{
      \includegraphics[width=0.9\linewidth]{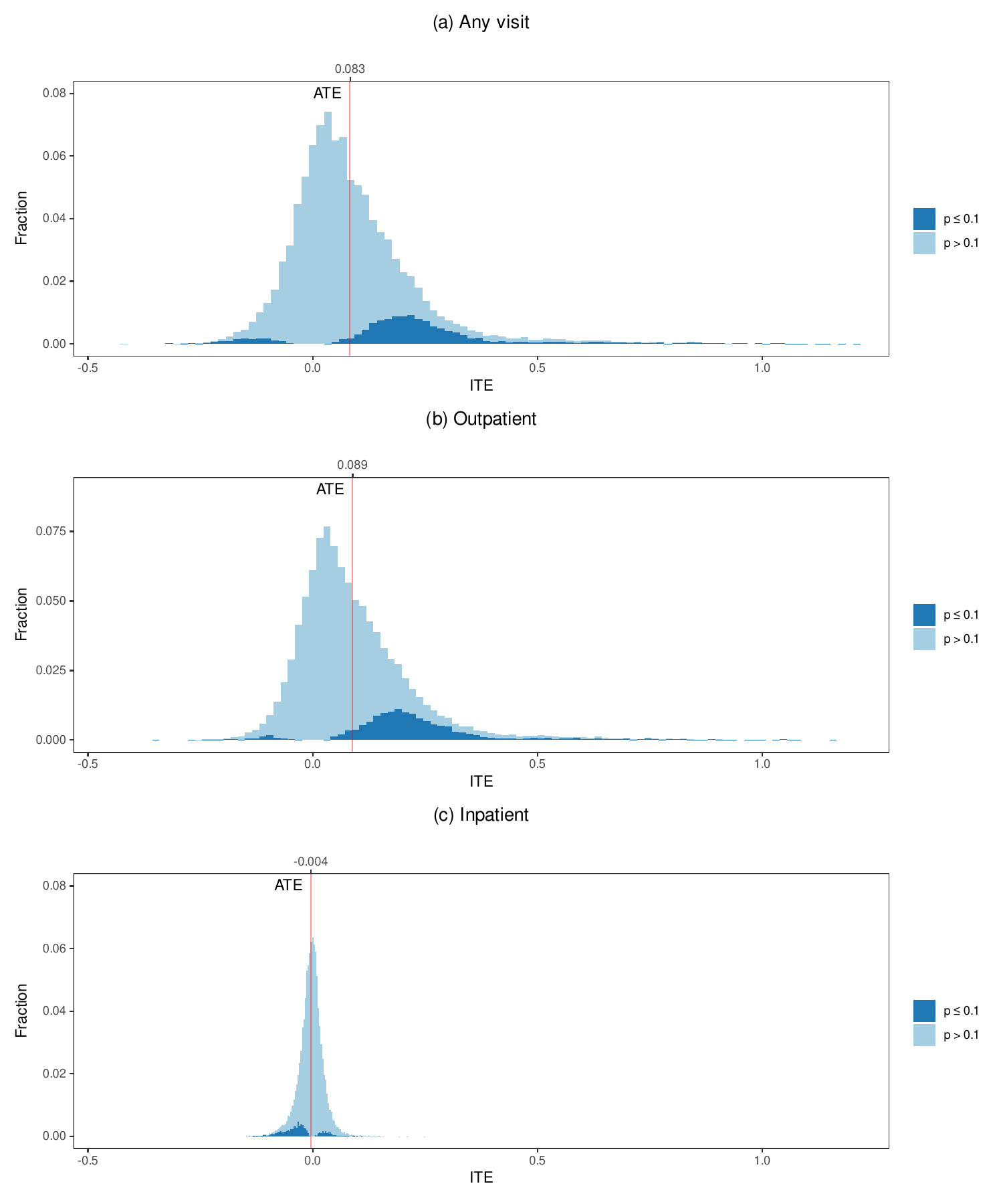}
    }
    \footnotesize
    \emph{Notes}: This figure plots the individualized treatment effects of winning the lottery (and being invited to apply for Medicaid) on the number of overall ED visit (panel a), the number of  outpatient ED visits (panel b), and the number of inpatient ED use (panel c) based on generalized random forests. The darker shade denotes statistical significance at the 10\% level. The red vertical line indicates the local average treatment effect. The baseline sample consists of 24,615 individuals in the \citeasnoun{taubman2014medicaid} sample with non-missing information on pre-lottery emergency department utilization and SNAP/TANF receipt. The estimates displayed exclude less than half a percentile at the top and bottom of the distribution, resulting in the axes corresponding approximately to the percentile range $[0.5\%, 99.5\%]$. Bin size is chosen according to the Freedman-Diaconis rule.
  \end{minipage}
\end{figure}

\begin{figure}
  \centering
  \caption{Distribution of individualized treatment effects  of winning the lottery on the number of ED visits by type of condition}
  \label{fig:num-visit-type-itt}
  \begin{minipage}{\linewidth}
    \makebox[\textwidth][c]{
      \includegraphics[width=\linewidth]{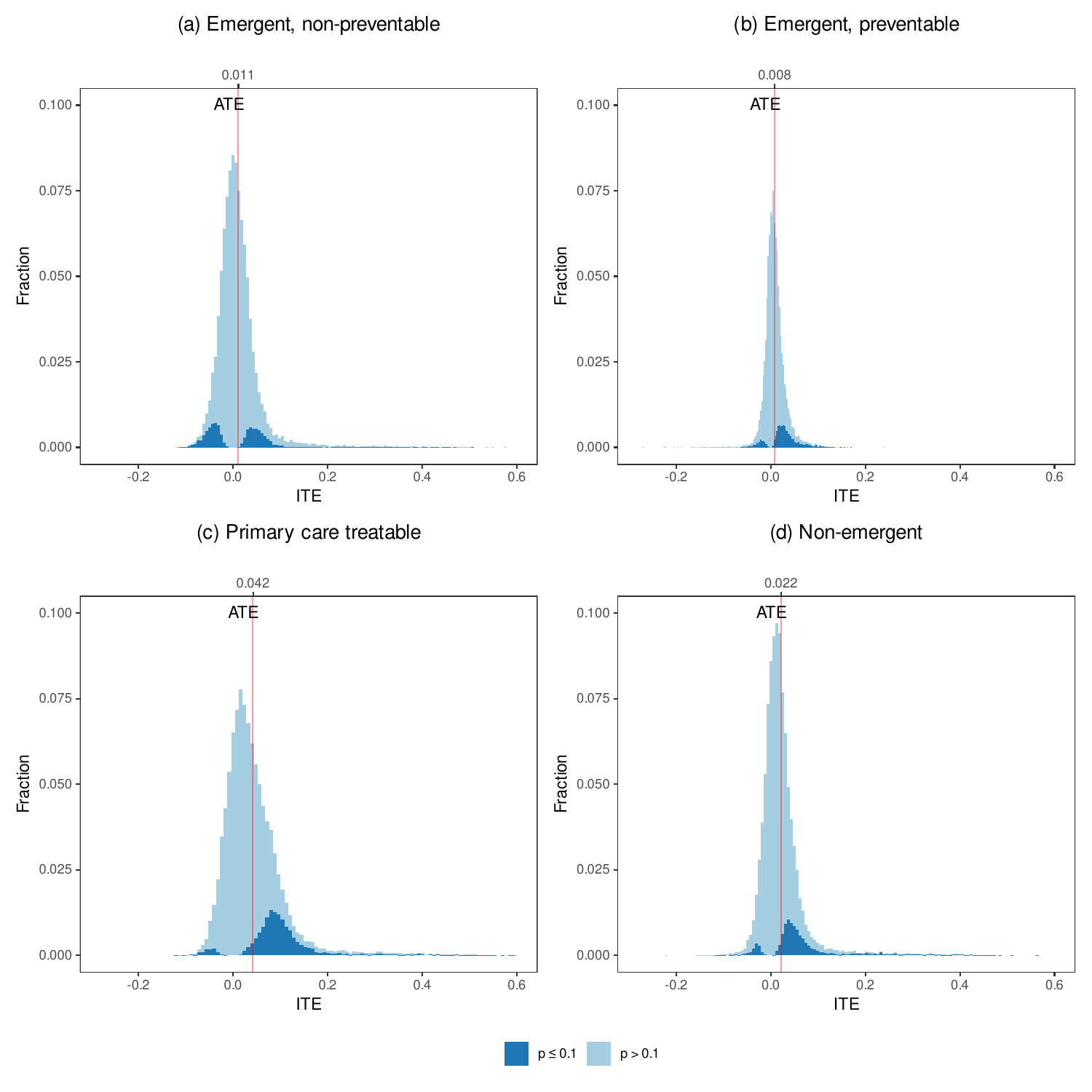}
    }
    \footnotesize
    \emph{Notes}: This figure plots the individualized treatment effects of winning the lottery (and being invited to apply for Medicaid) by type of ED visit based on generalized random forests for the number of emergent, non-preventable visit (panel a), the number of emergent, preventable visit (panel b), the number of primary care treatable visit (panel c), and the number of non-emergent visit (panel d). Measures of the type of ED visit are based on \citeapos{billings2000emergency} algorithm described in \citet{taubman2014medicaid}. The darker shade denotes statistical significance at the 10\% level. The red vertical line indicates the local average treatment effect. The baseline sample consists of 24,615 individuals in the \citeasnoun{taubman2014medicaid}  sample with non-missing information on pre-lottery emergency department utilization and SNAP/TANF receipt. The estimates displayed exclude less than half a percentile at the top and bottom of the distribution, resulting in the axes corresponding approximately to the percentile range $[0.5\%, 99.5\%]$. Bin size is chosen according to the Freedman-Diaconis rule.
  \end{minipage}
\end{figure}

%\begin{figure}
%  \centering
%  \caption{Group average treatment effects of winning the lottery (Number of visits)}
%  \label{fig:num-visit-gate-itt}
%  \begin{minipage}{\linewidth}
%    \makebox[\textwidth][c]{
%      \includegraphics[width=1.2\linewidth]{Figures/Figures_final/itt/fgate-num-visit-ed-itt.pdf}
%    }
%    \footnotesize
%      \emph{Notes}: This figure plots the group average treatment effects of  winning the lottery (and being invited to apply for Medicaid)  on the number of overall ED visit based on generalized random forests for binary indicators of gender, whether the individual provided a phone number, whether the individual requested English language materials, whether the individual signed up in the first week, pre-lottery SNAP receipt, an indicator of household size of at least two members, whether individual is at least 50 years old, and whether the individual had any pre-lottery primary care treatable ED visits. The darker tone colors denote statistical significance at the 10\% level. The  red horizontal line indicates the average effect. The baseline sample consists of 24,615 individuals in the \citeasnoun{taubman2014medicaid}  sample with non-missing information on pre-lottery emergency department utilization and SNAP/TANF receipt.
%  \end{minipage}
%\end{figure}

\begin{figure}
  \centering
  \caption{Variable importance scores in growing causal forest (Number of visits)}
  \label{fig:num-visit-ed-itt}
  \begin{minipage}{\linewidth}
    \makebox[\textwidth][c]{
      \includegraphics[width=\linewidth]{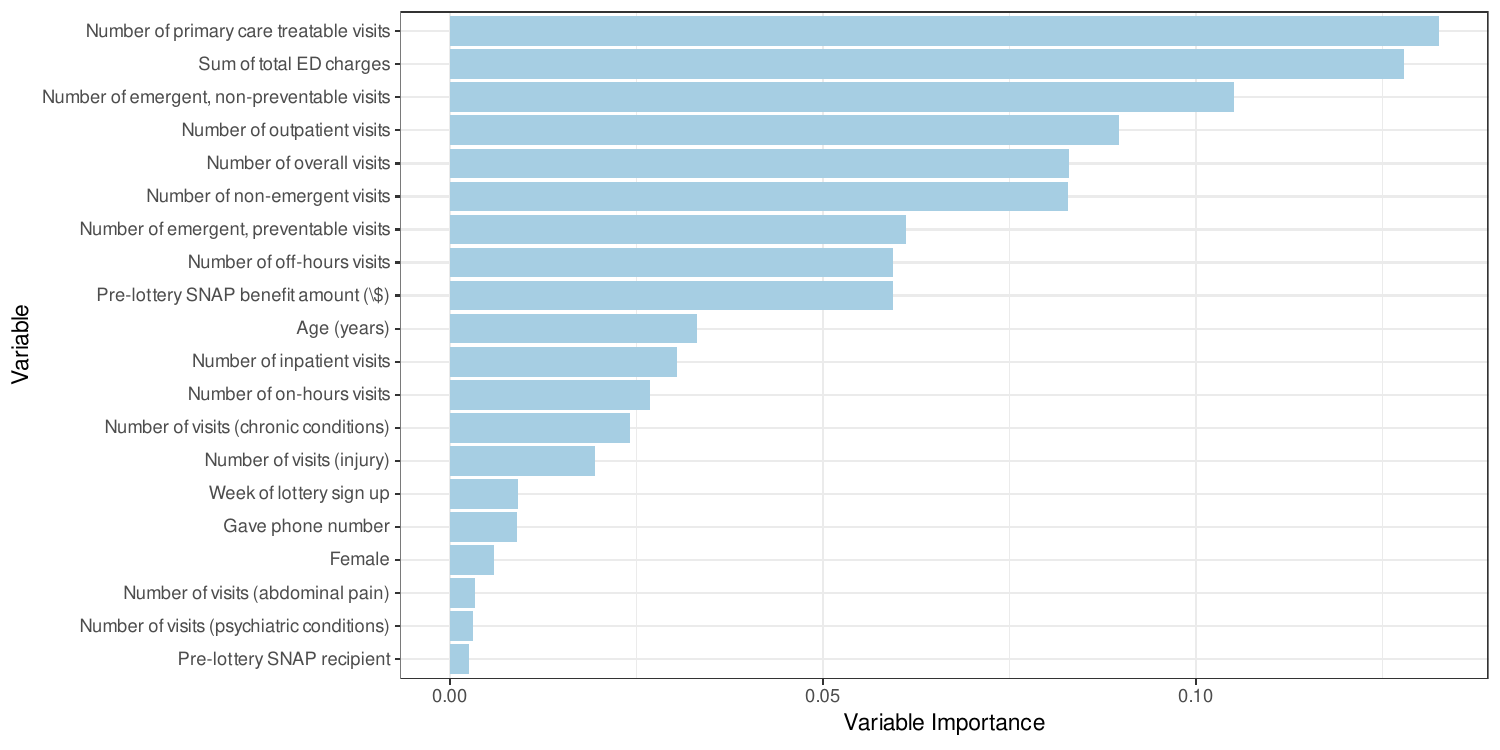}
    }
    \footnotesize
    \emph{Notes} This figure shows the estimates of variable importance for the top 20 characteristics used in growing the generalized random forests in estimating the CATE of winning the lottery (and being invited to apply for Medicaid) for the number of overall ED visits. The variable importance measure is a simple weighted sum of the proportion of times a variable is used in a splitting step at each depth in growing the causal forest, thus, capturing how important a variable is for driving treatment effect heterogeneity. The sample consists of 24,615 individuals in the \citeasnoun{taubman2014medicaid}  sample with non-missing information on pre-lottery emergency department utilization and SNAP/TANF receipt.
  \end{minipage}
\end{figure}

\clearpage

%==================================== Tables ===================================

\begin{singlespace}
  \input{Tables/Tables_final/itt/tmain-itt.tex}
  \input{Tables/Tables_final/itt/tqte-itt.tex}
  \clearpage
  \input{Tables/Tables_final/itt/tgates-any-visit-itt.tex}
  \clearpage
  \input{Tables/Tables_final/itt/tgates-num-visit-itt.tex}
  \clearpage
  \input{Tables/Tables_final/itt/tcomp-any-visit-itt.tex}
  \clearpage
  \input{Tables/Tables_final/itt/tcomp-num-visit-itt.tex}
  \clearpage
  \input{Tables/Tables_final/itt/tvarimp-visit-itt.tex}
\end{singlespace}

\end{document}

%% file: Tables/Tables_final/tdescriptives.tex
\begin{table}%[H]
\caption{\label{tab:desc.base}Descriptive Statistics}
\centering
\fontsize{9}{11}\selectfont
\begin{threeparttable}
\begin{tabular}[t]{lrrrrr}
\toprule
Variable & Mean & SD & Median & Min & Max\\
\midrule
\addlinespace[0.3em]
\multicolumn{6}{l}{\textbf{Lottery list and baseline characteristics}}\\
\hspace{1em}Age (years) & 39.60 & 12.05 & 39.0 & 20 & 63\\
\hspace{1em}Gave phone number & 0.87 & 0.34 & 1.0 & 0 & 1\\
\hspace{1em}English as preferred language & 0.86 & 0.34 & 1.0 & 0 & 1\\
\hspace{1em}Female & 0.55 & 0.50 & 1.0 & 0 & 1\\
\hspace{1em}Week of lottery sign up & 1.58 & 1.62 & 1.0 & 0 & 5\\
\hspace{1em}Provided P.O. box address & 0.03 & 0.16 & 0.0 & 0 & 1\\
\hspace{1em}Signed up self for lottery & 0.90 & 0.30 & 1.0 & 0 & 1\\
\hspace{1em}Pre-lottery SNAP recipient & 0.54 & 0.50 & 1.0 & 0 & 1\\
\hspace{1em}Pre-lottery SNAP benefit amount (\$) & 1331.94 & 1863.76 & 522.5 & 0 & 20745\\
\hspace{1em}Pre-lottery TANF recipient & 0.02 & 0.15 & 0.0 & 0 & 1\\
\hspace{1em}Pre-lottery TANF benefit amount (\$) & 96.16 & 694.70 & 0.0 & 0 & 16031\\
\addlinespace[0.3em]
\multicolumn{6}{l}{\textbf{Pre-randomization ED use}}\\
\hspace{1em}Number of overall visits & 0.77 & 1.86 & 0.0 & 0 & 17\\
\hspace{1em}Number of inpatient visits & 0.09 & 0.41 & 0.0 & 0 & 6\\
\hspace{1em}Number of outpatient visits & 0.69 & 1.71 & 0.0 & 0 & 16\\
\hspace{1em}Number of on-hours visits & 0.45 & 1.23 & 0.0 & 0 & 13\\
\hspace{1em}Number of off-hours visits & 0.33 & 0.92 & 0.0 & 0 & 10\\
\hspace{1em}Number of emergent, non-preventable visits & 0.16 & 0.50 & 0.0 & 0 & 9\\
\hspace{1em}Number of emergent, preventable visits & 0.06 & 0.28 & 0.0 & 0 & 6\\
\hspace{1em}Number of primary care treatable visits & 0.27 & 0.75 & 0.0 & 0 & 12\\
\hspace{1em}Number of non-emergent visits & 0.16 & 0.58 & 0.0 & 0 & 12\\
\hspace{1em}Number ambulatory-care-sensitive visits & 0.05 & 0.30 & 0.0 & 0 & 5\\
\hspace{1em}Number of visits (chronic conditions) & 0.14 & 0.64 & 0.0 & 0 & 9\\
\hspace{1em}Number of visits (injury) & 0.17 & 0.56 & 0.0 & 0 & 6\\
\hspace{1em}Number of visits (skin conditions) & 0.05 & 0.31 & 0.0 & 0 & 5\\
\hspace{1em}Number of visits (abdominal pain) & 0.04 & 0.27 & 0.0 & 0 & 5\\
\hspace{1em}Number of visits (back pain) & 0.03 & 0.26 & 0.0 & 0 & 5\\
\hspace{1em}Number of visits (chest pain) & 0.02 & 0.18 & 0.0 & 0 & 3\\
\hspace{1em}Number of visits (headache) & 0.02 & 0.23 & 0.0 & 0 & 4\\
\hspace{1em}Number of visits (mood disorders) & 0.02 & 0.23 & 0.0 & 0 & 5\\
\hspace{1em}Number of visits (psychiatric conditions) & 0.06 & 0.38 & 0.0 & 0 & 6\\
\hspace{1em}Sum of total ED charges & 894.85 & 2593.46 & 0.0 & 0 & 42315\\
\addlinespace[0.3em]
\multicolumn{6}{l}{\textbf{ED use outcomes}}\\
\hspace{1em}Any overall visit & 0.34 & 0.47 & 0.0 & 0 & 1\\
\hspace{1em}Any inpatient visit & 0.07 & 0.26 & 0.0 & 0 & 1\\
\hspace{1em}Any outpatient visit & 0.32 & 0.47 & 0.0 & 0 & 1\\
\hspace{1em}Any emergent, non-preventable visit & 0.16 & 0.37 & 0.0 & 0 & 1\\
\hspace{1em}Any emergent, preventable visit & 0.10 & 0.30 & 0.0 & 0 & 1\\
\hspace{1em}Any primary care treatable visit & 0.22 & 0.41 & 0.0 & 0 & 1\\
\hspace{1em}Any non-emergent visit & 0.16 & 0.37 & 0.0 & 0 & 1\\
\hspace{1em}Number of overall visits & 1.00 & 2.41 & 0.0 & 0 & 22\\
\hspace{1em}Number of inpatient visits & 0.11 & 0.53 & 0.0 & 0 & 7\\
\hspace{1em}Number of outpatient visits & 0.89 & 2.21 & 0.0 & 0 & 21\\
\hspace{1em}Number of emergent, non-preventable visits & 0.21 & 0.69 & 0.0 & 0 & 14\\
\hspace{1em}Number of emergent, preventable visits & 0.07 & 0.35 & 0.0 & 0 & 8\\
\hspace{1em}Number of primary care treatable visits & 0.35 & 0.96 & 0.0 & 0 & 16\\
\hspace{1em}Number of non-emergent visits & 0.20 & 0.70 & 0.0 & 0 & 13\\
\bottomrule
\end{tabular}
\begin{tablenotes}[para]
\item \emph{Notes}: This table presents descriptive statistics based on the Oregon Health Insurance Experiment data. The sample consists of 24,615 individuals in the \citet{taubman2014medicaid} sample with non-missing information on pre-lottery emergency department utilization and SNAP/TANF receipt. The pre-randomization SNAP and TANF benefits are total household benefits received from January 2007 through the individual's lottery notification date. The pre-randomization ED use measures are with respect to utilization between January 1, 2007 and March 9, 2008.
\end{tablenotes}
\end{threeparttable}
\end{table}

%% file: Tables/Tables_final/tqte.tex
\begin{singlespace}
\begin{table}[t]
\caption{\label{tab:qte.base}Empirical quantiles of the distribution of individualized treatment effects of Medicaid on ED use}
\centering
\fontsize{9}{11}\selectfont
\begin{threeparttable}
\begin{tabular}[t]{lrrrrrrr}
\toprule
{Variable} & {LATE} & {Min} & {25\%} & {50\%} & {75\%} & {Max}& {Share non-pos.}\\
\midrule
\addlinespace[0.3em]
\multicolumn{8}{l}{\textbf{Extensive margin}}\\
\hspace{1em}Any overall visit                   & 0.045  & -1.122 & -0.031 & 0.070  & 0.151 & 0.894 & 0.31 \\
\hspace{1em}Any inpatient visit                 & -0.003 & -0.415 & -0.049 & -0.010 & 0.031 & 0.619 & 0.57 \\
\hspace{1em}Any outpatient visit                & 0.057  & -1.121 & -0.013 & 0.078  & 0.160 & 0.800 & 0.28 \\
\hspace{1em}Any emergent, non-preventable visit & -0.007 & -0.622 & -0.075 & 0.003  & 0.072 & 0.583 & 0.49 \\
\hspace{1em}Any emergent, preventable visit     & 0.035  & -0.313 & -0.014 & 0.030  & 0.077 & 0.532 & 0.32 \\
\hspace{1em}Any primary care treatable visit    & 0.048  & -0.813 & -0.011 & 0.067  & 0.141 & 0.785 & 0.28 \\
\hspace{1em}Any non-emergent visit              & 0.073  & -0.607 & -0.004 & 0.060  & 0.132 & 0.604 & 0.27 \\
\addlinespace[0.3em]
\multicolumn{8}{l}{\textbf{Intensive margin}}\\
\hspace{1em}Number of overall visits                   & 0.354  & -1.503 & 0.006  & 0.273  & 0.567 & 5.939 & 0.24 \\
\hspace{1em}Number of inpatient visits                 & -0.012 & -0.941 & -0.074 & -0.013 & 0.046 & 1.072 & 0.56 \\
\hspace{1em}Number of outpatient visits                & 0.360  & -1.199 & 0.052  & 0.291  & 0.571 & 5.828 & 0.20 \\
\hspace{1em}Number of emergent, non-preventable visits & 0.028  & -0.590 & -0.066 & 0.020  & 0.109 & 3.025 & 0.44 \\
\hspace{1em}Number of emergent, preventable visits     & 0.038  & -1.533 & -0.015 & 0.026  & 0.077 & 0.724 & 0.34 \\
\hspace{1em}Number of primary care treatable visits    & 0.161  & -0.550 & 0.018  & 0.131  & 0.255 & 3.689 & 0.22 \\
\hspace{1em}Number of non-emergent visits              & 0.107  & -0.663 & -0.008 & 0.061  & 0.143 & 2.606 & 0.28 \\
\bottomrule
\end{tabular}
\begin{tablenotes}[para]
\item \emph{Notes}: This table reports selected quantiles of the individualized treatment effects of Medicaid on ED use based on generalized random forests. The first column reports the average effect. The final column reports the share of ITEs that are non-positive. The baseline sample consists of 24,615 individuals in the \citeasnoun{taubman2014medicaid} sample with non-missing information on pre-lottery emergency department utilization and SNAP/TANF receipt.
\end{tablenotes}
\end{threeparttable}
\end{table}
\end{singlespace}

%% file: Tables/Tables_final/tmain.tex
\begin{table}%[H]
\caption{\label{tab:main.base}Treatment effect estimates of Medicaid coverage on ED use}
\centering
\fontsize{9}{11}\selectfont
\makebox[\textwidth][c]{
\begin{threeparttable}
\begin{tabular}[t]{lrrrrrr}
\toprule
\multicolumn{1}{c}{ } & \multicolumn{3}{c}{GRF estimates} & \multicolumn{3}{c}{Linear estimates} \\
\multicolumn{1}{c}{ } & \multicolumn{3}{c}{} & \multicolumn{3}{c}{(\citet{taubman2014medicaid} Replication)} \\
\cmidrule(l{3pt}r{3pt}){2-4} \cmidrule(l{3pt}r{3pt}){5-7}
{Variable} & {LATE} & {SE} & {$p$-value} & {LATE} & {SE} & {$p$-value}\\
\midrule
\addlinespace[0.3em]
\multicolumn{7}{l}{\textbf{Extensive margin}}\\
\hspace{1em}Any overall visit & 0.045 & 0.028 & 0.111 & 0.069 & 0.024 & 0.004\\
\hspace{1em}Any inpatient visit & -0.003 & 0.015 & 0.855 & -0.011 & 0.013 & 0.424\\
\hspace{1em}Any outpatient visit & 0.057 & 0.028 & 0.037 & 0.082 & 0.024 & 0.001\\
\hspace{1em}Any emergent, non-preventable visit & -0.007 & 0.021 & 0.743 & 0.007 & 0.019 & 0.731\\
\hspace{1em}Any emergent, preventable visit & 0.035 & 0.016 & 0.029 & 0.039 & 0.015 & 0.009\\
\hspace{1em}Any primary care treatable visit & 0.048 & 0.024 & 0.040 & 0.069 & 0.021 & 0.001\\
\hspace{1em}Any non-emergent visit & 0.073 & 0.021 & 0.000 & 0.064 & 0.019 & 0.001\\
\addlinespace[0.3em]
\multicolumn{7}{l}{\textbf{Intensive margin}}\\
\hspace{1em}Number of overall visits & 0.354 & 0.114 & 0.002 & 0.375 & 0.106 & 0.000\\
\hspace{1em}Number of inpatient visits & -0.012 & 0.028 & 0.676 & -0.017 & 0.026 & 0.516\\
\hspace{1em}Number of outpatient visits & 0.360 & 0.106 & 0.001 & 0.387 & 0.098 & 0.000\\
\hspace{1em}Number of emergent, non-preventable visits & 0.028 & 0.034 & 0.414 & 0.040 & 0.032 & 0.223\\
\hspace{1em}Number of emergent, preventable visits & 0.038 & 0.018 & 0.039 & 0.036 & 0.017 & 0.034\\
\hspace{1em}Number of primary care treatable visits & 0.161 & 0.049 & 0.001 & 0.171 & 0.045 & 0.000\\
\hspace{1em}Number of non-emergent visits & 0.107 & 0.036 & 0.003 & 0.105 & 0.033 & 0.002\\
\bottomrule
\end{tabular}
\begin{tablenotes}[para]
\item \emph{Notes}: This table reports the estimates of Medicaid coverage on ED use based on generalized random forests and a linear IV model. The sample consists of 24,615 individuals in the \citeasnoun{taubman2014medicaid} sample with non-missing information on pre-lottery emergency department utilization and SP/TANF receipt.
\end{tablenotes}
\end{threeparttable}
}
\end{table}

% old, simple def
% \hspace{1em}Any emergent, non-preventable visit & -0.007 & 0.021 & 0.743 & 0.003 & 0.019 & 0.870\\
% \hspace{1em}Any emergent, preventable visit & 0.035 & 0.016 & 0.029 & 0.036 & 0.016 & 0.021\\
% \hspace{1em}Any primary care treatable visit & 0.048 & 0.024 & 0.040 & 0.064 & 0.021 & 0.002\\
% \hspace{1em}Any non-emergent visit & 0.073 & 0.021 & 0.000 & 0.067 & 0.019 & 0.000\\

%% file: Tables/Tables_final/tgates-any-visit.tex
\begin{table}%[H]
\caption{\label{tab:gates.any}Group average treatment effects of Medicaid on the propensity of ED use}
\centering
\fontsize{9}{11}\selectfont
\makebox[\textwidth][c]{
\begin{threeparttable}
\begin{tabular}[t]{lrrrHHHr}
\toprule
\multicolumn{1}{c}{ } & \multicolumn{3}{c}{GRF estimates} & \multicolumn{3}{H}{Panel B: Linear estimates} & \multicolumn{1}{c}{ } \\
\cmidrule(l{3pt}r{3pt}){2-4} \cmidrule(l{3pt}r{3pt}){5-7}
{Group}                                                 & {GATE} & {SE} & {$p$-value} & {GATE} & {SE} & {$p$-value} & {\%~N} \\
\midrule
%LATE                                                    & 0.04   & 0.03 & 0.12        & 0.08   & 0.03 & 0.00        & 100.00 \\
Female:                                                 & 0.00   & 0.04 & 0.97        & 0.03   & 0.04 & 0.34        & 0.55   \\
Male:                                                   & 0.10   & 0.04 & 0.01        & 0.12   & 0.03 & 0.00        & 0.45   \\
Gave phone number: No                                   & 0.00   & 0.08 & 0.99        & 0.03   & 0.08 & 0.66        & 0.13   \\
Gave phone number: Yes                                  & 0.05   & 0.03 & 0.11        & 0.08   & 0.03 & 0.00        & 0.87   \\
English as preferred language: No                       & 0.01   & 0.07 & 0.88        & 0.02   & 0.06 & 0.73        & 0.14   \\
English as preferred language: Yes                      & 0.05   & 0.03 & 0.12        & 0.08   & 0.03 & 0.01        & 0.86   \\
First week sign-up: No                                  & 0.04   & 0.04 & 0.31        & 0.09   & 0.03 & 0.01        & 0.62   \\
First week sign-up: Yes                                 & 0.06   & 0.04 & 0.20        & 0.06   & 0.04 & 0.16        & 0.38   \\
Pre-lottery SNAP recipient: No                          & -0.01  & 0.05 & 0.79        & 0.02   & 0.05 & 0.62        & 0.46   \\
Pre-lottery SNAP recipient: Yes                         & 0.09   & 0.03 & 0.01        & 0.08   & 0.03 & 0.01        & 0.54   \\
Pre-lottery TANF recipient: No                          & 0.04   & 0.03 & 0.17        & 0.07   & 0.03 & 0.00        & 0.98   \\
Pre-lottery TANF recipient: Yes                         & 0.05   & 0.18 & 0.79        & 0.08   & 0.37 & 0.82        & 0.02   \\
Age $\geq$ 50: No                                       & 0.08   & 0.03 & 0.02        & 0.10   & 0.03 & 0.00        & 0.75   \\
Age $\geq$ 50: Yes                                      & -0.06  & 0.05 & 0.24        & 0.01   & 0.04 & 0.77        & 0.25   \\
Two+ household members on lottery list: No              & 0.03   & 0.03 & 0.42        & 0.05   & 0.03 & 0.06        & 0.80   \\
Two+ household members on lottery list: Yes             & 0.09   & 0.06 & 0.15        & 0.19   & 0.07 & 0.00        & 0.20   \\
Any pre-lottery ED visit No                             & 0.03   & 0.03 & 0.34        & 0.07   & 0.03 & 0.02        & 0.69   \\
Any pre-lottery ED visit: Yes                           & 0.06   & 0.05 & 0.24        & 0.07   & 0.04 & 0.08        & 0.31   \\
Any pre-lottery on-hours ED visit: No                   & 0.03   & 0.03 & 0.35        & 0.07   & 0.03 & 0.01        & 0.77   \\
Any pre-lottery on-hours ED visit: Yes                  & 0.07   & 0.05 & 0.18        & 0.08   & 0.05 & 0.12        & 0.23   \\
Any pre-lottery off-hours ED visit: No                  & 0.04   & 0.03 & 0.21        & 0.06   & 0.03 & 0.03        & 0.81   \\
Any pre-lottery off-hours ED visit: Yes                 & 0.06   & 0.06 & 0.29        & 0.07   & 0.05 & 0.16        & 0.19   \\
Any pre-lottery emergent, non-preventable ED visit: No  & 0.04   & 0.03 & 0.19        & 0.07   & 0.03 & 0.01        & 0.87   \\
Any pre-lottery emergent, non-preventable ED visit: Yes & 0.06   & 0.07 & 0.40        & 0.07   & 0.06 & 0.28        & 0.13   \\
Any pre-lottery emergent, preventable ED visit: No      & 0.05   & 0.03 & 0.10        & 0.09   & 0.03 & 0.00        & 0.92   \\
Any pre-lottery emergent, preventable ED visit: Yes     & -0.05  & 0.08 & 0.55        & 0.02   & 0.07 & 0.83        & 0.08   \\
Any pre-lottery primary care treatable ED visit: No     & 0.02   & 0.03 & 0.59        & 0.06   & 0.03 & 0.02        & 0.81   \\
Any pre-lottery primary care treatable ED visit: Yes    & 0.13   & 0.06 & 0.03        & 0.12   & 0.05 & 0.03        & 0.19   \\
Any pre-lottery non-emergent ED visit: No               & 0.03   & 0.03 & 0.34        & 0.07   & 0.03 & 0.01        & 0.86   \\
Any pre-lottery non-emergent ED visit: Yes              & 0.08   & 0.07 & 0.21        & 0.09   & 0.06 & 0.17        & 0.14   \\
\bottomrule
\end{tabular}
\begin{tablenotes}[para]
\item \emph{Notes}: This table reports the group average treatment effects of Medicaid based on generalized random forests. The baseline sample consists of 24,615 individuals in the \citeasnoun{taubman2014medicaid} sample with non-missing information on pre-lottery emergency department utilization and SNAP/TANF receipt.
\end{tablenotes}
\end{threeparttable}
}
\end{table}

%\begin{tablenotes}[para]
%	\item \emph{Notes}: This table reports the group average treatment effects of Medicaid based on generalized random forests in Panel A and those based on a linear IV method's subsample analysis in Panel B. The first row reproduces the overall average effect. The baseline sample consists of 24,615 individuals in the \citeasnoun{taubman2014medicaid} sample with non-missing information on pre-lottery emergency department utilization and SNAP/TANF receipt.
%\end{tablenotes}

% old, simple def
% Any pre-lottery emergent, non-preventable ED visit: No  & 0.02   & 0.03 & 0.45        & 0.06   & 0.03 & 0.03        & 0.79   \\
% Any pre-lottery emergent, non-preventable ED visit: Yes & 0.08   & 0.06 & 0.17        & 0.10   & 0.05 & 0.06        & 0.21   \\
% Any pre-lottery emergent, preventable ED visit: No      & 0.04   & 0.03 & 0.18        & 0.08   & 0.03 & 0.00        & 0.90   \\
% Any pre-lottery emergent, preventable ED visit: Yes     & 0.08   & 0.08 & 0.29        & 0.10   & 0.07 & 0.13        & 0.10   \\
% Any pre-lottery primary care treatable ED visit: No     & 0.02   & 0.03 & 0.49        & 0.06   & 0.03 & 0.03        & 0.74   \\
% Any pre-lottery primary care treatable ED visit: Yes    & 0.09   & 0.05 & 0.08        & 0.09   & 0.05 & 0.05        & 0.26   \\
% Any pre-lottery non-emergent ED visit: No               & 0.04   & 0.03 & 0.25        & 0.07   & 0.03 & 0.01        & 0.86   \\
% Any pre-lottery non-emergent ED visit: Yes              & 0.08   & 0.07 & 0.22        & 0.08   & 0.06 & 0.18        & 0.14   \\

%% file: Tables/Tables_final/tcomp-any-visit.tex
\begin{table}%[H]
  \caption{\label{tab:comp.any}Characteristics of individuals who increased and decreased ED use}
  \centering
  \fontsize{9}{11}\selectfont
  \makebox[\textwidth][c]{
    \begin{threeparttable}
      \begin{tabular}[t]{>{\raggedright\arraybackslash}p{20em}rrr@{}l}
        \toprule
                                                   & \multicolumn{1}{c}{Increased} & \multicolumn{1}{c}{Decreased} &                                      \\
        Variable                                   & \multicolumn{1}{c}{ED use}    & \multicolumn{1}{c}{ED use}    & \multicolumn{2}{c}{Difference}       \\
                                                   & \multicolumn{1}{c}{(1)}       & \multicolumn{1}{c}{(2)}       & \multicolumn{2}{c}{(2)-(1)}       \\
        \midrule \addlinespace[1ex]
        \multicolumn{5}{l}{\hspace{-1ex}\textbf{Lottery list characteristics}}                                                                            \\
Age (years)                                & 38.76   & 41.48  & -2.72  & ***  \\
Gave phone number                          & 0.87    & 0.87   & 0.00   &    \\
English as preferred language              & 0.87    & 0.85   & 0.02   & ***   \\
Female                                     & 0.51    & 0.63   & -0.12  & ***  \\
Week of lottery sign up                    & 1.57    & 1.61   & -0.04  &    \\
Provided P.O. box address                  & 0.02    & 0.03   & -0.01  & ***  \\
Signed up self for lottery                 & 0.89    & 0.92   & -0.03  & ***  \\
Pre-lottery SNAP recipient                 & 0.62    & 0.35   & 0.27   & ***   \\
Pre-lottery SNAP benefit amount (\$)       & 1607.33 & 713.38 & 893.95 & *** \\
Pre-lottery TANF recipient                 & 0.03    & 0.01   & 0.02   & ***   \\
Pre-lottery TANF benefit amount (\$)       & 111.08  & 63.09  & 47.99  & ***  \\
        \multicolumn{5}{l}{\hspace{-1ex}\textbf{Pre-lottery ED usage}} \\
Number of overall visits                   & 0.88    & 0.49   & 0.39   & ***   \\
Number of inpatient visits                 & 0.09    & 0.07   & 0.02   & ***   \\
Number of outpatient visits                & 0.78    & 0.42   & 0.36   & ***   \\
Number of on-hours visits                  & 0.51    & 0.30   & 0.21   & ***   \\
Number of off-hours visits                 & 0.37    & 0.20   & 0.17   & ***   \\
Number of emergent, non-preventable visits & 0.18    & 0.09   & 0.09   & ***   \\
Number of emergent, preventable visits     & 0.07    & 0.04   & 0.03   & ***   \\
Number of primary care treatable visits    & 0.30    & 0.17   & 0.13   & ***   \\
Number of non-emergent visits              & 0.18    & 0.10   & 0.08   & ***   \\
Number ambulatory-care-sensitive visits    & 0.05    & 0.04   & 0.01   & ***   \\
Number of visits (chronic conditions)      & 0.14    & 0.11   & 0.03   & ***   \\
Number of visits (injury)                  & 0.20    & 0.09   & 0.11   & ***   \\
Number of visits (skin conditions)         & 0.05    & 0.03   & 0.02   & ***   \\
Number of visits (abdominal pain)          & 0.04    & 0.02   & 0.02   & ***   \\
Number of visits (back pain)               & 0.04    & 0.02   & 0.02   & ***   \\
Number of visits (chest pain)              & 0.02    & 0.02   & 0.00   & ***   \\
Number of visits (headache)                & 0.03    & 0.02   & 0.01   & ***   \\
Number of visits (mood disorders)          & 0.02    & 0.02   & 0.00   &    \\
Number of visits (psychiatric conditions)  & 0.06    & 0.05   & 0.01   & *     \\
Sum of total ED charges                    & 1013.28 & 584.41 & 428.87 & *** \\
        \addlinespace[0.3em]
        \hline
        \multicolumn{5}{l}{\textbf{}}                                                                                                                     \\
        N                                          & 16816                         & 7797                          & 24613                                \\
        \bottomrule
      \end{tabular}
      \begin{tablenotes}[para]
        \item  \emph{Notes}: This table reports the means of individual characteristics and pre-randomization ED use for those estimated to increase and decrease ED use upon receiving Medicaid coverage based on the generalized random forest ITE estimates. The sample consists of 24,613 individuals in the \citeasnoun{taubman2014medicaid} sample with non-missing information on pre-lottery emergency department utilization and SNAP/TANF receipt.
      \end{tablenotes}
    \end{threeparttable}
  }
\end{table}

%% file: Tables/Tables_final/tcomp-lp.tex
\begin{table}
\caption{\label{tab:comp.rule}Individuals selected by lottery and the optimal linear program}
\centering
\fontsize{9}{11}\selectfont
\makebox[\textwidth][c]{
\begin{threeparttable}
\begin{tabular}[t]{lcP{3cm}r@{}l}
\toprule
Variable & Selected by lottery & Selected by optimal linear program & \multicolumn{2}{c}{Difference}\\
         & (1) & (2) & \multicolumn{2}{c}{(2)-(1)}\\
\midrule
\addlinespace[0.3em]
\multicolumn{5}{l}{\textbf{Lottery list and baseline characteristics}}\\
\hspace{1em}Age (years)                                          & 39.50   & 40.16   & 0.66    & ***    \\
\hspace{1em}Gave phone number                                    & 0.88    & 0.87    & -0.01   &    \\
\hspace{1em}English as preferred language                        & 0.85    & 0.79    & -0.06   & ***   \\
\hspace{1em}Female                                               & 0.54    & 0.53    & -0.01   &    \\
\hspace{1em}Week of lottery sign up                              & 1.55    & 1.66    & 0.11    & ***    \\
\hspace{1em}Provided P.O. box address                            & 0.03    & 0.03    & 0.00    &    \\
\hspace{1em}Signed up self for lottery                           & 0.85    & 0.85    & 0.00    &    \\
\hspace{1em}Pre-lottery SNAP recipient                           & 0.53    & 0.43    & -0.10   & ***   \\
\hspace{1em}Pre-lottery SNAP benefit amount (\$) & 1376.53 & 1120.72 & -255.81 & *** \\
\hspace{1em}Pre-lottery TANF recipient                           & 0.02    & 0.02    & 0.00    & ***    \\
\hspace{1em}Pre-lottery TANF benefit amount (\$) & 100.09  & 72.82   & -27.27  & ***  \\
\hspace{1em}Number of overall visits                             & 0.71    & 0.44    & -0.27   & ***   \\
\addlinespace[0.3em]
\multicolumn{5}{l}{\textbf{Pre-randomization ED use}}                                          \\
\hspace{1em}Number of inpatient visits                           & 0.08    & 0.05    & -0.03   & ***   \\
\hspace{1em}Number of outpatient visits                          & 0.63    & 0.39    & -0.24   & ***   \\
\hspace{1em}Number of on-hours visits                            & 0.41    & 0.25    & -0.16   & ***   \\
\hspace{1em}Number of off-hours visits                           & 0.30    & 0.19    & -0.11   & ***   \\
\hspace{1em}Number of emergent, non-preventable visits           & 0.15    & 0.09    & -0.06   & ***   \\
\hspace{1em}Number of emergent, preventable visits               & 0.05    & 0.03    & -0.02   & ***   \\
\hspace{1em}Number of primary care treatable visits              & 0.25    & 0.15    & -0.10   & ***   \\
\hspace{1em}Number of non-emergent visits                        & 0.14    & 0.08    & -0.06   & ***   \\
\hspace{1em}Number ambulatory-care-sensitive visits              & 0.05    & 0.03    & -0.02   & ***   \\
\hspace{1em}Number of visits (chronic conditions)                & 0.12    & 0.08    & -0.04   & ***   \\
\hspace{1em}Number of visits (injury)                            & 0.17    & 0.10    & -0.07   & ***   \\
\hspace{1em}Number of visits (skin conditions)                   & 0.04    & 0.03    & -0.01   & ***   \\
\hspace{1em}Number of visits (abdominal pain)                    & 0.03    & 0.02    & -0.01   & ***   \\
\hspace{1em}Number of visits (back pain)                         & 0.03    & 0.02    & -0.01   & ***   \\
\hspace{1em}Number of visits (chest pain)                        & 0.02    & 0.01    & -0.01   & ***   \\
\hspace{1em}Number of visits (headache)                          & 0.02    & 0.01    & -0.01   & ***   \\
\hspace{1em}Number of visits (mood disorders)                    & 0.02    & 0.01    & -0.01   & **    \\
\hspace{1em}Number of visits (psychiatric conditions)            & 0.05    & 0.03    & -0.02   & ***   \\
\hspace{1em}Sum of total ED charges                              & 808.36  & 504.40  & -303.96 & *** \\
\hspace{1em}Number of overall visits                             & 0.98    & 0.43    & -0.55   & ***   \\
\addlinespace[0.3em]
\multicolumn{5}{l}{\textbf{ED use outcomes}}                                                   \\
\hspace{1em}Number of inpatient visits                           & 0.10    & 0.06    & -0.04   & ***   \\
\hspace{1em}Number of outpatient visits                          & 0.87    & 0.36    & -0.51   & ***   \\
\hspace{1em}Number of emergent, non-preventable visits           & 0.20    & 0.10    & -0.10   & ***   \\
\hspace{1em}Number of emergent, preventable visits               & 0.07    & 0.04    & -0.03   & ***   \\
\hspace{1em}Number of primary care treatable visits              & 0.35    & 0.17    & -0.18   & ***   \\
\hspace{1em}Number of non-emergent visits                        & 0.20    & 0.02    & -0.18   & ***   \\
\addlinespace[0.3em]
\hline
\multicolumn{4}{l}{\textbf{}}\\
N & 9607 & 9607 & \\
\bottomrule
\end{tabular}
\begin{tablenotes}[para]
\item \emph{Notes}: This table reports the means of individual characteristics, pre-randomization ED use and post-treatment outcomes for those selected by the lottery and selected by the optimal treatment assignment solution based on the optimal linear program which minimizes non-emergent ED use under the restriction that the same number of people are treated as within the lottery. The sample consists of 24,605 individuals in the \citeasnoun{taubman2014medicaid} sample with non-missing information on pre-lottery emergency department utilization and SNAP/TANF receipt.
\end{tablenotes}
\end{threeparttable}
}
\end{table}

%% file: Figures/Figures_final/optimal-policy-d2.tex
\begin{figure}
  \centering
  \caption{Decision rule for targeting outreach based on non-emergent ED use}
  \label{fig:optimal-policy-d2}
\fontsize{8}{10}\selectfont
  \makebox[\textwidth][c]{
\begin{threeparttable}

      % only admin/limited vars, binary type visits, emergent visit types aggregated, charges rounded to hundreds
      \begin{istgame}
    \sffamily
        \xtShowEndPoints
        \xtdistance{15mm}{60mm}
        \istroot(0)[initial node]{Prior outpatient hospital visits $\leq$ 2}
        \istb{\texttt{TRUE}}[al]
        \istb{\texttt{FALSE}}[ar]
        \endist
        \xtdistance{15mm}{29mm}
        \istroot(l)(0-1)<135>{Any prior emergent hospital visits $\leq$ 0}
        % \istb{\texttt{TRUE}}[l]{\text{No offer}}
        % \istb{\texttt{FALSE}}[r]{\text{Solicit application}}
        \istb{\texttt{TRUE}}[l]{\begin{matrix}\text{Select}\\\mathsf{\num{20,605}}\\\mathsf{\num{0.84}}\end{matrix}}
        \istb{\texttt{FALSE}}[r]{\begin{matrix}\text{Do not select}\\\mathsf{\num{2,114}}\\\mathsf{\num{0.09}}\end{matrix}}
        \endist
        \istroot(r)(0-2)<45>{Prior visits req.\ hospitalization $\leq$ 1}
        % \istb{\texttt{TRUE}}[l]{\text{No offer}}
        % \istb{\texttt{FALSE}}[r]{\text{Solicit application}}
        \istb{\texttt{TRUE}}[l]{\begin{matrix}\text{Select}\\ \mathsf{\num{1,742}}\\\mathsf{\num{0.07}}\end{matrix}}
        \istb{\texttt{FALSE}}[r]{\begin{matrix}\text{Do not select}\\\mathsf{\num{144}}\\\mathsf{\num{0.01}}\end{matrix}}
        \endist
      \end{istgame}

\begin{tablenotes}[para]
\footnotesize
  \item \emph{Notes}: This figure shows the optimal policy assignment rule based on the doubly robust scores of the conditional average treatment effect (CATE) of being randomly selected as a lottery winner in the Oregon Health Insurance Experiment using the efficient policy learning framework of \citet{athey2021policy}. The sample consists of 24,615 individuals in the \citet{taubman2014medicaid}  sample with non-missing information on pre-lottery emergency department utilization and SNAP/TANF receipt. 
\end{tablenotes}
\end{threeparttable}
}
\end{figure}

%%% Local Variables:
%%% mode: latex
%%% TeX-master: "../../medicaid_hte_paper"
%%% End:

%% file: Tables/Tables_final/tvarimp-visit.tex
\begin{table}[t]

\caption{\label{tab:varimp.visit}Variable importance for all variables in growing causal forest (overall ED use)}
\centering
\fontsize{9}{11}\selectfont
\makebox[\textwidth][c]{
\begin{threeparttable}
\begin{tabular}[t]{lrlr}
\toprule
\multicolumn{2}{c}{Any visit} & \multicolumn{2}{c}{Number of visits} \\
\cmidrule(l{3pt}r{3pt}){1-2} \cmidrule(l{3pt}r{3pt}){3-4}
Variable & Importance & Variable & Importance\\
\midrule
Pre-lottery SNAP benefit amount (\$) & 0.18 & Number of primary care treatable visits & 0.14\\
Age (years) & 0.18 & Sum of total ED charges & 0.13\\
Sum of total ED charges & 0.09 & Number of emergent, non-preventable visits & 0.10\\
Number of emergent, non-preventable visits & 0.09 & Number of non-emergent visits & 0.08\\
Number of primary care treatable visits & 0.08 & Number of outpatient visits & 0.08\\
Week of lottery sign up & 0.05 & Number of overall visits & 0.08\\
Number of non-emergent visits & 0.05 & Number of off-hours visits & 0.06\\
Female & 0.04 & Number of emergent, preventable visits & 0.06\\
Number of emergent, preventable visits & 0.02 & Pre-lottery SNAP benefit amount (\$) & 0.05\\
Number of visits (chronic conditions) & 0.01 & Age (years) & 0.04\\
Number of inpatient visits & 0.01 & Number of inpatient visits & 0.03\\
Pre-lottery SNAP recipient & 0.01 & Number of on-hours visits & 0.03\\
Number of outpatient visits & 0.01 & Number of visits (chronic conditions) & 0.03\\
Number of off-hours visits & 0.01 & Number of visits (injury) & 0.02\\
Number of on-hours visits & 0.01 & Week of lottery sign up & 0.01\\
Number of overall visits & 0.01 & Gave phone number & 0.01\\
Signed up self for lottery & 0.01 & Female & 0.01\\
Gave phone number & 0.01 & Number of visits (abdominal pain) & 0.00\\
Number of visits (injury) & 0.01 & Number of visits (psychiatric conditions) & 0.00\\
English as preferred language & 0.00 & Number of visits (chest pain) & 0.00\\
Pre-lottery TANF benefit amount (\$) & 0.00 & Pre-lottery TANF benefit amount (\$) & 0.00\\
Number of visits (psychiatric conditions) & 0.00 & Number of visits (back pain) & 0.00\\
Number of visits (mood disorders) & 0.00 & Number of visits (skin conditions) & 0.00\\
Number ambulatory-care-sensitive visits & 0.00 & Pre-lottery SNAP recipient & 0.00\\
Number of visits (skin conditions) & 0.00 & Number ambulatory-care-sensitive visits & 0.00\\
Number of visits (abdominal pain) & 0.00 & Number of visits (headache) & 0.00\\
Number of visits (back pain) & 0.00 & Number of visits (mood disorders) & 0.00\\
Pre-lottery TANF recipient & 0.00 & Signed up self for lottery & 0.00\\
Number of visits (chest pain) & 0.00 & Pre-lottery TANF recipient & 0.00\\
Number of visits (headache) & 0.00 & English as preferred language & 0.00\\
Provided P.O. box address & 0.00 & Provided P.O. box address & 0.00\\
\bottomrule
\end{tabular}
\begin{tablenotes}[para]
\item  \emph{Notes:} This table shows the top variable importance scores of all characteristics for growing the generalized random forests used to estimate the ITE of Medicaid coverage for overall ED visits. The variable importance measure is a simple weighted sum of the proportion of times a variable is used in a splitting step at each depth in growing the causal forest, thus, capturing how important a variable is for driving treatment effect heterogeneity. The sample consists of 24,615 individuals in the \citeasnoun{taubman2014medicaid} sample with non-missing information on pre-lottery emergency department utilization and SNAP/TANF receipt.
\end{tablenotes}
\end{threeparttable}
}
\end{table}

%% file: Tables/Tables_final/tgates-num-visit.tex
\begin{table}[t]

\caption{\label{tab:gates.num}Group average treatment effects of Medicaid on the number of ED visits}
\centering
\fontsize{9}{11}\selectfont
\makebox[\textwidth][c]{
\begin{threeparttable}
\begin{tabular}[t]{lrrrHHHr}
\toprule
\multicolumn{1}{c}{ } & \multicolumn{3}{c}{GRF estimates} & \multicolumn{3}{H}{Panel B: Linear estimates} & \multicolumn{1}{c}{ } \\
\cmidrule(l{3pt}r{3pt}){2-4} \cmidrule(l{3pt}r{3pt}){5-7}
{Group} & {GATE} & {SE} & {$p$-value} & {GATE} & {SE} & {$p$-value} & {\%~N}\\
\midrule
%LATE & 0.35 & 0.12 & 0.00 & 0.34 & 0.13 & 0.01 & 100.00\\
Female: & 0.27 & 0.16 & 0.08 & 0.23 & 0.18 & 0.21 & 0.55\\
Male: & 0.43 & 0.17 & 0.01 & 0.46 & 0.18 & 0.01 & 0.45\\
Gave phone number: No & -0.06 & 0.33 & 0.86 & -0.26 & 0.38 & 0.50 & 0.13\\
Gave phone number: Yes & 0.42 & 0.12 & 0.00 & 0.42 & 0.14 & 0.00 & 0.87\\
English as preferred language: No & 0.14 & 0.17 & 0.41 & 0.05 & 0.18 & 0.78 & 0.14\\
English as preferred language: Yes & 0.39 & 0.13 & 0.00 & 0.36 & 0.14 & 0.01 & 0.86\\
First week sign-up: No & 0.38 & 0.14 & 0.01 & 0.48 & 0.16 & 0.00 & 0.62\\
First week sign-up: Yes & 0.29 & 0.19 & 0.12 & 0.13 & 0.21 & 0.54 & 0.38\\
Pre-lottery SNAP recipient: No & 0.21 & 0.15 & 0.16 & 0.17 & 0.18 & 0.34 & 0.46\\
Pre-lottery SNAP recipient: Yes & 0.45 & 0.17 & 0.01 & 0.34 & 0.17 & 0.04 & 0.54\\
Pre-lottery TANF recipient: No & 0.34 & 0.11 & 0.00 & 0.32 & 0.13 & 0.01 & 0.98\\
Pre-lottery TANF recipient: Yes & 0.93 & 1.21 & 0.44 & 1.46 & 2.44 & 0.55 & 0.02\\
Age $\geq$ 50: No & 0.43 & 0.14 & 0.00 & 0.38 & 0.16 & 0.02 & 0.75\\
Age $\geq$ 50: Yes & 0.05 & 0.20 & 0.81 & 0.25 & 0.21 & 0.22 & 0.25\\
Two+ household members on lottery list: No & 0.32 & 0.14 & 0.02 & 0.27 & 0.15 & 0.07 & 0.80\\
Two+ household members on lottery list: Yes & 0.52 & 0.16 & 0.00 & 0.74 & 0.23 & 0.00 & 0.20\\
Any pre-lottery ED visit No & 0.22 & 0.09 & 0.01 & 0.25 & 0.08 & 0.00 & 0.69\\
Any pre-lottery ED visit: Yes & 0.67 & 0.32 & 0.03 & 0.45 & 0.31 & 0.15 & 0.31\\
Any pre-lottery on-hours ED visit: No & 0.19 & 0.09 & 0.04 & 0.26 & 0.09 & 0.00 & 0.77\\
Any pre-lottery on-hours ED visit: Yes & 0.87 & 0.41 & 0.03 & 0.67 & 0.42 & 0.11 & 0.23\\
Any pre-lottery off-hours ED visit: No & 0.22 & 0.09 & 0.02 & 0.21 & 0.09 & 0.03 & 0.81\\
Any pre-lottery off-hours ED visit: Yes & 0.95 & 0.45 & 0.04 & 0.56 & 0.45 & 0.21 & 0.19\\
Any pre-lottery emergent, non-preventable ED visit: No & 0.22 & 0.09 & 0.02 & 0.28 & 0.09 & 0.00 & 0.87\\
Any pre-lottery emergent, non-preventable ED visit: Yes & 1.15 & 0.63 & 0.07 & 0.69 & 0.63 & 0.27 & 0.13\\
Any pre-lottery emergent, preventable ED visit: No & 0.25 & 0.09 & 0.01 & 0.32 & 0.09 & 0.00 & 0.92\\
Any pre-lottery emergent, preventable ED visit: Yes & 1.50 & 0.96 & 0.12 & 1.31 & 0.90 & 0.15 & 0.08\\
Any pre-lottery primary care treatable ED visit: No & 0.18 & 0.09 & 0.04 & 0.26 & 0.09 & 0.00 & 0.81\\
Any pre-lottery primary care treatable ED visit: Yes & 1.03 & 0.46 & 0.03 & 0.72 & 0.47 & 0.13 & 0.19\\
Any pre-lottery non-emergent ED visit: No & 0.27 & 0.09 & 0.00 & 0.34 & 0.09 & 0.00 & 0.86\\
Any pre-lottery non-emergent ED visit: Yes & 0.82 & 0.61 & 0.18 & 0.26 & 0.63 & 0.68 & 0.14\\
\bottomrule
\end{tabular}
\begin{tablenotes}[para]
\item \emph{Notes}: This table reports the group average treatment effects of Medicaid based on generalized random forests. The baseline sample consists of 24,615 individuals in the \citeasnoun{taubman2014medicaid} sample with non-missing information on pre-lottery emergency department utilization and SNAP/TANF receipt.
\end{tablenotes}
\end{threeparttable}
}
\end{table}

%\begin{tablenotes}[para]
%	\item \emph{Notes}: This table reports the group average treatment effects of Medicaid based on generalized random forests (GRF) in Panel A and those based on the linear IV method's subsample analysis in Panel B. The first row reproduces the overall average effect. The baseline sample consists of 24,615 individuals in the \citeasnoun{taubman2014medicaid} sample with non-missing information on pre-lottery emergency department utilization and SNAP/TANF receipt.
%\end{tablenotes}

% old, simple def
% Any pre-lottery emergent, non-preventable ED visit: No & 0.23 & 0.09 & 0.02 & 0.29 & 0.09 & 0.00 & 0.79\\
% Any pre-lottery emergent, non-preventable ED visit: Yes & 0.80 & 0.42 & 0.06 & 0.43 & 0.44 & 0.32 & 0.21\\
% Any pre-lottery emergent, preventable ED visit: No & 0.23 & 0.10 & 0.03 & 0.29 & 0.11 & 0.01 & 0.90\\
% Any pre-lottery emergent, preventable ED visit: Yes & 1.53 & 0.67 & 0.02 & 1.03 & 0.65 & 0.11 & 0.10\\
% Any pre-lottery primary care treatable ED visit: No & 0.24 & 0.09 & 0.01 & 0.29 & 0.09 & 0.00 & 0.74\\
% Any pre-lottery primary care treatable ED visit: Yes & 0.69 & 0.36 & 0.05 & 0.41 & 0.36 & 0.25 & 0.26\\
% Any pre-lottery non-emergent ED visit: No & 0.28 & 0.10 & 0.00 & 0.36 & 0.10 & 0.00 & 0.86\\
% Any pre-lottery non-emergent ED visit: Yes & 0.76 & 0.56 & 0.18 & 0.28 & 0.57 & 0.62 & 0.14\\

%% file: Tables/Tables_final/tcomp-num-visit.tex
\begin{table}%[H]
  \caption{\label{tab:comp.num}Characteristics of individuals who increased and decreased ED use (Number of visits)}
  \centering
  \fontsize{9}{11}\selectfont
  \makebox[\textwidth][c]{
    \begin{threeparttable}
      \begin{tabular}[t]{>{\raggedright\arraybackslash}p{20em}rrr@{}l}
        \toprule
                                                   & \multicolumn{1}{c}{Increased} & \multicolumn{1}{c}{Decreased} &                                      \\
        Variable                                   & \multicolumn{1}{c}{ED use}    & \multicolumn{1}{c}{ED use}    & \multicolumn{2}{c}{Difference}       \\
        \midrule \addlinespace[1ex]
        \multicolumn{5}{l}{\hspace{-1ex}\textbf{Lottery list characteristics}}                                                                            \\
        Age (years)                                & 38.54                         & 42.88                         & -4.34                          & *** \\
        Gave phone number                          & 0.86                          & 0.89                          & -0.03                          & *** \\
        English as preferred language              & 0.87                          & 0.84                          & 0.03                           & *** \\
        Female                                     & 0.53                          & 0.58                          & -0.05                          & *** \\
        Week of lottery sign up                    & 1.59                          & 1.56                          & 0.03                           &     \\
        Provided P.O. box address                  & 0.02                          & 0.03                          & -0.01                          & *** \\
        Signed up self for lottery                 & 0.89                          & 0.92                          & -0.03                          & *** \\
        Pre-lottery SNAP recipient                 & 0.59                          & 0.36                          & 0.23                           & *** \\
        Pre-lottery SNAP benefit amount (\$)       & 1497.29                       & 817.99                        & 679.30                         & *** \\
        Pre-lottery TANF recipient                 & 0.02                          & 0.02                          & 0.00                           & *** \\
        Pre-lottery TANF benefit amount (\$)       & 96.79                         & 94.84                         & 1.95                           &     \\ \addlinespace[1ex]
        \multicolumn{5}{l}{\hspace{-1ex}\textbf{Pre-lottery ED usage}}                                                                                    \\
        Number of overall visits                   & 0.90                          & 0.30                          & 0.60                           & *** \\
        Number of inpatient visits                 & 0.10                          & 0.06                          & 0.04                           & *** \\
        Number of outpatient visits                & 0.80                          & 0.25                          & 0.55                           & *** \\
        Number of on-hours visits                  & 0.52                          & 0.19                          & 0.33                           & *** \\
        Number of off-hours visits                 & 0.38                          & 0.11                          & 0.27                           & *** \\
        Number of emergent, non-preventable visits & 0.18                          & 0.06                          & 0.12                           & *** \\
        Number of emergent, preventable visits     & 0.07                          & 0.03                          & 0.04                           & *** \\
        Number of primary care treatable visits    & 0.32                          & 0.08                          & 0.24                           & *** \\
        Number of non-emergent visits              & 0.18                          & 0.07                          & 0.11                           & *** \\
        Number ambulatory-care-sensitive visits    & 0.05                          & 0.03                          & 0.02                           & *** \\
        Number of visits (chronic conditions)      & 0.15                          & 0.07                          & 0.08                           & *** \\
        Number of visits (injury)                  & 0.20                          & 0.05                          & 0.15                           & *** \\
        Number of visits (skin conditions)         & 0.06                          & 0.01                          & 0.05                           & *** \\
        Number of visits (abdominal pain)          & 0.04                          & 0.01                          & 0.03                           & *** \\
        Number of visits (back pain)               & 0.04                          & 0.02                          & 0.02                           & *** \\
        Number of visits (chest pain)              & 0.02                          & 0.01                          & 0.01                           & *** \\
        Number of visits (headache)                & 0.03                          & 0.01                          & 0.02                           & *** \\
        Number of visits (mood disorders)          & 0.03                          & 0.01                          & 0.02                           & *** \\
        Number of visits (psychiatric conditions)  & 0.07                          & 0.03                          & 0.04                           & *** \\
        Sum of total ED charges                    & 1027.60                       & 387.86                        & 639.74                         & *** \\
        \addlinespace[0.3em]
        \hline
        \multicolumn{5}{l}{\textbf{}}                                                                                                                     \\
        N                                          & 18581.00                      & 6018.00                       & 24599                                \\
        \bottomrule
      \end{tabular}
      \begin{tablenotes}[para]
        \item \emph{Notes}: This table reports the means of individual characteristics and pre-randomization ED use for those estimated to increase and decrease ED use upon receiving Medicaid coverage based on the causal forest CATE estimates. ED use is measured as the number of ED visits. Panel A reports the means for the full sample while Panel B is limited to effects significant at the $10\%$ level. The sample consists of 24,615 individuals in the \citeasnoun{taubman2014medicaid} sample with non-missing information on pre-lottery emergency department utilization and SNAP/TANF receipt.
      \end{tablenotes}
    \end{threeparttable}
  }
\end{table}

%% file: Tables/Tables_final/itt/tmain-itt.tex
\begin{table}[t]

\caption{\label{tab:main.base.itt}Treatment effect estimates of winning the lottery on ED use}
\centering
\fontsize{9}{11}\selectfont
\makebox[\textwidth][c]{
\begin{threeparttable}
\begin{tabular}[t]{lrrrrrr}
\toprule
\multicolumn{1}{c}{ } & \multicolumn{3}{c}{GRF estimates} & \multicolumn{3}{c}{Linear estimates} \\
\cmidrule(l{3pt}r{3pt}){2-4} \cmidrule(l{3pt}r{3pt}){5-7}
{Variable} & {ATE} & {SE} & {$p$-value} & {ATE} & {SE} & {$p$-value}\\
\midrule
\addlinespace[0.3em]
\multicolumn{7}{l}{\textbf{Extensive margin}}\\
\hspace{1em}Any overall visit & 0.015 & 0.006 & 0.008 & 0.017 & 0.006 & 0.004\\
\hspace{1em}Any inpatient visit & -0.003 & 0.003 & 0.430 & -0.003 & 0.003 & 0.424\\
\hspace{1em}Any outpatient visit & 0.019 & 0.006 & 0.001 & 0.020 & 0.006 & 0.001\\
\hspace{1em}Any emergent, non-preventable visit & 0.000 & 0.005 & 0.940 & 0.002 & 0.005 & 0.731\\
\hspace{1em}Any emergent, preventable visit & 0.008 & 0.004 & 0.026 & 0.010 & 0.004 & 0.010\\
\hspace{1em}Any primary care treatable visit & 0.016 & 0.005 & 0.002 & 0.017 & 0.005 & 0.001\\
\hspace{1em}Any non-emergent visit & 0.015 & 0.005 & 0.001 & 0.016 & 0.005 & 0.001\\
\addlinespace[0.3em]
\multicolumn{7}{l}{\textbf{Intensive margin}}\\
\hspace{1em}Number of overall visits & 0.083 & 0.027 & 0.002 & 0.093 & 0.026 & 0.000\\
\hspace{1em}Number of inpatient visits & -0.004 & 0.006 & 0.519 & -0.004 & 0.006 & 0.516\\
\hspace{1em}Number of outpatient visits & 0.089 & 0.024 & 0.000 & 0.096 & 0.024 & 0.000\\
\hspace{1em}Number of emergent, non-preventable visits & 0.011 & 0.008 & 0.162 & 0.010 & 0.008 & 0.224\\
\hspace{1em}Number of emergent, preventable visits & 0.008 & 0.004 & 0.056 & 0.009 & 0.004 & 0.034\\
\hspace{1em}Number of primary care treatable visits & 0.042 & 0.011 & 0.000 & 0.042 & 0.011 & 0.000\\
\hspace{1em}Number of non-emergent visits & 0.022 & 0.008 & 0.007 & 0.026 & 0.008 & 0.002\\
\bottomrule
\end{tabular}
\begin{tablenotes}[para]
\item \emph{Notes}: This table reports the estimates of winning the lottery on ED use based on generalized random forests and a linear model. The sample consists of 24,615 individuals in the \citeasnoun{taubman2014medicaid} sample with non-missing information on pre-lottery emergency department utilization and SP/TANF receipt.
\end{tablenotes}
\end{threeparttable}
}
\end{table}

% old, simple def
% \hspace{1em}Any emergent, non-preventable visit & 0.000 & 0.005 & 0.940 & 0.001 & 0.005 & 0.870\\
% \hspace{1em}Any emergent, preventable visit & 0.008 & 0.004 & 0.026 & 0.009 & 0.004 & 0.022\\
% \hspace{1em}Any primary care treatable visit & 0.016 & 0.005 & 0.002 & 0.016 & 0.005 & 0.002\\
% \hspace{1em}Any non-emergent visit & 0.015 & 0.005 & 0.001 & 0.017 & 0.005 & 0.000\\

%% file: Tables/Tables_final/itt/tqte-itt.tex
\begin{table}[t]
\caption{\label{tab:qte.base.itt}Empirical quantiles of the distribution of individualized treatment effects of winning the lottery on ED use}
\centering
\fontsize{9}{11}\selectfont
\begin{threeparttable}
\begin{tabular}[t]{lrrrrrr}
\toprule
{Variable} & {ATE} & {Min} & {25\%} & {50\%} & {75\%} & {Max}\\
\midrule
\addlinespace[0.3em]
\multicolumn{7}{l}{\textbf{Extensive margin}}\\
\hspace{1em}Any overall visit & 0.015 & -0.112 & -0.007 & 0.017 & 0.039 & 0.159\\
\hspace{1em}Any inpatient visit & -0.003 & -0.072 & -0.012 & -0.002 & 0.007 & 0.076\\
\hspace{1em}Any outpatient visit & 0.019 & -0.107 & -0.003 & 0.018 & 0.041 & 0.159\\
\hspace{1em}Any emergent, non-preventable visit & 0.000 & -0.103 & -0.016 & 0.001 & 0.018 & 0.096\\
\hspace{1em}Any emergent, preventable visit & 0.008 & -0.061 & -0.003 & 0.007 & 0.019 & 0.095\\
\hspace{1em}Any primary care treatable visit & 0.016 & -0.089 & -0.003 & 0.016 & 0.035 & 0.137\\
\hspace{1em}Any non-emergent visit & 0.015 & -0.077 & -0.001 & 0.014 & 0.031 & 0.116\\
\addlinespace[0.3em]
\multicolumn{7}{l}{\textbf{Intensive margin}}\\
\hspace{1em}Number of overall visits & 0.083 & -0.427 & 0.000 & 0.061 & 0.145 & 1.205\\
\hspace{1em}Number of inpatient visits & -0.004 & -0.148 & -0.018 & -0.003 & 0.011 & 0.248\\
\hspace{1em}Number of outpatient visits & 0.089 & -0.347 & 0.009 & 0.065 & 0.146 & 1.154\\
\hspace{1em}Number of emergent, non-preventable visits & 0.011 & -0.116 & -0.015 & 0.004 & 0.026 & 0.576\\
\hspace{1em}Number of emergent, preventable visits & 0.008 & -0.270 & -0.003 & 0.006 & 0.018 & 0.238\\
\hspace{1em}Number of primary care treatable visits & 0.042 & -0.118 & 0.003 & 0.031 & 0.067 & 0.686\\
\hspace{1em}Number of non-emergent visits & 0.022 & -0.218 & -0.002 & 0.014 & 0.034 & 0.563\\
\bottomrule
\end{tabular}
\begin{tablenotes}[para]
\item \emph{Notes}: This table reports selected quantiles of the individualized treatment effects of winning the lottery (and being invited to apply for Medicaid) on ED use based on generalized random forests. The first column reports the average effect (intent-to-treat effect). The baseline sample consists of 24,615 individuals in the \citeasnoun{taubman2014medicaid} sample with non-missing information on pre-lottery emergency department utilization and SNAP/TANF receipt.
\end{tablenotes}
\end{threeparttable}
\end{table}

%% file: Tables/Tables_final/itt/tgates-any-visit-itt.tex
\begin{table}[H]

\caption{\label{tab:gates.any.itt}GATE estimates of winning the lottery on the propensity of ED use}
\centering
\fontsize{9}{11}\selectfont
\makebox[\textwidth][c]{
\begin{threeparttable}
\begin{tabular}[t]{lrrrrrrr}
\toprule
\multicolumn{1}{c}{ } & \multicolumn{3}{c}{Panel A: GRF estimates} & \multicolumn{3}{c}{Panel B: Linear estimates} & \multicolumn{1}{c}{ } \\
\cmidrule(l{3pt}r{3pt}){2-4} \cmidrule(l{3pt}r{3pt}){5-7}
{Group} & {GATE} & {SE} & {$p$-value} & {GATE} & {SE} & {$p$-value} & {\%~N}\\
\midrule
ATE & 0.02 & 0.01 & 0.01 & 0.02 & 0.01 & 0.00 & 100.00\\
Female: & 0.00 & 0.01 & 0.72 & 0.01 & 0.01 & 0.34 & 0.55\\
Male: & 0.03 & 0.01 & 0.00 & 0.03 & 0.01 & 0.00 & 0.45\\
Gave phone number: No & 0.01 & 0.02 & 0.54 & 0.01 & 0.02 & 0.66 & 0.13\\
Gave phone number: Yes & 0.02 & 0.01 & 0.01 & 0.02 & 0.01 & 0.00 & 0.87\\
English as preferred language: No & 0.01 & 0.01 & 0.59 & 0.00 & 0.01 & 0.73 & 0.14\\
English as preferred language: Yes & 0.02 & 0.01 & 0.01 & 0.02 & 0.01 & 0.01 & 0.86\\
First week sign-up: No & 0.01 & 0.01 & 0.06 & 0.02 & 0.01 & 0.01 & 0.62\\
First week sign-up: Yes & 0.02 & 0.01 & 0.06 & 0.01 & 0.01 & 0.16 & 0.38\\
Pre-lottery SNAP recipient: No & 0.00 & 0.01 & 0.94 & 0.00 & 0.01 & 0.62 & 0.46\\
Pre-lottery SNAP recipient: Yes & 0.03 & 0.01 & 0.00 & 0.02 & 0.01 & 0.01 & 0.54\\
Pre-lottery TANF recipient: No & 0.02 & 0.01 & 0.01 & 0.02 & 0.01 & 0.00 & 0.98\\
Pre-lottery TANF recipient: Yes & -0.01 & 0.04 & 0.78 & 0.01 & 0.04 & 0.82 & 0.02\\
Age $\geq$ 50: No & 0.02 & 0.01 & 0.00 & 0.02 & 0.01 & 0.00 & 0.75\\
Age $\geq$ 50: Yes & -0.01 & 0.01 & 0.67 & 0.00 & 0.01 & 0.77 & 0.25\\
Two+ household members on lottery list: No & 0.01 & 0.01 & 0.06 & 0.01 & 0.01 & 0.06 & 0.80\\
Two+ household members on lottery list: Yes & 0.03 & 0.01 & 0.02 & 0.04 & 0.01 & 0.00 & 0.20\\
Any pre-lottery ED visit No & 0.01 & 0.01 & 0.07 & 0.02 & 0.01 & 0.02 & 0.69\\
Any pre-lottery ED visit: Yes & 0.02 & 0.01 & 0.05 & 0.02 & 0.01 & 0.09 & 0.31\\
Any pre-lottery on-hours ED visit: No & 0.01 & 0.01 & 0.03 & 0.02 & 0.01 & 0.01 & 0.77\\
Any pre-lottery on-hours ED visit: Yes & 0.02 & 0.01 & 0.11 & 0.02 & 0.01 & 0.13 & 0.23\\
Any pre-lottery off-hours ED visit: No & 0.01 & 0.01 & 0.04 & 0.01 & 0.01 & 0.03 & 0.81\\
Any pre-lottery off-hours ED visit: Yes & 0.02 & 0.01 & 0.08 & 0.02 & 0.01 & 0.16 & 0.19\\
Any pre-lottery emergent, non-preventable ED visit: No & 0.01 & 0.01 & 0.07 & 0.01 & 0.01 & 0.03 & 0.79\\
Any pre-lottery emergent, non-preventable ED visit: Yes & 0.03 & 0.01 & 0.02 & 0.03 & 0.01 & 0.07 & 0.21\\
Any pre-lottery emergent, preventable ED visit: No & 0.01 & 0.01 & 0.02 & 0.02 & 0.01 & 0.00 & 0.90\\
Any pre-lottery emergent, preventable ED visit: Yes & 0.03 & 0.02 & 0.15 & 0.03 & 0.02 & 0.13 & 0.10\\
Any pre-lottery primary care treatable ED visit: No & 0.01 & 0.01 & 0.10 & 0.01 & 0.01 & 0.03 & 0.74\\
Any pre-lottery primary care treatable ED visit: Yes & 0.03 & 0.01 & 0.02 & 0.03 & 0.01 & 0.05 & 0.26\\
Any pre-lottery non-emergent ED visit: No & 0.01 & 0.01 & 0.04 & 0.02 & 0.01 & 0.01 & 0.86\\
Any pre-lottery non-emergent ED visit: Yes & 0.03 & 0.02 & 0.07 & 0.02 & 0.02 & 0.18 & 0.14\\
\bottomrule
\end{tabular}
\begin{tablenotes}[para]
\item \emph{Notes}: This table reports the GATE estimates of winning the lottery based on generalized random forests in Panel A and the linear model GATEs in Panel B. The overall effect is reproduced in the first row. The sample consists of 24,615 individuals in the \citeasnoun{taubman2014medicaid} sample with non-missing information on pre-lottery emergency department utilization and SNAP/TANF receipt.
\end{tablenotes}
\end{threeparttable}
}
\end{table}

%% file: Tables/Tables_final/itt/tgates-num-visit-itt.tex
\begin{table}[t]

\caption{\label{tab:gates.num.itt}GATE estimates of winning the lottery on the number of ED visits}
\centering
\fontsize{9}{11}\selectfont
\makebox[\textwidth][c]{
\begin{threeparttable}
\begin{tabular}[t]{lrrrrrrr}
\toprule
\multicolumn{1}{c}{ } & \multicolumn{3}{c}{Panel A: GRF estimates} & \multicolumn{3}{c}{Panel B: Linear estimates} & \multicolumn{1}{c}{ } \\
\cmidrule(l{3pt}r{3pt}){2-4} \cmidrule(l{3pt}r{3pt}){5-7}
{Group} & {GATE} & {SE} & {$p$-value} & {GATE} & {SE} & {$p$-value} & {\%~N}\\
\midrule
ATE & 0.08 & 0.03 & 0.00 & 0.08 & 0.03 & 0.01 & 100.00\\
Female: & 0.05 & 0.03 & 0.19 & 0.05 & 0.04 & 0.21 & 0.55\\
Male: & 0.13 & 0.04 & 0.00 & 0.12 & 0.05 & 0.01 & 0.45\\
Gave phone number: No & -0.04 & 0.08 & 0.62 & -0.06 & 0.09 & 0.50 & 0.13\\
Gave phone number: Yes & 0.10 & 0.03 & 0.00 & 0.10 & 0.03 & 0.00 & 0.87\\
English as preferred language: No & 0.03 & 0.03 & 0.43 & 0.01 & 0.04 & 0.78 & 0.14\\
English as preferred language: Yes & 0.09 & 0.03 & 0.00 & 0.09 & 0.04 & 0.01 & 0.86\\
First week sign-up: No & 0.09 & 0.03 & 0.00 & 0.11 & 0.04 & 0.00 & 0.62\\
First week sign-up: Yes & 0.06 & 0.05 & 0.17 & 0.03 & 0.06 & 0.54 & 0.38\\
Pre-lottery SNAP recipient: No & 0.04 & 0.03 & 0.17 & 0.03 & 0.03 & 0.34 & 0.46\\
Pre-lottery SNAP recipient: Yes & 0.12 & 0.04 & 0.01 & 0.11 & 0.05 & 0.04 & 0.54\\
Pre-lottery TANF recipient: No & 0.08 & 0.03 & 0.00 & 0.08 & 0.03 & 0.01 & 0.98\\
Pre-lottery TANF recipient: Yes & 0.07 & 0.27 & 0.78 & 0.17 & 0.29 & 0.55 & 0.02\\
Age $\geq$ 50: No & 0.10 & 0.03 & 0.00 & 0.09 & 0.04 & 0.02 & 0.75\\
Age $\geq$ 50: Yes & 0.03 & 0.05 & 0.50 & 0.07 & 0.06 & 0.23 & 0.25\\
Two+ household members on lottery list: No & 0.07 & 0.03 & 0.02 & 0.07 & 0.04 & 0.07 & 0.80\\
Two+ household members on lottery list: Yes & 0.12 & 0.03 & 0.00 & 0.15 & 0.05 & 0.00 & 0.20\\
Any pre-lottery ED visit No & 0.05 & 0.02 & 0.01 & 0.06 & 0.02 & 0.00 & 0.69\\
Any pre-lottery ED visit: Yes & 0.15 & 0.07 & 0.04 & 0.12 & 0.09 & 0.15 & 0.31\\
Any pre-lottery on-hours ED visit: No & 0.05 & 0.02 & 0.01 & 0.06 & 0.02 & 0.00 & 0.77\\
Any pre-lottery on-hours ED visit: Yes & 0.19 & 0.09 & 0.04 & 0.17 & 0.11 & 0.12 & 0.23\\
Any pre-lottery off-hours ED visit: No & 0.05 & 0.02 & 0.02 & 0.05 & 0.02 & 0.03 & 0.81\\
Any pre-lottery off-hours ED visit: Yes & 0.22 & 0.11 & 0.04 & 0.16 & 0.12 & 0.21 & 0.19\\
Any pre-lottery emergent, non-preventable ED visit: No & 0.06 & 0.02 & 0.01 & 0.07 & 0.02 & 0.00 & 0.87\\
Any pre-lottery emergent, non-preventable ED visit: Yes & 0.23 & 0.14 & 0.11 & 0.18 & 0.17 & 0.28 & 0.13\\
Any pre-lottery emergent, preventable ED visit: No & 0.06 & 0.02 & 0.00 & 0.08 & 0.02 & 0.00 & 0.92\\
Any pre-lottery emergent, preventable ED visit: Yes & 0.30 & 0.22 & 0.17 & 0.36 & 0.25 & 0.15 & 0.08\\
Any pre-lottery primary care treatable ED visit: No & 0.05 & 0.02 & 0.03 & 0.06 & 0.02 & 0.00 & 0.81\\
Any pre-lottery primary care treatable ED visit: Yes & 0.24 & 0.11 & 0.03 & 0.19 & 0.13 & 0.13 & 0.19\\
Any pre-lottery non-emergent ED visit: No & 0.07 & 0.02 & 0.00 & 0.08 & 0.02 & 0.00 & 0.86\\
Any pre-lottery non-emergent ED visit: Yes & 0.14 & 0.14 & 0.30 & 0.07 & 0.16 & 0.68 & 0.14\\
\bottomrule
\end{tabular}
\begin{tablenotes}[para]
\item \emph{Notes}: This table reports the GATE estimates of winning the lottery based on generalized random forests in Panel A and the GATE based on a linear model in Panel B. The overall effect is reproduced in the first row. The sample consists of 24,615 individuals in the \citeasnoun{taubman2014medicaid} sample with non-missing information on pre-lottery emergency department utilization and SNAP/TANF receipt.
\end{tablenotes}
\end{threeparttable}
}
\end{table}

%% file: Tables/Tables_final/itt/tcomp-any-visit-itt.tex
% \begin{landscape}
  \begin{table}[t]
    \caption{\label{tab:comp.any.itt}Characteristics of individuals who increased and decreased ED use upon winning the lottery}
    \centering
    \fontsize{8}{10}\selectfont
    \makebox[\textwidth][c]{
      \begin{threeparttable}
      \begin{tabular}[t]{>{\raggedright\arraybackslash}p{20em}rrr@{}l}
          \toprule
                                                   & \multicolumn{1}{c}{Increased} & \multicolumn{1}{c}{Decreased} &  &                                    \\
        Variable                                   & \multicolumn{1}{c}{ED use}    & \multicolumn{1}{c}{ED use}    & \multicolumn{2}{c}{Difference}       \\
        \midrule \addlinespace[1ex]
        \multicolumn{5}{l}{\hspace{-1ex}\textbf{Lottery list characteristics}}                                                                            \\
          Age (years)                                & 38.81                                    & 41.32                                               & -2.51      & ***                                              \\
          Gave phone number                          & 0.87                                     & 0.87                                                & 0.00       &                                                  \\
          English as preferred language              & 0.87                                     & 0.84                                                & 0.03       & ***                                              \\
          Female                                     & 0.51                                     & 0.63                                                & -0.12      & ***                                              \\
          Week of lottery sign up                    & 1.57                                     & 1.61                                                & -0.04      & **                                               \\
          Provided P.O. box address                  & 0.02                                     & 0.03                                                & -0.01      & ***                                              \\
          Signed up self for lottery                 & 0.89                                     & 0.92                                                & -0.03      & ***                                              \\
          Pre-lottery SNAP recipient                 & 0.62                                     & 0.35                                                & 0.27       & ***                                              \\
          Pre-lottery SNAP benefit amount (\$)       & 1613.82                                  & 715.39                                              & 898.43     & ***                                              \\
          Pre-lottery TANF recipient                 & 0.02                                     & 0.01                                                & 0.01       & ***                                              \\
          Pre-lottery TANF benefit amount (\$)       & 110.21                                   & 65.85                                               & 44.36      & ***                                              \\
         \addlinespace[1ex] \multicolumn{5}{l}{\hspace{-1ex}\textbf{Pre-lottery ED usage}}                                                                                    \\
          Number of overall visits                   & 0.87                                     & 0.52                                                & 0.35       & ***                                              \\
          Number of inpatient visits                 & 0.09                                     & 0.07                                                & 0.02       & ***                                              \\
          Number of outpatient visits                & 0.78                                     & 0.45                                                & 0.33       & ***                                              \\
          Number of on-hours visits                  & 0.50                                     & 0.32                                                & 0.18       & ***                                              \\
          Number of off-hours visits                 & 0.37                                     & 0.21                                                & 0.16       & ***                                              \\
          Number of emergent, non-preventable visits & 0.18                                     & 0.10                                                & 0.08       & ***                                              \\
          Number of emergent, preventable visits     & 0.07                                     & 0.05                                                & 0.02       & ***                                              \\
          Number of primary care treatable visits    & 0.30                                     & 0.18                                                & 0.12       & ***                                              \\
          Number of non-emergent visits              & 0.18                                     & 0.11                                                & 0.07       & ***                                              \\
          Number ambulatory-care-sensitive visits    & 0.05                                     & 0.04                                                & 0.01       & ***                                              \\
          Number of visits (chronic conditions)      & 0.14                                     & 0.12                                                & 0.02       & ***                                              \\
          Number of visits (injury)                  & 0.20                                     & 0.10                                                & 0.10       & ***                                              \\
          Number of visits (skin conditions)         & 0.05                                     & 0.03                                                & 0.02       & ***                                              \\
          Number of visits (abdominal pain)          & 0.04                                     & 0.02                                                & 0.02       & ***                                              \\
          Number of visits (back pain)               & 0.04                                     & 0.02                                                & 0.02       & ***                                              \\
          Number of visits (chest pain)              & 0.02                                     & 0.02                                                & 0.00       & ***                                              \\
          Number of visits (headache)                & 0.03                                     & 0.02                                                & 0.01       & ***                                              \\
          Number of visits (mood disorders)          & 0.02                                     & 0.03                                                & -0.01      &                                                  \\
          Number of visits (psychiatric conditions)  & 0.06                                     & 0.05                                                & 0.01       &                                                  \\
          Sum of total ED charges                    & 1002.57                                  & 615.50                                              & 387.07     & ***                                              \\
          \addlinespace[0.3em]
          \hline
          %\multicolumn{7}{l}{\textbf{}}                                                                                                                                                                               \\
          N                                          & 16871                                    & 7742                                                & 24613                 &                                        \\
          \bottomrule
        \end{tabular}
        \begin{tablenotes}[para]
          \item \emph{Notes}: This table reports the means of individual characteristics and pre-randomization ED use for those estimated to increase and decrease ED use upon winning the lottery based on the causal forest CATE estimates. ED use is measured as the propensity to use the ED. Panel A reports the means for the full sample while Panel B is limited to individuals with effects significant at the $10\%$ level. The sample consists of 24,615 individuals in the \citeasnoun{taubman2014medicaid} sample with non-missing information on pre-lottery emergency department utilization and SNAP/TANF receipt.
        \end{tablenotes}
      \end{threeparttable}
    }
  \end{table}
% \end{landscape}

%% file: Tables/Tables_final/itt/tcomp-num-visit-itt.tex
\begin{table}[H]
  \caption{\label{tab:comp.num.itt}Characteristics of individuals who increased and decreased ED use upon winning the lottery (Number of visits)}
  \centering
  \fontsize{9}{11}\selectfont
  \makebox[\textwidth][c]{
    \begin{threeparttable}
      \begin{tabular}[t]{>{\raggedright\arraybackslash}p{20em}rrr@{}l}
          \toprule
                                                   & \multicolumn{1}{c}{Increased} & \multicolumn{1}{c}{Decreased} &   &                                   \\
        Variable                                   & \multicolumn{1}{c}{ED use}    & \multicolumn{1}{c}{ED use}    & \multicolumn{2}{c}{Difference}       \\
        \midrule \addlinespace[1ex]
        \multicolumn{5}{l}{\hspace{-1ex}\textbf{Lottery list characteristics}}                                                                            \\
        Age (years)                                & 38.52                                    & 42.89                                               & -4.37  &***    \\
        Gave phone number                          & 0.87                                     & 0.88                                                & -0.01  &***    \\
        English as preferred language              & 0.87                                     & 0.84                                                & 0.03   &***     \\
        Female                                     & 0.53                                     & 0.59                                                & -0.06  &***    \\
        Week of lottery sign up                    & 1.59                                     & 1.55                                                & 0.04   &     \\
        Provided P.O. box address                  & 0.02                                     & 0.03                                                & -0.01  &***    \\
        Signed up self for lottery                 & 0.89                                     & 0.92                                                & -0.03  &***    \\
        Pre-lottery SNAP recipient                 & 0.59                                     & 0.36                                                & 0.23   &***     \\
        Pre-lottery SNAP benefit amount (\$)       & 1502.03                                  & 813.33                                              & 688.70 &***   \\
        Pre-lottery TANF recipient                 & 0.02                                     & 0.02                                                & 0.00   &**      \\
        Pre-lottery TANF benefit amount (\$)       & 95.68                                    & 98.23                                               & -2.55  &     \\
         \addlinespace[1ex] \multicolumn{5}{l}{\hspace{-1ex}\textbf{Pre-lottery ED usage}}                                                                   \\
        Number of overall visits                   & 0.90                                     & 0.32                                                & 0.58   &***     \\
        Number of inpatient visits                 & 0.10                                     & 0.06                                                & 0.04   &***     \\
        Number of outpatient visits                & 0.80                                     & 0.26                                                & 0.54   &***     \\
        Number of on-hours visits                  & 0.52                                     & 0.21                                                & 0.31   &***     \\
        Number of off-hours visits                 & 0.38                                     & 0.12                                                & 0.26   &***     \\
        Number of emergent, non-preventable visits & 0.18                                     & 0.07                                                & 0.11   &***     \\
        Number of emergent, preventable visits     & 0.07                                     & 0.03                                                & 0.04   &***     \\
        Number of primary care treatable visits    & 0.32                                     & 0.08                                                & 0.24   &***     \\
        Number of non-emergent visits              & 0.18                                     & 0.08                                                & 0.10   &***     \\
        Number ambulatory-care-sensitive visits    & 0.05                                     & 0.03                                                & 0.02   &***     \\
        Number of visits (chronic conditions)      & 0.15                                     & 0.08                                                & 0.07   &***     \\
        Number of visits (injury)                  & 0.20                                     & 0.06                                                & 0.14   &***     \\
        Number of visits (skin conditions)         & 0.06                                     & 0.01                                                & 0.05   &***     \\
        Number of visits (abdominal pain)          & 0.04                                     & 0.01                                                & 0.03   &***     \\
        Number of visits (back pain)               & 0.04                                     & 0.02                                                & 0.02   &***     \\
        Number of visits (chest pain)              & 0.02                                     & 0.01                                                & 0.01   &***     \\
        Number of visits (headache)                & 0.03                                     & 0.01                                                & 0.02   &***     \\
        Number of visits (mood disorders)          & 0.03                                     & 0.02                                                & 0.01   &***     \\
        Number of visits (psychiatric conditions)  & 0.07                                     & 0.03                                                & 0.04   &***     \\
        Sum of total ED charges                    & 1022.87                                  & 411.30                                              & 611.57 &***   \\
        \addlinespace[0.3em]
        \hline
       % \multicolumn{7}{l}{\textbf{}}                                                                                                                                                                               \\
        N                                          & 18494.00                                 & 6105.00                                             & 24599      &       \\
        \bottomrule
      \end{tabular}
      \begin{tablenotes}[para]
        \item \emph{Notes}: This table reports the means of individual characteristics and pre-randomization ED use for those estimated to increase and decrease ED use upon winning the lottery based on the causal forest CATE estimates. ED use is measured as the number of total visits. Panel A reports the means for the full sample while Panel B is limited to individuals with effects significant at the $10\%$ level. The sample consists of 24,615 individuals in the \citeasnoun{taubman2014medicaid} sample with non-missing information on pre-lottery emergency department utilization and SNAP/TANF receipt.
      \end{tablenotes}
    \end{threeparttable}
  }
\end{table}

%% file: Tables/Tables_final/itt/tvarimp-visit-itt.tex
\begin{table}[t]
\caption{\label{tab:varimp.visit.itt}Variable importance for all variables in growing causal forest (overall ED use)}
\centering
\fontsize{9}{11}\selectfont
\makebox[\textwidth][c]{
\begin{threeparttable}
\begin{tabular}[t]{lrlr}
\toprule
\multicolumn{2}{c}{Any visit} & \multicolumn{2}{c}{Number of visits} \\
\cmidrule(l{3pt}r{3pt}){1-2} \cmidrule(l{3pt}r{3pt}){3-4}
Variable & Importance & Variable & Importance\\
\midrule
Pre-lottery SNAP benefit amount (\$) & 0.23 & Number of primary care treatable visits & 0.13\\
Age (years) & 0.16 & Sum of total ED charges & 0.13\\
Sum of total ED charges & 0.09 & Number of emergent, non-preventable visits & 0.11\\
Number of emergent, non-preventable visits & 0.08 & Number of outpatient visits & 0.09\\
Number of primary care treatable visits & 0.08 & Number of overall visits & 0.08\\
Female & 0.04 & Number of non-emergent visits & 0.08\\
Number of non-emergent visits & 0.04 & Number of emergent, preventable visits & 0.06\\
Week of lottery sign up & 0.04 & Number of off-hours visits & 0.06\\
Pre-lottery SNAP recipient & 0.02 & Pre-lottery SNAP benefit amount (\$) & 0.06\\
Number of emergent, preventable visits & 0.02 & Age (years) & 0.03\\
Number of visits (chronic conditions) & 0.01 & Number of inpatient visits & 0.03\\
Number of inpatient visits & 0.01 & Number of on-hours visits & 0.03\\
Number of outpatient visits & 0.01 & Number of visits (chronic conditions) & 0.02\\
Number of off-hours visits & 0.01 & Number of visits (injury) & 0.02\\
Number of overall visits & 0.01 & Week of lottery sign up & 0.01\\
Number of on-hours visits & 0.01 & Gave phone number & 0.01\\
Gave phone number & 0.01 & Female & 0.01\\
Signed up self for lottery & 0.01 & Number of visits (abdominal pain) & 0.00\\
Number of visits (injury) & 0.01 & Number of visits (psychiatric conditions) & 0.00\\
Pre-lottery TANF benefit amount (\$) & 0.00 & Pre-lottery SNAP recipient & 0.00\\
English as preferred language & 0.00 & Number of visits (chest pain) & 0.00\\
Number of visits (mood disorders) & 0.00 & Pre-lottery TANF benefit amount (\$) & 0.00\\
Number of visits (psychiatric conditions) & 0.00 & Number of visits (skin conditions) & 0.00\\
Number ambulatory-care-sensitive visits & 0.00 & Number of visits (back pain) & 0.00\\
Number of visits (abdominal pain) & 0.00 & Number of visits (headache) & 0.00\\
Number of visits (skin conditions) & 0.00 & Number of visits (mood disorders) & 0.00\\
Pre-lottery TANF recipient & 0.00 & Number ambulatory-care-sensitive visits & 0.00\\
Number of visits (chest pain) & 0.00 & Signed up self for lottery & 0.00\\
Number of visits (back pain) & 0.00 & Pre-lottery TANF recipient & 0.00\\
Number of visits (headache) & 0.00 & English as preferred language & 0.00\\
Provided P.O. box address & 0.00 & Provided P.O. box address & 0.00\\
\bottomrule
\end{tabular}
\begin{tablenotes}[para]
\item  \emph{Notes:} This table shows the top variable importance scores of all characteristics for growing the generalized random forests used to estimate the ITE of winning the lottery for overall ED visits. The variable importance measure is a simple weighted sum of the proportion of times a variable is used in a splitting step at each depth in growing the causal forest, thus, capturing how important a variable is for driving treatment effect heterogeneity. The sample consists of 24,615 individuals in the \citeasnoun{taubman2014medicaid} sample with non-missing information on pre-lottery emergency department utilization and SNAP/TANF receipt.
\end{tablenotes}
\end{threeparttable}
}
\end{table}